\documentclass[a4paper,11pt]{article}

\usepackage{jheppub} 

\usepackage[T1]{fontenc} 

\title{\boldmath Differentiating Dilatons from Axions by their mixing
with photons}

\author[a]{Ankur Chaubey,}
\author[a]{Manoj K. Jaiswal,}
\author[a]{Damini Singh,}
\author[b]{Venktesh Singh,}
\author[a,1]{and Avijit K. Ganguly}\note{Corresponding author.}

\affiliation[a]{Institute of Science, Dept. of  Physics, \\Banaras Hindu University, Varanasi-221005, U.P, India }
\affiliation[b]{Central University of south Bihar (Gaya), \\ Bihar 824236, India}

\emailAdd{ankur.chaubey@bhu.ac.in}
\emailAdd{manojau.87@gmail.com}
\emailAdd{damini.songh13@bhu.ac.in}
\emailAdd{venktesh@cusb.ac.in}
\emailAdd{avijitk@hotmail.com}

\abstract{According  to the model ($\Lambda$CDM), based on deep cosmological observations, the current universe
is constituted of 5$\%$ baryonic matter and 25 $\%$ non-baryonic cold dark matter
(of speculative origin).
These include quanta of scalar filed like dilaton($\phi$) of 
 scale  symmetry   origin and quanta of pseudoscalar field of  extra standard model
symmetry ( Peccei-Quinn) origin, like axion ($\phi'$). 
These  fields    couple  to di-photons through dim-5 operators. In  magnetized medium,  they in principle can interact  with the  three degrees of freedom (two transverse ($A_{\parallel,\perp}$) and one longitudinal ($A_{L}$))
of  photon($\gamma$)  as long as the total spin is conserved. Because  of  intrinsic spin being  zero,  both $\phi$ and $\phi'$ could in principle have interacted with   $A_{L}$,
 (having $s_{z}=0$).
However,  out of $\phi$ and $\phi'$ only one interacts with  $A_{L}$. Furthermore, the ambient  external magnetic field and media, breaks the intrinsic Lorentz symmetry of the system
 invoking  Charge conjugation, Parity and Time reversal symmetries, we analyse
 the mixing dynamics of $\phi\gamma$ and $\phi'\gamma$ systems  and the  structural  {\it difference} of  their mixing pattern. 
 The strength  of electromagnetic  (EM) signals due to $\phi\gamma$ and $\phi'\gamma$ mixing as a result would be {\it different}. 
We conclude by  commenting  on the possibility of  detecting  this {\it difference} --in  polarimetric observables the EMS-- using the existing  space-borne detectors.}

\begin{document} 
\maketitle
\flushbottom

\section{Introduction}
\begin{figure}[h!]
\begin{center}
\includegraphics[scale=.55]{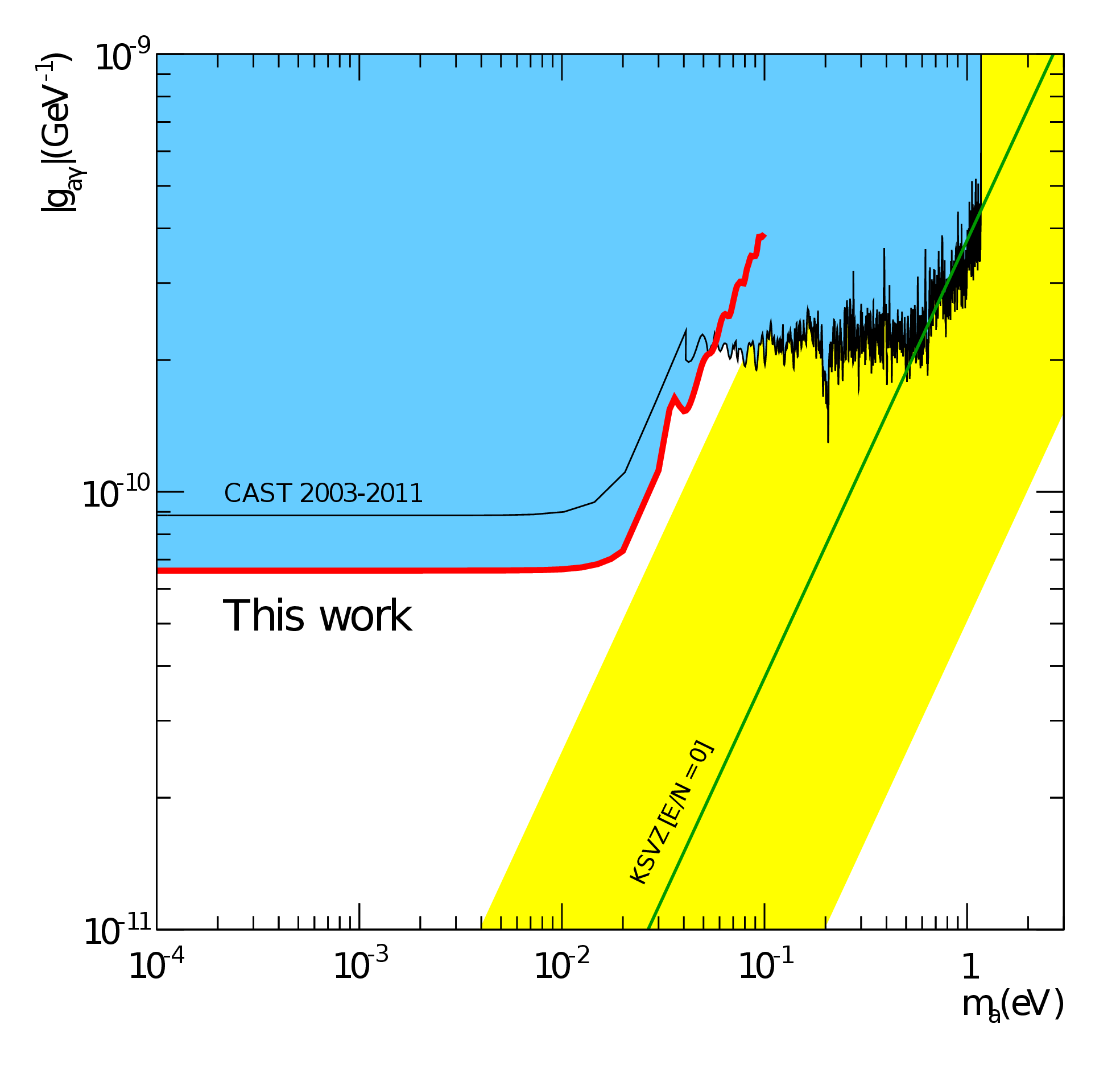}
\end{center}
\caption{Exclusion plot of $g_{\phi_i \gamma\gamma}$ vs $m_{\phi_i }$. The 
unshaded regions are yet to be observationally excluded \cite{CAST_nat}.}   
\label{excu_plot}
\end{figure}

Scalar or pseudoscalar bosons like dilaton ($\phi(x)$) \cite{Kim, donoghue}  or  axion 
($\phi'(x)$) \cite{Peccei,Weinberg,Wilczek,Mohapatra}  are  postulated to arise out
of symmetry breaking  through quantum effects in field theory.
That apart, unified theories like string theory \cite{Witten, arvan, Conlon, Marsh, Maharana, Conlon-Quevedo} also predict existence of similar particles, that may  appear through moduli compactification.  
Out of the two bosons mentioned above,  dilaton  is supposed to restores the scaling symmetry and  QCD axion is supposed to cure the $U_{A}$(1)  anomaly, lastly-- string theory axion is supposed to break the shift symmetry of the theory. Both of these particles, though are still illusive to experimental verification, however offer remarkable solutions to the outstanding problems of physics. This makes any effort towards their verification a highly sought after 
activity. Thus detection of EM signals of their presence has turned out to be a coveted activity all over the globe.\\
\indent
In terms of electromagnetic  field strength tensors $f_{\mu\nu}$ (defined in terms of gauge potentials $A_{\mu}$, as,  $f_{\mu\nu}(k)= k_{\mu}A_{\nu}-k_{\nu}A_{\mu}$) and its dual ${\tilde{f}}^{\mu\nu}= \frac{1}{2}\epsilon^{\lambda\sigma\mu\nu}f_{\lambda\sigma}$, the interaction of photons $(\gamma)$ with $\phi$ or $\phi'$ is  governed by dimension-five anomalous  interaction  Lagrangian of the form,
\begin{eqnarray}
L_{int,\phi}= g_{\phi\gamma\gamma} \phi f^{\mu\nu}f_{\mu\nu},  \mbox { ~~ is for scalar photon and  ~~~} 
L_{int,\phi'}=g_{\phi'\gamma\gamma} \phi' {\tilde{f}}^{\mu\nu}f_{\mu\nu},
\label{LintPhi}
\end{eqnarray} 
 \noindent
is for pseudoscalar photon interaction, where  symbols  $g_{\phi\gamma\gamma}$ or $g_{\phi'\gamma\gamma}$   appearing in  eqn. (\ref{LintPhi}) are the  anomaly induced coupling   constants between $\phi$ or $\phi'$ with photons. 
We denote  $\phi$ and $\phi'$ collectively  with a subscript $i$  as $\phi_i$ ( such that for scalars $\phi_{i}= \phi$ and for pseudo-scalars $\phi_{i}= \phi'$ ).
As a result of this anomalous coupling, the life time of these particles, having mass $m_{\phi_i}$, against decay into    two photons,  
 is given by, 
\begin{eqnarray}
\tau_{\phi_i \gamma\gamma} \sim \frac{1}{g^2_{\phi_i \gamma\gamma} m^3_{\phi_i}}.
\label{lifetime}
\end{eqnarray}
\indent
As an aside we note that,
except  for the axions of  quantum chromodynamics (QCD), the mass  $m_{\phi'}$ and the 
coupling constant $g_{\phi' \gamma\gamma}$ for axions of other theories are not related to each other. 
The same is true 
for dilatons too.  In this work they (scalar and pseudoscalar) are referred, collectively as axion like particles (ALP).\\
\indent
 Ever since the cosmological production mechanisms of these particles were realised \cite {FDS}, the numerical estimates of the parameters ($m_{\phi_i}$ and $g_{\phi_i \gamma \gamma}$),  started being constrained  from various astronomical observational facts. One such  fact involves  the  number density of  photons $n_{\gamma}$ 
at different epochs of cosmological evolution.   For instance  cosmological big bang nucleosynthesis 
puts a very restrictive bound on the number density of photons $n_{\gamma}$ during  
primordial deuterium synthesis  \cite{subir, kolb-turner, weinberg}. Therefore 
processes that lead to a change in $n_{\gamma}$,
e.g. $\phi_{i}  \to  \gamma \gamma$,  have to be fine-tuned  -- to 
fulfil this criteria. This  in turn puts a bound on parameters  
$m_{\phi_i}$ and $g_{\phi_i \gamma \gamma}$ so that there's no change in $n_{\gamma}$ due to (pseudo)scalar decay
into diphotons -- during cosmic deuterium synthesis.\\
\indent
In the figure below (fig. [\ref{excu_plot}]) an exclusion plot of parameters $g_{\phi_{i}\gamma\gamma}$ vs $m_{\phi_{i}}$, has been provided. This plot came into existence after considering similar such constraints--those followed from various astrophysical 
and laboratory based experiments. The relevant point about this plot is, apart from identifying the excluded regions, the  same also shows the allowed regions of parameters $g_{\phi_{i}\gamma\gamma}$ and $m_{\phi_{i}}$, i.e.,  their possible 
numerical size.\footnote{We would like to emphasise
here that, these bounds are  generally insensitive to the nature of the particle($\phi$ or $\phi'$) and are applicable 
to any ALP candidate ( particle )  in general.}\\
\indent
The emergence of $\phi_{i}$'s  as  possible candidates of dark matter (DM)
followed from the realisation that-- out of the allowed  regions available 
in the  $g_{\phi_i \gamma\gamma}$$m_{\phi_i} $ exclusion plot ( of fig.[\ref{excu_plot}] )-- --there are still regions, for which 
the life time $\tau_{\phi_i\gamma\gamma}$ of (pseudo)scalars can be large enough to be comparable to the age of the universe. That too without disturbing the bound on baryon ($n_{B}$) to photon ($n_{\gamma}$) density-ratio observed 
at the current epoch (i.e., $\frac{n_{B}}{n_{\gamma}}  \sim 10^{-11}$).\\  
\indent
Coming back to the main objective of this investigation,  we recall  that, 
the identification of ALP like dark matter candidates--
would provide an elegant solution to dual problems; one  in the realms of 
{\it cosmology} and the other in the areas of {\it physics beyond the standard model of  particle physics}.\\
\indent
 However this task of ALP identification becomes difficult due to the two fold objectives associated with it, (i) identification of their type; i.e., ALP or not  and (ii)  identifying their nature, i.e.,  scalar or pseudoscalar.
There are two possible ways to search for them, those are (a) collider based 
or (b) spectro-polarimetry based. Collider based search is 
difficult, because of the colliding beam luminosity 
required to have high signal to noise ratio. But that apart, 
in some channels it becomes even more difficult, if the final state
decay products involve   scalar or  
pseudoscalar interaction --in the initial or intermediate states--have  identical interaction parameters. For instance, the life time observations (estimates), in the 
diphoton or dilepton decay channels-- for $\phi$ and  $\phi'$ 
for nearly same numerical strength and size of $m_{\phi_i}$ and 
$g_{{\phi_i}\gamma\gamma}$, would offer 
very limited scope to distinguish one from the other. \\
\indent
Discussion on collider based search is out of scope for 
this work and we will focus on the second method. The same was initiated
 in the seminal work of \cite{Raffelt} and  then followed by many others -- 
over a considerable period of time.\\
\indent
Depending on the (astrophysical) objects --i.e., their environments 
and physical situations-- {\it detecting-- 
and distinguishing} signals due to $\phi$ from $\phi'$, using spectro-polarimetric techniques may also turn out to be equally difficult as their collider based counterparts. However, one can make the same achievable  with lesser  efforts if one appropriates the symmetry properties of the DM 
candidates (i.e., $\phi$ and  $\phi'$ ) one is looking for and 
the background media that is under consideration: before analysing
the spectro-polarimetric signals from their collective interactions.    
However to fully appreciate our approach and  understand it's distinction 
(from that of available ones), a general symmetry  based analysis of the 
system is necessary; we  present a deliberation on these issues in the next
few sections of this work.

\subsection{Motivation : Symmetries and their consequences: \label {Motivation}}

\noindent
We start by recollecting the obvious fact that, any Lagrangian transforms like a scalar under Lorentz transformation. Hence, the sum of the spin angular momentum assignments 
of the fields ( background and dynamical) constituting each term of the same   must add up to zero.Therefore the interacting Lagrangian in an ambient background EM field  ($\bar{F}^{\mu\nu}$) -- for  scalar   or   pseudo-scalar photon --  system should also conform accordingly  with this observation. That is, 
(pseudo)scalar-photon interaction Lagrangian in ambient  
$\bar{F}_{\mu\nu}$ field should also respect the same. 
Stated differently, the  $\phi\gamma$  interaction Lagrangian (IL)  
and $\phi'\gamma$  IL are obtained by making the following  replacement 
$f_{\mu\nu} \to \bar{F}_{\mu\nu} + f_{\mu\nu}$ in  the Lagrangians given in 
eqn. (\ref{LintPhi}) giving the ILs 
 $g_{\phi\gamma\gamma}\phi {\bar{F}}_{\mu\nu}f^{\mu\nu}$  
and $g_{\phi'\gamma\gamma}\phi' \tilde{{\bar{F}}}_{\mu\nu}f^{\mu\nu}$, should   also respect the same.\\
\indent
The characteristics of the scalar-photon ($\phi\gamma$) or 
pseudoscalar -photon ($\phi'\gamma$) IL, in any ambient background 
electromagnetic (EM) field  depends on the 
nature of the background field itself, i.e. whether it is electric or magnetic type and the discrete symmetries of the dynamical fields, i.e., that of the 
photons and the scalars/pseudoscalars  involved.\\ 
\indent
Furthermore, all the terms present in a Lagrangian endowed with gauge fields 
do not contribute entirely to the equations of motion, unless gauge fixed, 
due to  the associated gauge ambiguities.  Expressing  the  gauge degrees of freedom of a gauge fixed Lagrangian 
 in terms of a set of  orthonormal 4-vectors representing the  modes or degrees of freedom, one obtains a 
transformed  Lagrangian -- such that each term of the field equations -- 
obtained from this transformed Lagrangian -- has a set of  definite 
discrete symmetry assignments. That is {\em each term} in every equation 
of motion, must transform identically either individually-- under charge 
({\bf C}) conjugation,  parity ({\bf P})  and  time({\bf T}) reversal transformations; or identically  under the action of  a set of their combinations.
If the Lagrangian has some additional unitary internal symmetries associated 
with it then  each term in the field equations should also transform 
identically under these unitary symmetry transformations; thus preserving 
the covariance of the field equations.\\
\indent 
This results in the 
following consequences:
i.e., in a magnetized vacuum (${\bf MV}$)(i.e., in an ambient magnetic field $B_3$, such that, only  $B_3=B_{z}= \bar{F}^{12}\ne 0$), only one dof describing a particular polarisation state of the photons out of the two possible states  (polarized transversely - along and orthogonal to the ambient (external) magnetic field $B$)
interacts  with the scalars or pseudoscalars-- at {\it this}  vertex; 
 the other one doesn't. This interaction depends on the {\bf CP} symmetry of the polarization state of the photon   and that of the  interacting fields $\phi$ or $\phi'$.
Dynamics of the two {\it states } or the transverse degrees of freedom (dof) of the  
photons can be described in two ways one in terms of gauge fields, two in terms of gauge invariant form-factors,   
$\Psi$  and $\tilde{\Psi}$;  defined as $\Psi= \bar{F}^{\mu\nu}f_{\mu\nu}$ and  $\tilde{\Psi}=\tilde{\bar{F}}^{\mu\nu}f_{\mu\nu}$, in  the notations of \cite{Ganguly-jaiswal}. 
In this notation the dof of the photons having  plane of polarization (POP) orthogonal to the ambient magnetic field $B$ is 
denoted by   $\Psi$. Similarly  the same for the photons having POP along ${B}$, will be 
described by  $\tilde{\Psi}$.\\ 
\indent
Under the operation of charge conjugation {\bf C} and parity transformation
{\bf P}, both the tensors  $\bar{F}^{\mu\nu}$ and $f^{\mu\nu}$  turn out to be odd and even (see  section two for more details)\cite{cherkas}. Hence, the combinations $\bar{F}^{\mu\nu}f_{\mu\nu}$, turns out to be   
{\bf{CP}} even and the other one, i.e., $\tilde{\bar{F}}^{\mu\nu}f_{\mu\nu}$  
turns out to be  {\bf PT} even, therefore {\bf{CP}} odd \cite{Raffelt,Adler}.
Since the scalar $\phi(x)$  is always {\bf{CP}} even, 
therefore at the level of the equation of motion (EOM)   in a ${\bf MV}$,  for  $\phi \gamma  $ interaction  the
{\bf{CP}} even $\Psi$ couples  only to {\bf{CP}} even $\phi(x)$, 
  remaining {\bf{CP}} odd form-factor $\tilde{\Psi}$ propagates freely.
In  similar situation, when   interaction  of   $\phi'(x)$  with $\gamma$  is considered, the former being  
(a  pseudoscalar  i.e.,  {\bf{P}} odd but  {\bf{C}} even )  {\bf{CP}} 
odd  will couple   only  to  the {\bf{CP}} odd form-factor $\tilde{\Psi}$ of the photon. The {\bf{CP}} even $\Psi$ would 
 propagate freely. This is just the reverse of what happens for $\phi \gamma  $ interaction in ${\bf MV}$.
That is, in ${\bf MV}$, the role of  $\Psi $  in $\phi'\gamma$ interaction
turns out to be  be  similar to that of $\tilde{\Psi} $ in $\phi\gamma$ interaction
and vice-versa.  Hence one can conclude that-- in a ${\bf MV}$--as the form of the  interaction changes from $\phi\gamma    \to  \phi'\gamma$, the role of $\Psi $  and  
$\tilde{\Psi} $  interchanges  with  each other.
Hence, the mixing dynamics of both 
systems, $\phi\gamma$ or $\phi'\gamma$,  are governed by  
2$\times$2  mixing matrix.  
This was originally pointed out by  Raffelt and Stodolosky 
in their seminar paper \cite{Raffelt}.  \\
\indent
 The propagation modes of the $\phi\gamma$ system in ${\bf MV}$,  are obtained by diagonalising the the mixing  matrix. The {\bf CP} even  propagating modes for the same,
  in the mass-less limit, turn out to be $\left(\Psi \pm \Phi\right)$ 
where $\Phi=  \left( |k_{\perp}|B\phi  \right)$ when $k_{\perp} = ksin\theta$ and $\theta$ is the angle between propagation vector ($\vec{k}$) and the magnetic field ($\vec{B}$).
Their corresponding dispersion relations satisfy, $k^2 \pm g_{\phi\gamma\gamma}|k_{\perp}|B=0$,
\cite{Ganguly-jaiswal}. \\
\noindent
 Remarkably,  the   propagating modes for  $\phi'\gamma$ system, in similar situation,  though
is   {\bf CP} odd and  given by  $\left(\tilde{\Psi} \pm \Phi\right) $,
(but) their dispersion relation 
 remains the same as   $\phi\gamma$ system;  that  is, $k^2 \pm g_{\phi'\gamma\gamma}|k_{\perp}|B=0$.
As can be seen that  the  dispersion relations (DRs)  for 
both cases are indeed invariant under boost  and rotation about  the  direction 
of the magnetic field. 
This happens because,  the presence of the external field breaks the Lorentz symmetry of the system. 
It has been argued  in  section two that,  the action  integral 
( $S=\int L d^4x$) remains invariant only under the action of  the generators   of boost and rotation along and around the 3rd direction of the Lorentz group, so as to keep $S$ invariant all other generators for the remaining Lorentz symmetry transformations must vanish. The dispersion relation mentioned above is a manifestation
 of the same. As a result, a  ${\bf MV}$  
turns out to  be  optically active  and dichroic in presence of the  $ g_{\phi'\gamma\gamma} \phi' {\tilde{f}}^{\mu\nu}f_{\mu\nu}$  or 
 $g_{\phi\gamma\gamma}\phi f^{\mu\nu}f_{\mu\nu}$  interactions.\footnote{It is note worthy that an additional possibility of subluminal or superluminal 
motion of the photons with such interaction, in some energy range, have also been reported 
in the literature\cite{Kahniashvili,Ganguly-parthasarathy,Parthamajumdar, Ganguly-jaiswal}}\\
\indent
Therefore, for near degenerate magnitudes of the coupling constants and 
masses of the candidate particles, proper identification
of one from the other may become an arduous task, for 
unfavourably oriented magnetic field, even for some astrophysical 
situations.
 One possible way  to  get out of this impasse and  refine the study of mixing physics is through  incorporation of matter effects
\cite{Raffelt-magnetized,Pankaj-Ralston,Ganguly-ann,Ganguly-jain,Tercas,subir-axion,choi-kim}, by
 including $A^{\mu}(k)\Pi_{\mu\nu}(k,\mu, T)A^{\nu}(k)$, in 
the effective  Lagrangian ($L_{eff}$) of  the system;  with $\Pi_{\mu\nu}(k,\mu, T)$ being 
the in-medium photon self-energy tensor,  that depends on 
 temperature (T), chemical potential ($\mu$) and  4-momentum ($k^{\mu}$).
  We recall chemical potential ($\mu$)  brings effect of {\bf C} symmetry into the  system. \\
  \indent
The introduction of matter effects bring additional  subtlety in the mixing dynamics, by  introducing   additional spin zero dof
 called longitudinal degree of freedom ($A_{L}$) for photons.  Both  Scalars and pseudoscalars,   having spin 
zero  can now  mix with $A_{L}$ because of spin conservation,  thus having a $4\times4$ mixing matrix.
And thereby spoiling the cure that one was looking for. \\
\indent
However a {\bf PT} symmetry  analysis of the terms of the field equations for both the  systems --- expressed
 -- in terms of  the photon form-factors, as was  introduced in \cite{Ganguly-jain}, shows that  because of {\bf {PT}} symmetry,
 pseudo-scalar photon mixing matrix is   $3\times3$, and scalar photon mixing matrix is   $2\times2$.  This however
 leaves the question unanswered, that is,  if  all the elements of the $3\times3$ pseudo-scalar  mixing matrix are finite  or  its just some of them are? The answer is no not all the elements are nonzero.   This can be understood from the fact, that
 all the nonzero elements, those  present in the mixing matrix,  are due to  interaction of the one transverse degree of photon with the pseudo-scalar.  Because, since the form factors were expressed in an orthogonal
 basis, therefore  a medium  undergoing only EM interaction,  the longitudinal one wouldn't mix with the transverse ones,
 neither the transverse ones will mix with each other.  More about it is detailed in  section two.\\
\indent
The refined study, narrated before, accounts only for mixing of two degrees of freedom for scalar-photon and 
three degrees of freedom for pseudo-scalar  photon systems. Leaving two degrees of  freedom for scalar-photon and 
one degree of freedom for pseudo-scalar  photon system free.  In other words, though the mixing pattern for both
have changed,  but we yet to achieve maximal mixing ( for either of the two).This issue  can further  be cured by incorporation of the 
parity violating part of photon self-energy tensor in the system,  following the steps mentioned  before;  the necessary details 
of which, like before, are  also relegated in the next section.

\section{Interaction Dynamics}

\subsection{Tree level Lagrangian}
The non-minimally coupled  $\phi_{i}\gamma\gamma$   interacting Lagrangian 
 in an ambient  magnetic field ( $\phi_{i}$ stands for  either a scalar ($\phi$) 
 or a pseudoscalar ($\phi'$) field )   can be expressed  as a sum over three parts;
 \begin{equation}
 L_{\phi_{i}}=L_{free,\phi_{i}} + L^{B}_{int,\phi_{i}}+ L_{int,\phi_{i}}.
 \label{Leff1}
\end{equation}
When $ L_{free,\phi_{i}}  $in   eqn. (\ref{Leff1}) stands for free field Lagrangian, given by,
\begin{eqnarray}
L_{free,\phi_{i}} = \frac{1}{2}\phi_{i}(-k)\left(  k^2-m_{\phi_{i}}^2 \right) \phi_{i}(k)-\frac{1}{4}f_{\mu\nu}f^{\mu\nu},
\end{eqnarray}
  $L^{B}_{int,\phi_{i}}$ ,  in the same equation stands  for  the interactive  part,  due to ambient magnetic field -- and is given by
\begin{eqnarray}
L^{B}_{int,\phi_{i}}=-\frac{1}{4}g_{\phi_{i}\gamma\gamma}\phi_{i}{\cal F}_{\mu\nu}f^{\mu\nu}
,   \mbox{~~where,~ ${ \cal{F}}^{\mu\nu}$ =~~}
\left\{
\begin{array}{c} 
 \bar{F}^{\mu\nu} \,$when$ \, \phi_{i}=\phi \\
\tilde{\bar{F}}^{\mu\nu}\,$when$ \, \phi_{i}= \phi',
\end{array}\label{Lint}
\right. 
\end{eqnarray}
and  lastly  the   Lagrangian for the $\phi_{i}\gamma\gamma$ interactive  part $ L_{int,\phi_{i}} $-- the last term  in the same equation  is given by,
\begin{eqnarray}
L_{int,\phi_{i}}= - \frac{1}{4}g_{\phi_{i}\gamma\gamma}\phi_{i}{\mathbf f}_{\mu\nu}f^{\mu\nu} , \mbox{~~~~~~where,~ ${\mathbf {f}}_{\mu\nu}$ =~~}
\left\{
\begin{array}{c} 
 {{f}}_{\mu\nu} \,$when$ \, \phi_{i}=\phi \\
\tilde{{ {f}}}_{\mu\nu}\,$when$ \, \phi_{i}= \phi'. 
\end{array}\label{Lint}
\right. 
\end{eqnarray}

\subsection{Symmetries}

\subsubsection{Discrete symmetries}
The equations of motion for (pseudo)scalar-photon system in ${\bf MV}$, following from the Lagrangian (\ref{Leff1}), can be cast  in matrix form, in terms of the gauge invariant variables,
\begin{eqnarray}
\Psi = \bar{F}^{\mu\nu}f_{\mu\nu} \mbox{~~and~~ } \tilde{\Psi} = \tilde{\bar{F}}^{\mu\nu}f_{\mu\nu}.
\label{GIV}
\end{eqnarray}
as; 
\begin{eqnarray} 
   \left(\begin{matrix}
     & k^2   \,\,\,   &0  \,\,\, & 0  \cr
     & 0      \,\,\,    & k^2  \,\,\,  &-g_{\phi'\gamma\gamma}\omega
                                                                      B_{T}  \cr
&0 \,\,\,   & -g_{\phi'\gamma\gamma}  \omega B_{T}\,\,\,  & k^2-m_{\phi'}^2
\end{matrix}
\right)
\left(
\begin{matrix}
{\Psi} \cr
\tilde{\Psi} \cr
\Phi'
\end{matrix}
\right)= 0,
\label{m1}
\end{eqnarray}
\begin{eqnarray}
 \left(\begin{matrix}
     & k^2   \,\,\,   &0  \,\,\, & 0  \cr
     & 0      \,\,\,    & k^2  \,\,\,  &-g_{\phi\gamma\gamma}\omega
                                                                      B_{T}  \cr
&0 \,\,\,   & -g_{\phi\gamma\gamma}  \omega B_{T}\,\,\,  & k^2-m_{\phi}^2
\end{matrix}
\right)
\left(
\begin{matrix}
{\tilde{\Psi}} \cr
{\Psi} \cr
\Phi
\end{matrix}
\right)= 0.
\label{m2}
\end{eqnarray}
Here $\Phi = |k_{\perp}| B\phi$, $\Phi^{\prime}= |k_{\perp}|B\phi^{\prime}$ and  $B_{T} = Bsin\theta$. The matrix eqn. (\ref{m1}) stands for the equation of motion (EOM) for  photon-pseudoscalar system and the one given by eqn. (\ref{m2})
 stands for  the EOM of  photon-scalar system.
Under ${\bf CP}$  transformation, the  bilinear field variables $\Psi $  and ${\tilde{\Psi}} $  and the fields $\Phi$ and $\Phi^{\prime}$ transforms in the  following manner,
\begin{eqnarray}
({\bf CP}) \bar{F}_{\mu\nu}f^{\mu\nu} ({\bf CP})^{-1}&=\bar{F}_{\mu\nu}f^{\mu\nu} ; \hspace{0.1 cm} ({\bf CP}) \Phi(x)({\bf CP})^{-1}& = \Phi(x) \\
 ({\bf CP}) \tilde{\bar{F}}_{\mu\nu}f^{\mu\nu} ({\bf CP})^{-1}&=-\tilde{\bar{F}}_{\mu\nu}f^{\mu\nu} ;  \hspace{0.1 cm} ({\bf CP}) \Phi^{\prime}(x)({\bf CP})^{-1}&=-\Phi^{\prime}(x) 
\end{eqnarray}

\noindent
 From  the equations of motion  (\ref{m1}) and (\ref{m2}), it is easy to see  that ${\bf CP}$  even $\Phi$ (or $\phi$) couples to  ${\bf CP}$ even ${\bar{F}_{\mu\nu}f^{\mu\nu}}$ and {\bf CP} odd   $\Phi^{\prime}$ (or $\phi^{\prime}$ )  couples to    {\bf CP} odd $ \tilde{\bar{F}}_{\mu\nu}f^{\mu\nu}$. \\
\indent 
 We can also find how individual EM form factors, representing each degree of freedom  of the photon couple to $\phi$ or $\phi^{\prime}$, in an ambient external field  by  expressing  the gauge field $A_{\mu}(k)$,
 in a  orthogonal 4-vector basis,   in terms of the electromagnetic  form factors  ($A_{\parallel}, A_{\perp}, A_{L}$ and $A_{gf}$)   following \cite{Ganguly-jain}, as,
\begin{eqnarray}
A_{\alpha}(k)= A_{\parallel}(k)\rm{N}_1 b^{(1)}_{\alpha}+ A_{\perp}(k) \rm{N}_2I_{\alpha}+ A_{L}(k)
\rm{N}_L\tilde{u}_{\alpha} + \frac{k_{\alpha}}{k^2}A_{gf}(k),
\label{gauge-pot.}
\end{eqnarray}
and substituting  the same in  the expression for field strength tensor $f_{\mu\nu}$  appearing in the field strength bilinear  variables ($\tilde{\bar{F}}_{\mu\nu}f^{\mu\nu}$) and (${\bar{F}_{\mu\nu}f^{\mu\nu}}$) both. The expressions for the normalization constants ($N_{1}, N_{2}, N_{L}$) and form the factors ($A_{\parallel}, A_{\perp}, A_{L}$ and $A_{gf}$) are to be found in the supplemental document \cite{Supp} of this article. Upon doing so we obtain,

\begin{eqnarray}
\tilde{\bar{F}}_{\mu\nu}f^{\mu\nu}&=& A_{\perp}N_2I_{\nu}b^{(2)\nu} + A_{L}N_{L}\tilde{u}_{\nu}b^{(2)\nu}.
\label{vertex} \\
\bar{F}_{\mu\nu}f^{\mu\nu}& =& A_{\parallel}N_{1} b^{(1)\nu}b^{(1)}_{\nu},   
\label{vertex1}  
\end{eqnarray}

\noindent
Using equation (\ref{vertex})-- in pseudo-scalar-photon interaction Lagrangian, $L^{B}_{int,\phi'} = g_{\phi'\gamma\gamma} \phi' \tilde{\bar{F}}^{\mu\nu}f_{\mu\nu}$, we obtain; 
\begin{eqnarray}
L^{B}_{int,\phi'}=g_{\phi'\gamma\gamma}\phi' 
A_{\perp}N_2I_{\nu}b^{(2)\nu} + g_{\phi'\gamma\gamma}  \phi'  A_{L}N_{L}\tilde{u}_{\nu}b^{(2)\nu}.
\label{red1}
\end{eqnarray} 
 Equation (\ref{red1}) shows that,   in principle,  $A_L $  and  $A_{\perp}$  
  can mix    with $\phi'$  in a background magnetic field.  
 But   since $A_L$ comes in to existence  only  in presence of medium; therefore
 in {magnetized vacuum},  $\phi'$ will  mix only with  $A_{\perp}$, $A_{\parallel}$  would   remain  free.  
 Similar exercise  for $\phi\gamma$ system,  with the interaction Lagrangian
\begin{eqnarray} 
 L^{B}_{int,\phi}= g_{\phi\gamma\gamma} \phi {\bar{F}}^{\mu\nu}f_{\mu\nu},  
\end{eqnarray}
 yields
\begin{eqnarray} 
 L^{B}_{int,\phi} = ig_{\phi\gamma\gamma} \phi  A_{\parallel}N_{1} b^{(1)\nu}b^{(1)}_{\nu},
\label{scpht}
\end{eqnarray} 
 implying  interaction of $ A_{\parallel}$ with   $\phi$ only in ambient magnetic field.   In other words,
 mixing is possible only between $A_{\parallel}$ and $\phi$;   $A_{\perp}$ remaining free. 
 Hence mixing matrix   for both  systems remain $2\times2$   in ${\bf MV}$. 
 Furthermore,  the mixing matrix   for $\phi\gamma$ system,   even in a medium  would  remain $2\times2$ 
 because  of non-existent $A_{L}-\phi$ interaction at the Lagrangian level,  as was verified in  eqn.(\ref{scpht}).
However,  since the   form factor  $A_{L}$, for  longitudinal  degree of freedom  is physically  realised only in a medium,
the mixing matrix for  $\phi'\gamma$ system  in a medium  turn out to be   $3\times3$.
This happens  because,   $\phi'$  couples directly to the  form factors $A_{\perp}$ and $A_L$ at the 
level of interaction Lagrangian  (eqn. (\ref{red1})), so at the level of EOM  they all would 
drive each other through their coupling with $\phi$.   More about the same will be clear when we discuss consequences of  magnetized
 matter effects.\\

\subsubsection{Lorentz symmetry}
\label{LT}
\indent
In  the Lagrangian,  $L_{\phi_{i}}$except for $L^{B}_{int,\phi_{i}}$, all other terms respect: Lorentz and gauge symmetry and remain invariant under charge 
conjugation ({\bf C}) along with parity ({\bf P}) and 
time-inversion ({\bf T}) transformations. Re-normalizability of the theory 
however gets compromised due to the presence of dim-5 operators 
$L^{B}_{int, \phi_{i}} $ and $ L_{int,\phi_{i}}$.\\
\indent
 Response of this system  to any symmetry transformation can be investigated
 by subjecting the  dynamical fields constituting  the action,
 \begin{eqnarray}
S= \int d^4x L_{\phi_{i}} = \int d^4x( L_{free,\phi_{i}} + L^{B}_{int,\phi_{i}}+ L_{int,\phi_{i}} ).
\label{lb}
\end{eqnarray} 
undergo  the symmetry   transformations - that one intends them to.  For instance  under an infinitesimal Lorentz transformation 
($\Lambda$) around the identity (${\bf I}$) -- given by:
\begin{eqnarray}
\Lambda_{\mu\nu} = {\bf \delta_{\mu\nu}} + \omega_{\mu\nu}, \mbox{~~~( when the parameter,
$\omega_{\mu\nu}= - \omega_{\nu\mu} $)},   
\label{lt}
\end{eqnarray}
 the dynamical fields strength  tensor, $f_{\mu\nu}$  would change in the following  way, 
 \begin{eqnarray}
 f'_{\mu\nu}=\Lambda_{\mu}^{\lambda} \Lambda_{\nu}^{\rho} f_{\lambda\rho} 
 \label{dnf}
 \end{eqnarray}
 and  the  background  electromagnetic field strength tensor  
  ${\mathcal {F }}^{\mu\nu} $ held constant. 
 It can further be shown that, under an infinitesimal Lorentz transformation  given by (\ref{lt})--  in a background  EM field (such that,  $ \bar{F}^{21} =\tilde{\bar{F}}^{03} \ne 0$ )
-- the change in the action (\ref{lb})--turns out to be
\begin{eqnarray} 
\delta S = \int d^4x   {\mathcal {F }}_{\mu\nu}  \omega^{\nu\lambda}f_{\lambda\mu} .
\label{changed-action}
\end{eqnarray}
 Having  the action (given by equation) (\ref{lb}) invariant, under transformation (\ref{lt})
 i.e., $\delta S  =0$;  is possible
if two out of the six components of the antisymmetric parameters $\omega_{\mu\nu}$, 
(only $\omega_{03}$ and $\omega_{12}$) are non-zero \cite{Ferrer} . Rest four are identically zero. Implying that
 the action given by equation (\ref{lb})
remains invariant only under the action of two generators of the Lorentz group; (i) 
the generators of boost 
(${\bf K_3}$) and (ii) rotation (${\bf J_3}$) along the third direction. These two generators of the Lorentz symmetry are unbroken, rest four generators are broken.
Therefore  the only space-time symmetries -- those remain  preserved --
are boost along third ( i.e. z )-direction and rotation in the one-two (xy) plane. 
Rest of the symmetries are all broken. 
Hence, the, equations of motion and the 
dispersion relations, following thereof, should respect these conditions. 
Therefore it is not surprising that the dispersion relations  --obtained  from the mixing matrices appearing  in eqns. (\ref{m1}) and (\ref{m2}), when  expressed in terms of the
 four vector  $k $ and  the component of  the same
orthogonal to B,  $k_{\perp}$  i.e., 
\begin{eqnarray}
k^2=0,  \nonumber \\
(k^2 \pm g_{\phi_{i}\gamma\gamma}k_{\perp}B_{T} ) = 0 ;
\label{disp}
\end{eqnarray}
do respect this conclusion.
That is,  these  relations remain invariant under boost and rotation around z-axis, a consequence claimed -- on the basis of Lorentz symmetry.\\
\indent
 This results in,  ${\bf MV}$ being birefringent and dichroic. In remaining part of this section, though we may not explicitly  discuss the fate of the 
 space-time/Lorentz  symmetry    of the Lagrangian,  upon incorporation of other  corrections due to material or magnetized material effects--but the same however can be shown to remain  compromised due to the appearance of velocity 4-vector of the  centre of mass of the material medium    $u^{\mu}$,   or the external field $B_z$ or both, 
 in the description of the effective Lagrangian.

\section {Solutions of field equation}
\subsection{Solutions in magnetized vacuum }

Since the gauge symmetry remains intact, so in the following we'll study
the dynamics of these systems  in terms of gauge invariant variables noted earlier  in  
\cite{Ganguly-jaiswal}.   To find  the solutions of the variables appearing in  either of the  matrix equations(\ref{m1})   and  (\ref{m2}), 
 for $\phi\gamma$   or $\phi'\gamma$  system  in ${\bf MV}$,  its sufficient  to solve for first of the  two;
  the solution for the second  can be obtained from the first  due to the symmetry $\Psi  \leftrightarrow  \tilde{\Psi}$. 
 Hence
we start with the solns of (\ref{m2}).
The solutions of the same \cite{JKPS}, in terms of constants $A_{0} $,  $A_{1} $ and $A_{2} $   are given by
\begin{eqnarray}
\tilde{\Psi}(t,x) &=& A_{0} e^{i(\omega t - k.x)},  \nonumber \\ 
\Psi(t,x) &=& A_{1} \cos(\theta) e^{i(\omega_{+} t - k.x)}- A_{2}\sin(\theta) e^{(\omega_{-}t-k.x)},     \nonumber \\
\Phi(t,x) &=& A_{1} \sin(\theta) e^{i(\omega_{+} t - k.x)}+ A_{2}\cos(\theta) e^{(\omega_{-}t-k.x)}.   
\label{soln-vac}
\end{eqnarray}
Where $\omega$, $\omega_+$, $\omega_{-}$ appearing in eqns. (\ref{soln-vac})  follows from the dispersion relations and are given by,
\begin{eqnarray}
\omega &=& K,\\
\omega_{+}&=&\pm \sqrt{K^2+ \frac{m^2_{\phi_{i}}}{2}+\left(\frac{m^4_{\phi_{i}}}{4}+ g^{2}_{\phi_{i}\gamma\gamma}B^{2}_{T}\omega^{2}\right)^\frac{1}{2}},\\
\omega_{+}&=&\pm \sqrt{K^2+ \frac{m^2_{\phi_{i}}}{2}-\left(\frac{m^4_{\phi_{i}}}{4}+ g^{2}_{\phi_{i}\gamma\gamma}B^{2}_{T}\omega^{2}\right)^\frac{1}{2}}.
\label{vac-disp-reln.}
\end{eqnarray}
\indent
The absence of  (pseudo)scalar at the initial stage,  (according to  the physics of curvature radiation), 
favours  the following boundary conditions, $\Phi(0,0)= 0 $ and $\Psi(0,0)=1$ and $\tilde{\Psi}(0,0)=1$.  These boundary conditions yield  $A_{0}=1$,   $A_{1}= \cos(\theta) $ and $  A_{2} = -\sin(\theta)$. The angle $\theta$ in equation
(\ref{soln-vac}) is given by $\theta = \frac{1}{2}\tan^{-1}(\frac{2g_{\phi\gamma\gamma} B_{T}\omega}{m^2}) $.  With these conditions the solutions for $\Psi$ finally turns out to be,
\begin{eqnarray}
\Psi(t,x) &=& \cos^{2}(\theta) e^{i(\omega_{+} t - k.x)}+ \sin^{2}(\theta) e^{i(\omega_{-}t-k.x)}, \\
\Psi(t,x)& = & a_{x}(t)e^{i(\tan^{-1}(\frac{\cos^{2}\theta \sin \omega_{+} t + \sin^{2} \theta \sin \omega_{-}t}{\cos^{2}\theta \cos \omega_{+} t + \sin^{2} \theta \cos\omega_{-}t})-k.x)},
\end{eqnarray}
when, $a_x^2(t)=  1+ 2\sin^2\theta\cos^2\theta\left( \cos(\omega_+-\omega_-\right)t - 1)$. 

\section{Observables}
\subsubsection{Polarimetric observables }
The Stokes variables, as obtained from the coherency matrix can be  expressed
in terms of the solutions, as: 
 \begin{eqnarray}
{\bf I} &=& <\tilde{\Psi}^{*}(z)\tilde{\Psi}(z)>+<\Psi^{ *}(z)\Psi(z)>, \nonumber \\
{\bf Q}&=& <\tilde{\Psi}^{*}(z)\tilde{\Psi}(z)>-<\Psi^{ *}(z)\Psi(z)>,\nonumber\\
{\bf U}&=&2 Re <\tilde{\Psi}^{*}(z)\tilde{\Psi}(z)>,\nonumber\\
{\bf V}&=& 2Im <\tilde{\Psi}^{*}(z)\tilde{\Psi}(z)>.
\label{Stokes-formal} 
  \end{eqnarray}
It may be noted that {\bf V}  appearing in the equation (\ref{Stokes-formal}) is measure of   circular polarisation.  Other polarimetric observables like  ellipticity angle, polarization angle, degree of 
linear polarization, follows from  the expressions of  {\bf I}, {\bf U}, 
{\bf Q} and {\bf V}. Polarization angle  (represented by $\psi_{\phi_{i}}$),   is defined  in terms of ${\bf U}$ and ${\bf Q}$, 
as:, 
\begin{eqnarray}
tan(2\psi_{\phi_{i}}) = \frac{{\bf U}(\omega,z)}{{\bf Q}(\omega,z)}.
\label{psi}
\end{eqnarray}
 The ellipticity angle (denoted by $\chi_{\phi_{i}}$), is defined as:
\begin{eqnarray}
tan(2\chi_{\phi_{i}}) = \frac{{\bf V}(\omega,z)}{\sqrt{{\bf Q}^2(\omega,z) + {\bf U}^2(\omega,z)}}.
\label{chi}
\end{eqnarray}
\noindent
The polarization fraction $\Pi^{P}_{\phi_{i}}$ of the radiation in terms of parameters ${\bf Q}$ and ${\bf I}$ is given by,

\begin{eqnarray}
\Pi^{P}_{\phi_{i}} = \frac{{\bf Q}(\omega,z) }{{\bf I}(\omega,z)}.
\label{Pip}
\end{eqnarray}
And lastly, the degree of linear polarization (represented as $P_{L}$) is given by, 
\begin{eqnarray}
P_{L\phi_{i}} = \frac{\sqrt{{\bf Q}^2(\omega,z) + {\bf U}^2(\omega,z)}}{{\bf I}(\omega,z)}.
\label{pl}
\end{eqnarray}

\subsubsection{Polaremetric observables for magnetized vacuum}
\noindent
So  using the solutions given in equations (\ref{soln-vac}), in  equation ({\ref{Stokes-formal}}) one can obtain  the expressions for the Stokes parameters; for scalar 
photon system. And  they are:

\begin{eqnarray}
{\bf I}(\omega ; z) &=&\sin^4(\theta) + \cos^4(\theta) + 0.5 \sin^2(2 \theta) \cos(\omega_{+}+\omega_{-})z + 1,\\
{\bf Q}(\omega ; z)&=&-[\sin^4(\theta) + \cos^4(\theta) + 0.5 \sin^2(2 \theta) \cos(\omega_{+}-\omega_{-})z - 1],\\
{\bf U}(\omega ; z) &=&2 \sin^2( \theta) \cos(\omega_{-}-\omega)z + 2 \cos^2( \theta) \cos(\omega_{+}-\omega)z ,\\
{\bf V}(\omega ; z) &=&2 \sin^2( \theta) \sin(\omega_{-}-\omega)z + 2 \cos^2( \theta) \sin (\omega_{+}-\omega)z .
\end{eqnarray}

\noindent
The same for pseudo-scalar photon system  can be obtained by interchanging the solutions of  $\Psi$ with $\tilde{\Psi}$ and vice-versa  in the equations  (\ref{soln-vac}). As a result, the Stokes parameters for $\phi\gamma$  (expressed with suffix s) can be related to that of  $\phi'\gamma$ 
system in the following form: 
\begin{eqnarray}
&{\bf I_{s} } \rightarrow  {\bf I_{ps}} , \hspace{0.2 cm} {\bf Q_s} \rightarrow {\bf -Q_{ps}}&\\
&{\bf U_{s} } \rightarrow  {\bf U_{ps}} , \hspace{0.1 cm} {\bf V_s} \rightarrow {\bf -V_{ps}}&
\end{eqnarray}

\noindent
As a result, Stokes ${\bf Q}$, ${\bf V}$,  polarisation angle $\psi_{\phi_{i}}$ and ellipticity parameter $\chi_{\phi_{i}}$ picks up a sign as one moves from $\phi\gamma$  to $\phi'\gamma$  system, provided other parameters remain the same.  As a result  identification of one from the other becomes  difficult, when $\chi_{\phi_{i}}$ and  $\psi_{\phi_{i}}$ both tend to zero.
Making degree of linear polarisation ${ P_{L\phi_{i}}}$  work to distinguish $\phi$ from $\phi'$ is even more difficult since $P_{L\phi_{i}}$
remain same.
\section{Matter effects}
\subsection{Matter and magnetized matter effect}
Effect of medium on such a system is usually incorporated through the inclusion of $\Pi_{\mu\nu}(k,\mu,T)$, the photon self energy tensor evaluated in  a medium. 
The magnetic field independent part of the polarization tensor, $ \Pi_{\mu\nu}(k,\mu,T)$ can further be expressed 
in terms of the scalar form factors times the tensors constructed out of the vectors available in hand 
for the system under consideration.  They are as follows,
\begin{eqnarray}
\Pi_{\mu\nu}(k,\mu,T) &=& \Pi_{T}R_{\mu\nu}+\Pi_{L} Q_{\mu\nu} , 
\mbox{~~~~where,~~~~}
\left\{
\begin{array}{c} 
\!\!\!\!\!\!\!\!\!\!\!\!\!\! Q_{\mu\nu}= \frac{{\tilde{u}_{\mu}}{\tilde{u_{\nu}}}}{{{\tilde{u}^2}}}
\\ 
R_{\mu\nu} =\tilde{g}_{\mu\nu} -Q_{\mu\nu}. \\
\,\,\,\,\,\,\,\,\,\tilde{g}_{\mu\nu}= \left( g_{\mu\nu}  -\frac{k_{\mu}k_{\nu}}{k^2} \right), \\
\!\!\!  \tilde{u}_{\mu}= \tilde{g}_{\mu\nu} u^{\nu}  
\end{array}
\right.
\label{pit1}
\end{eqnarray}

\noindent 
where, $\Pi_T$ and $\Pi_{L}$ are the transverse and the longitudinal form 
factors for photon polarization tensor evaluated in an unmagnetised media.
In the long wave length  limit, $\Pi_T$  can be approximated to be equal 
to the square of the plasma frequency $\omega^2_p$. Explicit expression of 
the same is provided later. The vectors $u^{\mu}$ and $k^{\mu}$
are the centre of mass velocity of the system and four momentum of the photon. 
The same for magnetized medium to linear order in magnetic field strength is taken into account by inclusion of the parity violating part of photon-self energy tensor \cite{ganguly-konar-pal}.
Following the same notations of (\ref{Lint}), the corresponding effective Lagrangian for $\phi_{i}\gamma$ system, are given by $L_{eff.\phi_{i}} = L_{eff,\phi_{i}}^{free} + L_{eff,\phi_{i}}^{int}$
when,
\begin{eqnarray}
L_{eff,\phi_{i}}^{free} &=& \frac{1}{2}\phi_{i}(-k)\left(k^2-m_{\phi_{i}}^2\right)\phi_{i}(k) -\frac{1}{4}f_{\mu\nu}f^{\mu\nu} + \frac{1}{2}A_{\mu}(-k)\Pi^{\mu\nu}(k,\mu,T)A_{\nu}(k) \nonumber \\ 
                             &+ &\frac{1}{2}A_{\mu}(-k)\Pi^{\mu\nu}(k,\mu,T,eB)A_{\nu}(k)   \label{Lef}   \\
L_{eff,\phi_{i}}^{int}&=& -\frac{1}{4}g_{\phi_{i}\gamma\gamma}\phi_{i}{\cal F}^{\mu\nu}f_{\mu\nu},
\end{eqnarray}

%
\indent
The one loop polarization tensor $\Pi_{\mu\nu}(k, \mu, T, eB)$ in equation (\ref{Lef})
in general contains the magnetized-medium induced corrections to photon self energy to 
all orders in $eB$    \cite{shahbad, ganguly-konar-pal, nieves}. It can further be 
decomposed into three pieces, (a) medium induced correction due to temperature $T$ and 
chemical potential $\mu$ only (b) medium plus magnetic field induced correction to all 
even order terms in $eB$ and (c) medium and magnetic field induced correction to all odd order
terms in $eB$. In the limit $\frac{eB}{m^2_{e}} <1$, i.e., weak field and  $\mu \ne 0$, the 
leading contribution to  photon self energy comes from the $O(eB)$ piece in 
$\Pi^{\mu\nu}(k, \mu, T, eB)$. This piece happens to be {\bf PT} symmetric hence {\bf CP} 
violating or {\bf P} violating. Other pieces being even to all orders in $eB$, are
blind to any {\bf CP} transformations.   
In most of the astrophysical situations, this is the limit
that is realised, hence, in our effective Lagrangian, we have retained terms up-to
$O(eB)$ in $\Pi^{\mu\nu}(k, \mu, T, eB)$. Henceforth the $eB$ independent piece is to be 
denoted by $\Pi_{\alpha\nu}(k)$ and the $O(eB)$ parity violating piece by $\Pi^{p}_{\alpha\nu}(k)$.
The parity violating part can similarly be  written as,
 $ \Pi^{p}_{\mu\nu}(k)= \Pi_p(k)P_{\mu\nu}$;
where the tensor $P_{\mu\nu} = i\epsilon_{\mu\nu\delta\beta}\frac{k^{\beta}}
{\mid k \mid}u^{\tilde{\delta}_\parallel}$ is the projection operator and 
$\Pi_p(k)$ is the associated form factor. Unsubscripted Greek indices  
in these expressions can have values lying between $0$ to $3$, however the same
with subscript $\parallel$, (e.g., $\mu_{\parallel}$) means that the same can assume 
only two values, either $0$ or $3$.
The form factor  $\Pi_p(k)$ to order  $eB $, turns out to be
\begin{eqnarray}  
\Pi_p(k) = \frac{\omega \omega_B \omega^2_p}{\omega^2 - \omega^{2}_B} \sim
 \frac{\omega_B \omega^2_p}{\omega},
\end{eqnarray}
and is  also called the Faraday term in the literature. We may do the same at 
places in this work.\\ 

%

\section{Systems with dim-5 interactions}

\subsection{Scalar-photon  system in magnetized medium}
\indent
 The equations of motion for the coupled $\phi\gamma$ interacting system, including Faraday term, can be cast
in a compact matrix form, as:\\  
\indent
\begin{eqnarray}
\left[\begin{array}{c} k^2 {\bf I} - {\bf M } 
 \end{array} \right]
\left(\begin{array}{c} A_{\parallel}(k) \\ A_{\perp}(k)\\ A_{L}(k) \\\phi(k) \end{array} \right)=0,
\label{photon-scalar-mixing-matixx}\nonumber\\
\end{eqnarray}


\noindent
where {\bf I} is $4\times4$ identity matrix, and {\bf M} is a $4\times4$ mixing matrix given by, \\

\begin{eqnarray} 
{\bf M}&=&\left[ 
  \begin{matrix}
  \Pi_T     &      - N_{1}\,N_{2}\, \Pi^{p}(k)P_{\mu\nu}  b^{(1)\mu} I^{\nu}      &       0        &     - ig_{\phi\gamma\gamma}{N_2b^{(2)}_\mu I^{\mu}}   \cr     
     N_{1}\,N_{2}\, \Pi^{p}(k)P_{\mu\nu}  b^{(1)\mu} I^{\nu}          & \Pi_T  &      0        &       0    \cr
     0             &   0          &  \Pi_L     &       0     \cr
     ig_{\phi\gamma\gamma}{N_2b^{(2)}_\mu I^{\mu}}             &   0         &      0         &  m_{\phi}^2 
  \end{matrix}
\right] 
\label{mat}
\end{eqnarray}
It is necessary to  note that, $A_{L}$  are not getting mixed up with other degrees of freedom in magnetized medium. Therefore we would not be considering the longitudinal degree of freedom of photon in further calculations.\\
\indent
Solutions for the field equations can be obtained by diagonalising
the matrix provided in eqn. (\ref{mat}).
Performing the same and considering the 
initial condition that at the origin there are no scalars (i.e., $\phi(\omega,0)= 0$); the  final solutions for the electromagnetic form factors can  be 
written  (in terms of  the elements of the orthonormal  eigenvectors
of the mixing matrix) as: 
\begin{eqnarray}
{\bf A_{\parallel}(\omega,z)}&=& 
\left( e^{-i\Omega_\parallel z }{\bar u}_1 { \bar u}^{*}_1  +  e^{-i\Omega_\perp z } {\bar u}_2{ \bar u}^{*}_2   
+ e^{-i\Omega_\phi z } {\bar u}_3{ \bar u}^{*}_3 \right){\bf A_{\parallel}(\omega, 0)}  \nonumber \\
&+& \left(
e^{-i\Omega_\parallel z }{\bar u}_1 { \bar v}^{*}_1  +  e^{-i\Omega_\perp z } {\bar u}_2{ \bar v}^{*}_2   
+ e^{-i\Omega_\phi z } {\bar u}_3{ \bar v}^{*}_3 \right){\bf A_{\perp}(\omega, 0)}. 
\label{soln-a-parallel}
\end{eqnarray}  
\noindent 
Similarly the perpendicular $A_{\perp}(\omega,z) $ component turns out to be, 
\begin{eqnarray}
{\bf A_{\perp}(\omega,z)}&=& 
\left( e^{-i\Omega_\parallel z }{\bar v}_1 { \bar u}^{*}_1  +  e^{-i\Omega_\perp z } {\bar v}_2{ \bar u}^{*}_2   
+ e^{-i\Omega_\phi z } {\bar v}_3{ \bar u}^{*}_3 \right){\bf A_{\parallel}(\omega, 0)}  \nonumber \\
&+& \left(
e^{-i\Omega_\parallel z }{\bar v}_1 { \bar v}^{*}_1  +  e^{-i\Omega_\perp z } {\bar v}_2{ \bar v}^{*}_2   
+ e^{-i\Omega_\phi z } {\bar v}_3{ \bar v}^{*}_3 \right){\bf A_{\perp}(\omega, 0)}.  
\label{soln-a-perp}
\end{eqnarray}

The variables $ \Omega_\parallel$, $ \Omega_\perp $ and $\Omega_\phi $ introduced in eqns. (\ref{soln-a-perp}) and (\ref{soln-a-parallel}), 
are functions of the roots of the $3 \times 3$ matrix {\bf M}. They are given by:\\
\begin{eqnarray} 
\Omega_\parallel = \left(\omega -\frac{\lambda_1}{2\omega}\right),~~ 
\Omega_\perp = \left(\omega -\frac{\lambda_2}{2\omega}\right)
\mbox{~~and~~} 
\Omega_\phi =\left( \omega -\frac{\lambda_3}{2\omega}\right). 
\label{capomega}
\end{eqnarray}\\
\noindent
The energy of the photon in equation (\ref{capomega}) is given by  $\omega$. In the final solutions, i.e., 
in eqns. (\ref{soln-a-parallel}) and (\ref{soln-a-perp}), the initial 
conditions for the two form factors of the photons at the origin are denoted by
${\bf A_{\perp}(\omega, 0)}$  and ${\bf A_{\parallel}(\omega, 0)}$. 
Their magnitude can be estimated  according to the process under consideration. 
The expressions for the Stokes parameter turns out to be : 
\begin{eqnarray}
{\bf I}(\omega,z) = {\cal {I}_{\parallel}}
 {\bf |A_{\parallel}(\omega,0)|^2} 
+ {\cal {I}_{\perp}}
 {\bf |A_{\perp}(\omega,0)|^2} 
-2{\cal {I}_{\parallel\perp}}
 {\bf |A_{\parallel}(\omega,0)}{\bf A_{\perp}(\omega,0)|},
\label{I}
\end{eqnarray}
\begin{eqnarray}
{\bf Q}(\omega,z) = {\cal{Q}_{\parallel}}
  {\bf |A_{\parallel}(\omega,0)|^2}
- {\cal {Q}_{\perp}} 
{\bf |A_{\perp}(\omega,0)|^2} 
+2{\cal {Q}_{\parallel \perp}} 
 {\bf |A_{\parallel}(\omega,0)}{\bf A_{\perp}(\omega,0)|},
\label{Q}
\end{eqnarray}
\begin{eqnarray}
{\bf U}(\omega;z) = 2 {\cal {U}_{\parallel}}
 {\bf |A_{\parallel}(\omega,0)|^2} 
+2{\cal {U}_{\perp}}
 {\bf |A_{\perp}(\omega,0)|^2}
+2 {\cal {U}_{\parallel\perp}}
 {\bf |A_{\parallel}(\omega,0)}{\bf A_{\perp}(\omega,0)|},
\label{U} 
\end{eqnarray}
\begin{eqnarray}
{\bf V}(\omega;z) =  2{\cal {V}_{\parallel}}
{\bf |A_{\parallel}(\omega,0)|^2}
+2 {\cal {V}_{\perp}}
  {\bf |A_{\perp}(\omega,0)|^2}
+2 {\cal {V}_{\parallel \perp}}
 {\bf |A_{\parallel}(\omega,0)A_{\perp}(\omega,0)|}.
\label{V}
\end{eqnarray}

\subsection{Pseudoscalar-photon system in magnetized medium}

\noindent
In this section we study  $\phi'\gamma$ mixing in a magnetized medium
when the parity violating part of photon self-energy or polarization tensor 
$\Pi^{p}_{\mu\nu} ( \mu, T, eB)$, is included in  $\phi'\gamma$ effective Lagrangian, $L_{eff,\phi'}$. \\
\indent
In this work, we first point out the distinct analytic features of the mixing matrix    
that makes the $\gamma\phi'$ mixing dynamics different
from the $\gamma\phi$  case. Following that, we obtain the estimates of 
polarimetric observables arising out this mixing--for compact astrophysical objects.\\
\indent
Next we evaluate the same observables, for $\phi\gamma$  system, for same 
physical situation, in the same physical parameter range and point out the difference 
in their magnitude  from that for the $\phi'\gamma$ system, arising due to 
the difference in their respective mixing dynamics.\\ 
\indent
The equations of motion for $\phi'\gamma$ system obtained from Lagrangian --$L_{eff,\phi'}$, as before, can be expressed in compact matrix notations as :
 
\begin{equation}
\left[k^2{\bf I} -{\bf M^{\prime} }\right]
\left( \begin{array}{c}
A_{\parallel}(k)   \\
A_{\perp}(k)   \\
A_{L}(k)  \\
\phi' (k)  \\
\end{array} \right)=0,
\label{axion_field}
\end{equation}
where ${\bf I}$ is an identity matrix and   matrix ${\bf M^{\prime}}$ is 
the  $4\times 4$ mixing matrix. The same, in terms of its
elements is given by,
\begin{equation}
{\bf M^{\prime}} = \left( \begin{array}{cccc}
 \Pi_T & - \Pi_p N_1 N_2 P_{\mu\nu}b^{(1)\mu}I^{\nu} & 0 & 0 \\
\Pi_p N_1 N_2 P_{\mu\nu} b^{(1)\mu}I^{\nu}  &  \Pi_T
& 0 & -ig_{\phi'\gamma\gamma}{N_2b^{(2)}_\mu I^{\mu}} \\
0 & 0 &  {\Pi_L}     &   -ig_{\phi'\gamma\gamma}{N_Lb^{(2)}_\mu \tilde{u}^{\mu}}\\
0 & ig_{\phi'\gamma\gamma}{N_2b^{(2)}_\mu I^{\mu}} &  i
g_{\phi'\gamma\gamma}{N_L b^{(2)}_\mu \tilde{u}^{\mu}}  & m^2_{\phi'}
\end{array} \right).
\label{mat-axion-photon}
\end{equation}
\noindent
 Note that, projection operator $P_{\mu\nu}$, appearing in the ${\bf M^{\prime}_{12}}$ and   ${\bf M^{\prime}_{21}}$ 
elements of the mixing matrix ${\bf M^{\prime}}$ is a complex one, that makes the matrix, ${\bf M^{\prime}}$, a hermitian 
matrix, that is expected even otherwise on general grounds.\\
\indent
In order to obtain the solutions in terms of form factors, we apply the boundary conditions that at origin there is no pseudoscalar field i.e.,
$\phi'(0)=0$ 
Necessary steps to arrive at the solution has been provided in \cite{Supp}, i.e,
\begin{eqnarray}
A_{\parallel}(\omega,z)&=& 
\left( e^{i\Omega_\parallel z}{\hat u}_1 { \hat u}^{*}_1  +  e^{i\Omega_\perp z } {\hat u}_2{ \hat u}^{*}_2 +  e^{i\Omega_L z } {\hat u}_3{ \hat u}^{*}_3 
+ e^{i\Omega_{\phi'} z } {\hat u}_4{ \hat u}^{*}_4 \right)A_{\parallel}(\omega, 0)  \nonumber \\
&+& \left(
e^{i\Omega_\parallel z }{\hat u}_1 { \hat v}^{*}_1  +  e^{i\Omega_\perp z } {\hat u}_2{ \hat v}^{*}_2   
+ e^{i\Omega_L z } {\hat u}_3{ \hat v}^{*}_3 + e^{i\Omega_{\phi'} z } {\hat u}_4{ \hat v}^{*}_4     \right)A_{\perp}(\omega, 0)\nonumber\\
&+& \left(
e^{i\Omega_\parallel z }{\hat u}_1 { \hat w}^{*}_1  +  e^{i\Omega_\perp z } {\hat u}_2{ \hat w}^{*}_2   
+ e^{i\Omega_L z } {\hat u}_3{ \hat w}^{*}_3 + e^{i\Omega_{\phi'} z } {\hat u}_4{ \hat w}^{*}_4     \right)A_{L}(\omega, 0).
\label{soln-parallel1}
\end{eqnarray}  
\noindent 
The  component $A_{\perp}(\omega,z)$ is given by, 
\begin{eqnarray}
A_{\perp}(\omega,z)&=& 
\left( e^{i\Omega_\parallel z}{\hat v}_1 { \hat u}^{*}_1  +  e^{i\Omega_\perp z } {\hat v}_2{ \hat u}^{*}_2 +  e^{i\Omega_L z } {\hat v}_3{ \hat u}^{*}_3 
+ e^{i\Omega_{\phi'} z } {\hat v}_4{ \hat u}^{*}_4 \right)A_{\parallel}(\omega, 0)  \nonumber \\
&+& \left(
e^{i\Omega_\parallel z }{\hat v}_1 { \hat v}^{*}_1  +  e^{i\Omega_\perp z } {\hat v}_2{ \hat v}^{*}_2   
+ e^{i\Omega_L z } {\hat v}_3{ \hat v}^{*}_3 + e^{i\Omega_{\phi'} z } {\hat v}_4{ \hat v}^{*}_4     \right)A_{\perp}(\omega, 0)\nonumber\\
&+& \left(
e^{i\Omega_\parallel z }{\hat v}_1 { \hat w}^{*}_1  +  e^{i\Omega_\perp z } {\hat v}_2{ \hat w}^{*}_2   
+ e^{i\Omega_L z } {\hat v}_3{ \hat w}^{*}_3 + e^{i\Omega_{\phi'} z } {\hat v}_4{ \hat w}^{*}_4     \right)A_{L}(\omega, 0).
\label{soln-perp1}
\end{eqnarray}  
when
$ \Omega_\parallel = \left(\omega -\frac{\lambda^{0}_1}{2\omega}\right)$, 
$ \Omega_\perp = \left(\omega -\frac{\lambda^{0}_2}{2\omega}\right)$, 
$ \Omega_L = \left(\omega -\frac{\lambda^{0}_3}{2\omega}\right)$, and 
$ \Omega_{\phi'} =\left(\omega -\frac{\lambda^{0}_4}{2\omega}\right)$. 
 The difference between $\phi\gamma$ and this system, is the appearance of contribution due to $A_L$ 
 or longitudinal component of photon.
Using the solutions obtained in (\ref{soln-parallel1}) and (\ref{soln-perp1}) for $\phi\gamma$  interaction, we provide the polarimetric variables for preudoscalar-photon interaction taking magnetized matter effect into account, as follows: .
\begin{eqnarray}
{\bf I}(\omega,z) =  \hskip -0.5 cm && I_{\parallel} {\bf |A_{\parallel}(\omega,0)|^2} 
+ I_{\perp} {\bf |A_{\perp}(\omega,0)|^2} 
- I_{L} {\bf |A_{L}(\omega,0)|^2} \nonumber\\
+ \hskip -0.5 cm && I_{\parallel \perp} {\bf |A_{\parallel}(\omega,0) \bf A_{\perp}(\omega,0)|} 
+ I_{\parallel L} {\bf |A_{\parallel}(\omega,0) \bf A_{L}(\omega,0)|} \nonumber\\
+ \hskip -0.5 cm && I_{\perp L} {\bf |A_{\parallel}(\omega,0) \bf A_{L}(\omega,0)|}, 
\label{I-axion}
\end{eqnarray}
\begin{eqnarray}
{\bf Q}(\omega,z) = \hskip -0.5 cm && Q_{\parallel} {\bf |A_{\parallel}(\omega,0)|^2}
- Q_{\perp} {\bf |A_{\perp}(\omega,0)|^2} - Q_{L}{\bf |A_{L}(\omega,0)|^2} \nonumber\\
+ \hskip -0.5 cm && Q_{\parallel \perp} {\bf |A_{\parallel}(\omega,0) \bf A_{\perp}(\omega,0)|} 
+ Q_{\parallel L}{\bf |A_{\parallel}(\omega,0) \bf A_{L}(\omega,0)|} \nonumber\\
+ \hskip -0.5 cm && Q_{\perp L} {\bf |A_{\perp}(\omega,0) \bf A_{L}(\omega,0)|}, 
\label{Q-axion}
\end{eqnarray}
\begin{eqnarray}
{\bf U}(\omega,z)= \hskip -0.5 cm &&U_{\parallel}
 {\bf |A_{\parallel}(\omega,0)|^2} 
+ U_{\perp}
{\bf |A_{\perp}(\omega,0)|^2} 
+ U_{L}
{\bf |A_{L}(\omega,0)|^2} \nonumber\\
+ \hskip -0.5 cm && U_{\parallel \perp}
{\bf |A_{\parallel}(\omega,0) \bf A_{\perp}(\omega,0)|}
+ U_{\parallel L}
{\bf |A_{\parallel}(\omega,0) \bf A_{L}(\omega,0)|} \nonumber\\
+ \hskip -0.5 cm && U_{\perp L}
{\bf |A_{\perp}(\omega,0) \bf A_{L}(\omega,0)|}. 
\label{U-axion}
\end{eqnarray}
\begin{eqnarray}
{\bf V}(\omega,z)= \hskip -0.5 cm && V_{\parallel}
 {\bf |A_{\parallel}(\omega,0)|^2}
+ V_{\perp}
 {\bf |A_{\perp}(\omega,0)|^2} 
+ V_{L}
 {\bf |A_{L}(\omega,0)|^2} \nonumber\\
+ \hskip -0.5 cm &&V_{\parallel \perp}
{\bf |A_{\parallel}(\omega,0) \bf A_{\perp}(\omega,0)|}   
+ V_{\parallel L} 
{\bf |A_{\parallel}(\omega,0) \bf A_{L}(\omega,0)|}\nonumber\\
+   \hskip -0.5 cm &&V_{\perp L}
 {\bf |A_{\perp}(\omega,0) \bf A_{L}(\omega,0)|}.
\label{V-axion}
\end{eqnarray}
The complete expressions for the coefficients of $A_{\parallel}, A_{\perp}, A_{L},A_{\parallel }A_{\perp},A_{\parallel }A_{L}$ 
and $A_{\perp}A_{ L}$ are explicitly defined in \cite{Supp}.

\section{Mixing}
\subsection{Analysis of mixing.}
In this section, we deliberate on the  origin of the couplings between various 
degrees of freedom -- on the mixing dynamics -- from the point of view of  
symmetry.  We begin our discussion first on the consequences of including 
just the matter effects, followed by, the same upon including the matter
plus the magnetized  matter effects -- on the mixing dynamics.\\
\indent
We begin with the demonstration of the claim made in  
sec.(\ref{Motivation}), that, with  the incorporation of matter effects  
by including  $A^{\mu}(k)\Pi_{\mu\nu}(k,\mu, T)A^{\nu}(k)$ in the effective Lagrangian  $L_{eff, \phi_i}$ -- the mixing matrix of scalar photon system  turns out to be 2 $\times$ 2  but 
the same for pseudoscalar photon system turns out to be 3 $\times$ 3.\\
 \indent
This feature can easily  be verified  by setting  the parity violating 
part $\Pi_p=0$ in  eqns. (\ref{mat}) and  (\ref{mat-axion-photon}) for 
$\phi\gamma$ and $\phi'\gamma$ system. This is due to the fact that  
eqns. (\ref{mat}) has both the effects incorporated in it, so by setting 
$\Pi_p=0$, one retains only the matter effects.\\
\indent
Next we address the question of the coupling between different degrees 
of freedom  and their discrete symmetry assignments.  The  emerging
 2 $\times$ 2 mixing matrix structure for $\phi\gamma$ system is due 
to mixing  between   $A_{\parallel}$ and $\phi$.
This follows from
the  expression of the Lagrangian   ($L^{B}_{int,\phi}$) given by  
eqn. (\ref{scpht}), denoting $\phi\gamma$ interaction  
in an external magnetic field. It follows from there  that  the only 
form factor that  has nonzero interaction with the  scalar $\phi$, is 
 $A_{\parallel}$.  No other form factor has any interaction  with 
$\phi$ at the level of $L^{B}_{int,\phi}$. Hence the mixing  matrix is
 2 $\times$ 2.\\
\indent
The other interesting issue in this mixing is: although, naively one would have 
expected the longitudinal degree of 
freedom of the photon and the scalar $\phi$ to mix with each other for having 
same spin assignments $s_z=0 $, but that remains  forbidden because 
at the level of the Lagrangian $L^{B}_{eff,\phi_i}$,
they have no  interaction. So  the mixing  matrix for $\phi\gamma$ system  
turns out to be $2\times2$.\\
\indent
Lastly, we examine another subtlety in this mixing dynamics, following from 
the discrete symmetry arguments. We begin with the observation that, the discrete 
symmetries of the associated degrees of freedom of  electromagnetic form factors, appearing in the description of the 
gauge potential, given by (\ref{gauge-pot.}) -- including $A_{\parallel}$ -- are odd under Time ({\bf T}) reversal  and   even under Parity ({\bf P}) inversion symmetry  transformations ( for details consult table [1]. 
And $\phi$, being a  scalar remains even under {\bf PT} transformation. 
So at the level of field equations coupling of {\bf PT } odd  $A_{\parallel}$  with {\bf PT } even scalar  $\phi$, 
is a bit  bizarre. \\
\indent 
However, a careful inspection of the fields-equations reveal 
that, these two  degrees of freedom with uneven {\bf PT} symmetry,
appears in  the field -equations after being multiplied by specific
{\bf PT} dependent factors. It so happens that  upon {\bf PT} transformation  
the products in field equations have retained  same {\bf PT} symmetry.
Stated differently, if we consider the IL of equation ({\ref{scpht}) i.e., 
\begin{eqnarray} 
 L^{B}_{int,\phi} = ig_{\phi\gamma\gamma} \phi  A_{\parallel}N_{1} b^{(1)\nu}b^{(1)}_{\nu}
\label{scpht2}
\end{eqnarray} 
the product, $ b^{(1)\nu}b^{(1)}_{\nu}$ is even under Time reversal symmetry transformation  ({\bf T})
and so are the transformation properties of $N_{1}$ and $\phi$, but $A_{\parallel}$ is odd under ${\bf T}$.
Therefore, the eqn. (\ref{scpht2}) wouldn't have been {\bf T} symmetric {\it unless the multiplicative factor $i$ was absent}.
Therefore, here the factor $i$ is the  {\bf T} dependent multiplicative factor, that was discussed in preceding paragraph. \\
\indent
Next we consider the issue of the effect of magnetized medium  on  
$\phi\gamma$  mixing dynamics. 
The effect of the same is obtained through the  incorporation of
the parity violating part  $\Pi_p$, in the effective Lagrangian. Since the 
same makes the two transverse degrees of freedom of the photon $A_{\parallel}$ with $A_{\perp}$ couple directly with 
each other  and   $\phi$ too couples directly with $A_{\parallel}$, thus the appearance of $\Pi^{p}$ makes $A_{\perp}$ coupled indirectly to  $\phi$. The longitudinal degree of 
freedom  of photon however remains decoupled. 
Thus the mixing matrix for $\phi\gamma$  system turns out
to be $ 3 \times 3 $ on incorporation of the same.\\
\indent
Next we turn our attention to pseudoscalars (axion) photon system. The same  
under similar situation, has non-zero interaction with  $A_{\perp}$ and 
$A_{L}$ at  the level of the interaction Lagrangian, as can be checked from
eqn. (\ref{red1}). Therefore,  the inclusion of matter effects, makes 
direct mixing possible between $\phi'$, $A_L$ and $A_{\perp}$, thus 
turning the mixing matrix  a $3\times3$ one.\\
\indent
On inclusion of magnetized-matter effects  along with unmagnetized matter-effects, 
the mixing matrix  for the $\phi'\gamma$  becomes $ 4 \times 4$ , due 
to the following reason:  because of matter effects alone , $\phi'$ couples 
directly to $A_L$ and $A_{\perp}$,  at  the level of $L^{B}_{int, \phi'}$.  
As we already have pointed out, upon inclusion of the parity violating 
$\Pi_p$ part,  $A_{\perp}$ further couples directly  with $A_{\parallel}$, hence
there is an indirect mixing between  $i\phi'$ and photon EM form-factor for transverse polarization ($A_{\parallel}$); thus, making all the degrees of 
freedom for pseudoscalar photon system interact with each other. The essence of the same is captured  through the elements of the $4 \times 4 $ mixing matrix.\\
\indent
Coming to  question of the discrete symmetry assignments of the degrees of freedom and their coupling in the field equations for the pseudoscalar photon system, we note the  existence coupling between 
degrees of freedom with uneven {\bf PT} symmetry assignments.
This can be understood as follows. Due to  odd
{\bf  PT} transformation assignment, $A_L$ (see  table [\ref{table:1}] in \cite{Supp} )  can 
couple only to the product  $i \times\phi'$ in any equation;  
because, as noted before, though $\phi'$ is  even  under {\bf PT} but  $i$ 
(i.e., $\sqrt{-1}$) being odd under {\bf T} , and even under {\bf P} makes  $i \times \phi'$ odd under {\bf PT}; thus favouring  the coupling between the two at the level of field equations. This can be verified from the last equation of the equations  presented at (\ref{photon-eom})  in  the supplementary section \cite{Supp}. Similar argument follows for other equations too.  We conclude this discussion with the observation that, the difference in the  mixing pattern between $\phi\gamma$ and $\phi^{\prime}\gamma$ system in  an environment having magnetized-matter can  have far reaching
consequences on the polarimetric signals produced by each. Hence, the same 
can serve as an ideal astrophysical laboratory, to distinguish one type of interaction from the 
other. A  detailed discussion on the possible nonzero elements of the mixing matrix, is
however, left for a future communication  \cite{ganguly-mixing}.\\

\subsection{  Searches so far:}
\noindent 
Based on the physics principles mentioned 
above, systematic studies to find ALP, through Laboratory or 
astrophysics based experiments has been going on, for 
some time now.  The ongoing and proposed laboratory based searches for ALP
are  
\cite{Cast,15, 16,17,18, 18a,19,20,21}. Among those, the CAST collaboration of CERN
\cite{Cast} is a prominent one. This collaboration is engaged in 
detecting ALP, produced at the interior of the Sun
via primakoff process \cite{primakoff, palash-prima},
based on energy loss arguments from stellar interior. 
Earlier they had reported the following bounds on ALP parameters, 
$m_{\phi'} \leq 0.02 $ eV and 
$g_{\phi'\gamma\gamma} < 1.16 \times 10^{-10}$ GeV$^{-1}$ \cite{Cast}; 
and their last improvement to this result reads,
$g_{\phi'\gamma\gamma} < 0.66 \times 10^{-10}$ GeV$^{-1}$ 
in the same mass range \cite{Castnature}.\\
\indent
In the astrophysical front,  bound on ALP parameters also obtained by study the
cooling rate of stellar objects,  e.g., super-giants, red giants, 
helium core  burning stars, white dwarfs, and neutron stars 
\cite{9, 10, 11, 12}etc., due to the same. An up to date introduction to the 
relevant issues can be found  in \cite{Giannotti}. The interesting part is
that the astrophysical bound on axion photon coupling, based on 
 measuring the ratio of the number of horizontal branch stars to 
that of red giant branch stars in the globular cluster, also provide
 $g_{\phi'\gamma\gamma} < 0.66 \times 10^{-10}$ GeV$^{-1}$ at $95\%$ C.L. 
\cite{ayla} as that of \cite{Castnature}.  Most of these bounds were obtained  by  taking just the matter effects
into account in the axion-photon effective Lagrangian. That is  mixing occurs  between only
three degrees of freedom($\phi^{\prime}$, $A_{\perp}$ and $A_{L}$), leaving  $A_{\parallel}$
free.\\
\indent
However  according to  current observations, the  data  gives a hint of additional
cooling at various stages of their evolution, termed as cooling anomaly. This  has renewed
the interest in a model dependent axion induced cooling \cite{yanagida}.
However a possible  of way to achieve this  is through
{\it mixing of all the degrees of freedom of photon with
ALPs}  that will open {\it additional channels of energy transport} via
conversion of photons into axions.\footnote{Incorporation of $\Pi$ does not do  that }
Same can be achieved by incorporation of the parity violating part of 
photon self energy tensor $\Pi^{P}_{\mu\nu}$   that emerges  from  magnetized-matter effects,
in the  pseudoscalar photon effective-Lagrangian 
$L_{eff}$. This  happens to changes the mixing dynamics, hence, the {\emph{mixing matrix}}  for $\phi'\gamma$ system. The  mixing matrix  for $\phi' \gamma$ system, turns out to be  
4$\times$4, implying, complete mixing between all the three degrees 
of freedom of (in-medium) photon with the single degree of freedom of
$\phi'$ .\\
\indent 
 For $\phi\gamma$ system,  the same  procedure changes  the mixing matrix {\emph{ from $2\times 2$ to a  $3\times3$ one}} 
\cite{manoj}. That is, the mixing is only between the two transverse 
polarization states of the photon and $\phi$. The longitudinal 
degree of freedom of the photon gets decoupled and propagates freely.
The spectro-polarimetric  signatures of these  features, are the ones  investigated  in the subsequent sections of this work.\\ 

\section{Astrophysical application.}
In this section we would focus on a possible astrophysical scenario, that is likely 
to provide the distinct signatures that would distinguish  scalar photon mixing from pseudoscalar photon mixing using magnetized media effects
 The choice of the astrophysical system  
considered here for getting a reasonably strong signal is:
an eclipsing compact binary system. This choice is due to the 
strong surface magnetic field ($10^{9}- 10^{14}$ Gauss ) associated with them.
We expect that the $\phi_{i}\gamma$ interaction in such strong field would 
generate strong enough signal to be detected by the  upcoming space-borne X-ray observatories. \\
\indent
Since polarization angle $(\psi_{\phi_{i}})$ and ellipticity angle $(\chi_{\phi_{i}})$ have emerged as
a reliable observable for polarimetric signatures, therefore
we would consider estimating their contribution for signals, for the kind of situation discussed here. 

\begin{figure}[ht!!]
\begin{center}
\includegraphics[scale=.25]{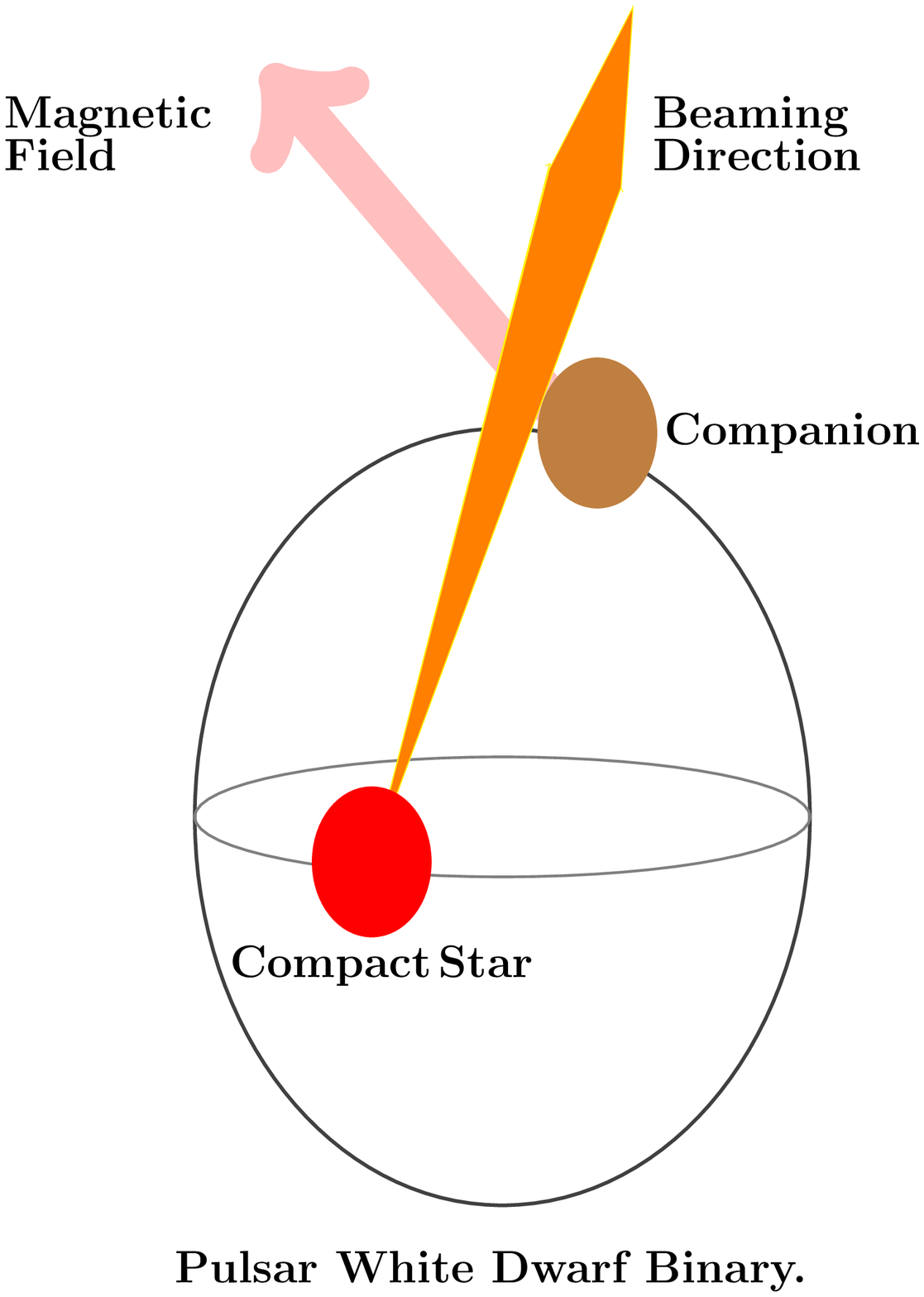}
\end{center}
\vskip -1.52cm
\caption{Schematic diagram of the binary configuration, as considered in the
text.}   
\label{f2}
\end{figure}

\indent
The binary system considered here, is assumed to be composed of a neutron star 
(primary) and it's ``companion'', that may be a white dwarf or a 
neutron star. The primary is imagined to be an axisymmetric rotor and the companion to be  an orthogonal rotor, having their individual dipole field direction nearly orthogonal to each other. At the onset 
of the eclipsing phase, electromagnetic radiation beam from the primary grazes past the polar 
cap of the companion -- to the observer. This happens during a finite fraction of their 
orbital cycle  [fig.[{\ref{f2}}]]. For rest of the  cycle, the primary would  
beam its radiation directly to the observer.  
In this scenario, any difference in the observed values of $(\Psi_{\phi_{i}})$ and  
$(\chi)$ between the {\it direct beaming phase} to the {\it grazing phase} 
can be attributed to propagation in the  magnetized environment 
of the companion. If scalars or pseudo-scalars do exist in nature, their 
interaction in the magnetized environment of the companion would also 
contribute to the difference in the measured values of $\psi_{\phi_{i}}$ and  $ \chi_{\phi_{i}}$;
the study of which happens to be the main goal of this investigation, as 
mentioned before.\\
\indent
Since the EM beam, according to our physical picture,  grazes past 
the polar cap of the companion, hence the magnetized path length that 
photons travel would be $\sim$ twice the polar cap radius $(R_{PC})$ 
of the companion.
Assuming the magnetic field to be dipolar, the polar cap
opening angle for the last open field lines at the surface of 
the compact star can be estimated 
from dipole field equations; the same in terms of light-cylinder radius
$R_{LC}$ turns out to be,
\begin{eqnarray}
\theta_{pc} = sin^{-1}\left[
\left(\frac{R}{R_{LC}}\right)^{\frac{1}{2}} \right] 
\sim \left(\frac{R}{R_{LC}}\right)^{\frac{1}{2}}, \mbox{~for~}\left(\frac{R}{R_{LC}}\right) << 1.
\end{eqnarray}
\noindent
Therefore, the polar-cap radius of the companion, in terms of  its 
radius $R$  and period $P$ turns out to be,
\begin{eqnarray}
R_{pc}= R \left(\frac{2 \pi R}{P}\right)^{\frac{1}{2}},
\label{rpc}
\end{eqnarray}
while writing equation (\ref{rpc}) we have used the relation $R_{pc}= \frac{P}{2\pi}$.
\noindent
Assuming  period $P \sim 100$ sec, radius $R=10$ km,  for the companion,
 the polar cap radius turns out to be $R_{pc}=6.5$ meter. 
We further assume the primary and the companion to be separated enough, 
not to pass through each others light cylinder during their orbital 
motion. 
The companion considered here is cold enough to have no charge particles in its magnetosphere. Hence, the availability of plasma would be around the polar-cap where the plasma frequency will be  given by 
$\omega_{p} = \frac{n_{GJ}}{m_e}$. Goldreich Julien number density  {\cite{Goldreich}} given by $n_{GJ} = \frac{\vec{\Omega}.\vec{B}}{2\pi}$ for an orthogonal rotor can in principle be very small, depending on the angle between $\vec{\Omega}$ and $\vec{B}$ ;  So  we consider the  plasma frequency  to be of the order of $10^{-9}$ eV and surface
 magnetic field, $B_{S} =  1.25\times 10^{12}$G. The choice of these 
fiducial parameters ensure that, the particle density  in the stellar magnetosphere  
will be negligible and the narrow EM beam
will just pass through the plasma close to the surface of the companion.
This scenario implies that, the interaction with the magnetized media
is confined only in the polar cap region for the propagating EM beam. \\

\section{Results}
In this section we would like to discuss the signatures of $\gamma\phi$ and $\gamma\phi^{\prime}$ interactions on the 
spectro-polarimetric observables {\bf I},{\bf Q},{\bf U},{\bf V},$\chi_{\phi_{i}},\psi_{\phi_{i}}$, $P_{L\phi_{i}}$ and polarization fraction $\Pi^{P}_{\phi_{i}}$ of the EM radiations coming from the astrophysical sources. We would also like to establish whether the contributions to the polarimetric variables  are same or different for these two types of interactions. Further an attempt would   be made to  understand the
origin of the difference in the size of their contributions based on the number of participating dof in mixing.\\
\indent
The current and proposed space based polarimetric experiments to 
detect polarimetric signals from astrophysical sources include,
APEX \cite{APEX}, IXPE \cite{IXPE}, e-ASTOGRAM \cite{e-ASTROGAM},
ASTRO-2020 \cite{ASTRO-2020} to name a few. They are expected
to cover the soft X-ray band of the EM spectra. Furthermore
since the  energy distribution of the observed cosmic $\gamma$-ray 
spectra  is seen to peak in the 1-10 KeV range \cite{e-ASTROGAM},   
we have confined our analysis in a slightly broader energy domain of 1-100 KeV that covers the peak emission range.
\begin{figure}[!ht]
\begin{center}
\hskip -1 cm\includegraphics[scale=1.5]{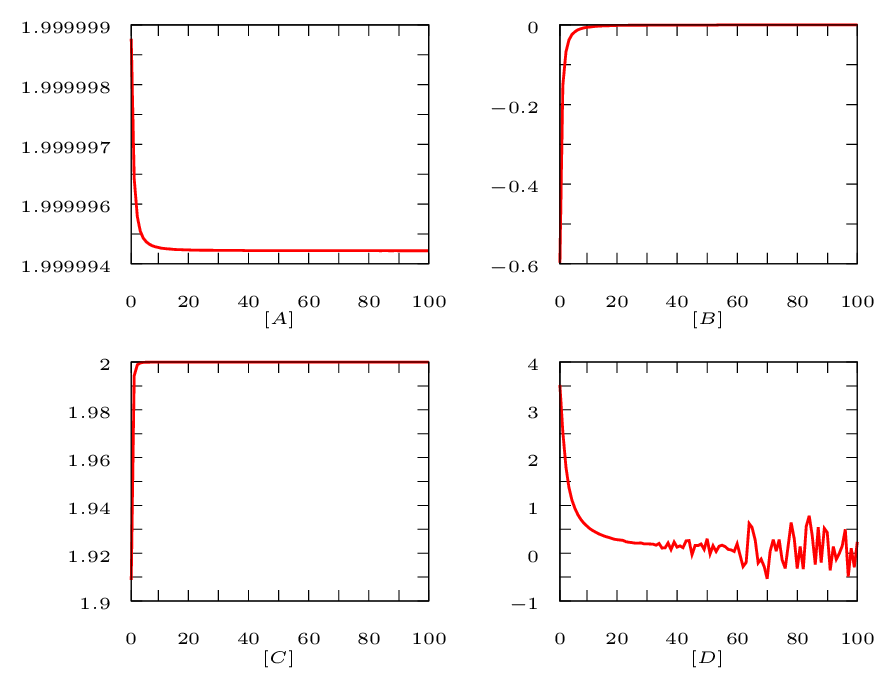}
\end{center}
\caption{Stokes parameters {\bf I}, {\bf Q},  {\bf U} and {\bf V} (in ordinates) for $\phi\gamma$ interaction vs photon energy $\omega$ (in abscissas) plotted in panel $A$, $B$, $C$ and $D$ respectively.
The parameter {\bf V} in $D$ is scaled by the factor of $10^{6}$. The parameters choosen for this estimations are: Mass of scalar and pseudo scalar $m_{\phi,\phi^{\prime}} = 1.0 \times 10^{-11}$ GeV, coupling constant $g_{\phi\gamma\gamma} = 1.0 \times 10^{-11}$ GeV, magnetic field $\sim 1.0 \times 10^{12}$ Gauss and plasma frequency = $1.0 \times10^{-12} $ GeV, photon pathlength = 12 meter.}
\label{Stokes_SC}\
\end{figure}

\begin{figure}[!ht]
\begin{center}
\hskip -1 cm\includegraphics[scale=1.5]{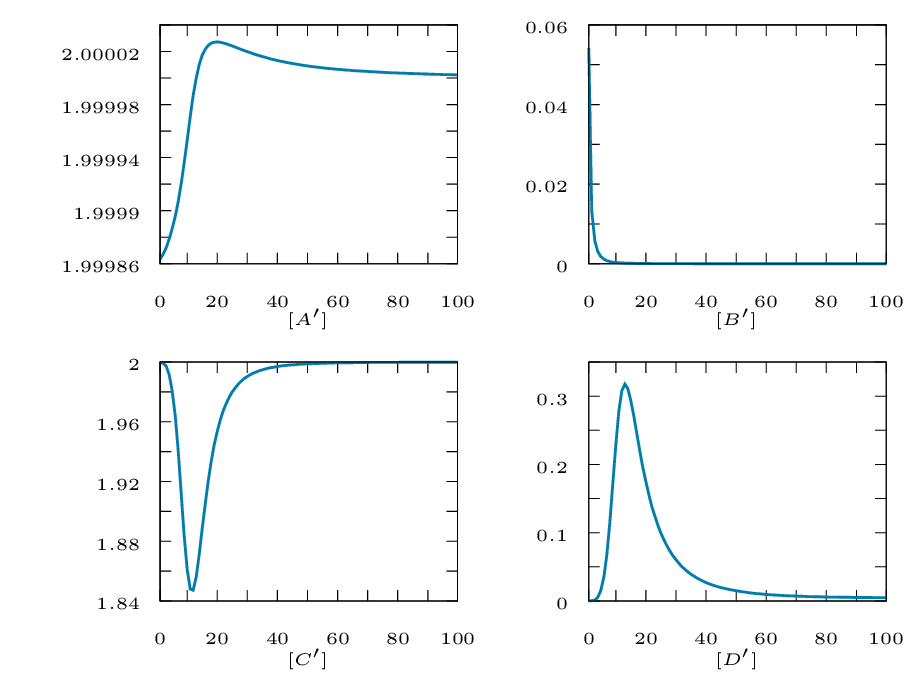}
\end{center}
\caption{Stokes parameters {\bf I}, {\bf Q},  {\bf U} and {\bf V} (in ordinates)  for $\phi^{\prime}\gamma$ interaction vs photon energy $\omega$ (in abscissas)  plotted in panel $A^{\prime}$, $B^{\prime}$, $C^{\prime}$ and $D^{\prime}$ (in the same parameter space as of fig. [\ref{Stokes_SC}]) respectively. }
\label{Stokes_AX}\
\end{figure}

\begin{figure}[!h]
\begin{center}
\hskip -1 cm\includegraphics[scale=1.5]{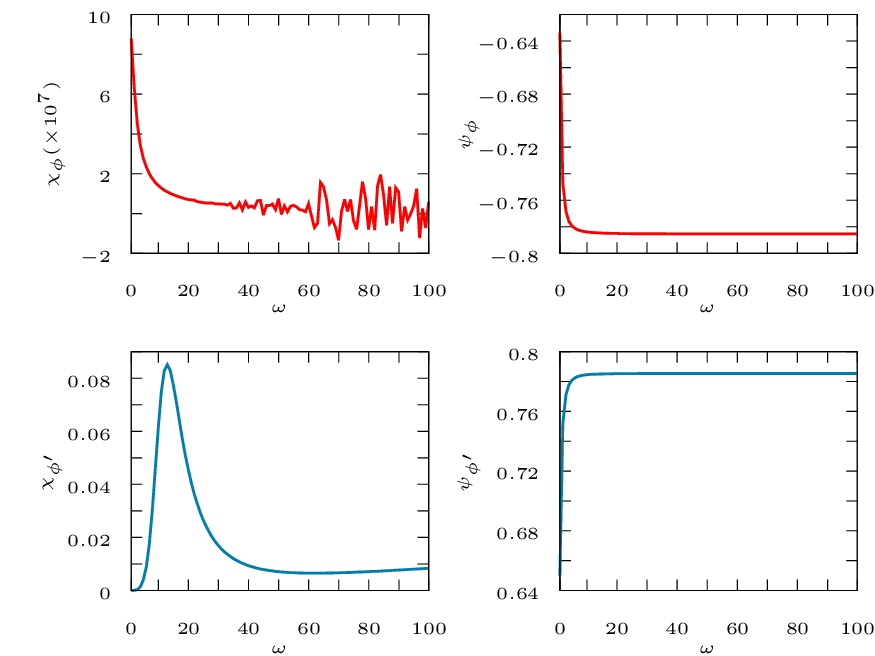}
\end{center}
\caption{The ellipticity angle, polarization angle  of dilaton are in the top panel and the same for axion are in bottom panel in the same parameter space as of fig. [\ref{Stokes_SC}]. The x-axes are in unit of KeV.  
}   
\label{f312m}
\end{figure}

\begin{figure}[!ht]
\begin{minipage}[b]{.48\textwidth}  
\hskip 0.6 cm\includegraphics[width=1\linewidth]{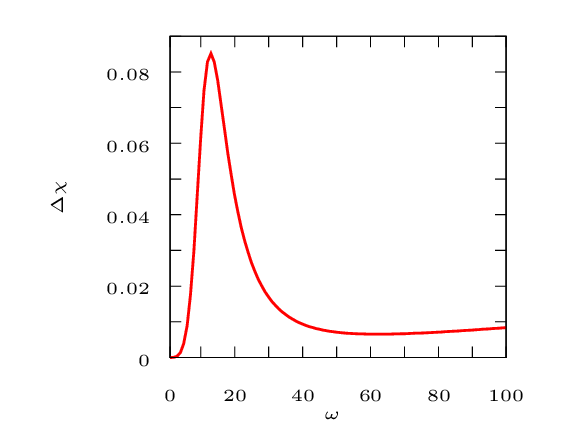}
\end{minipage}
\begin{minipage}[b]{.48\textwidth}  
\hskip -0.6 cm \includegraphics[width=1\linewidth]{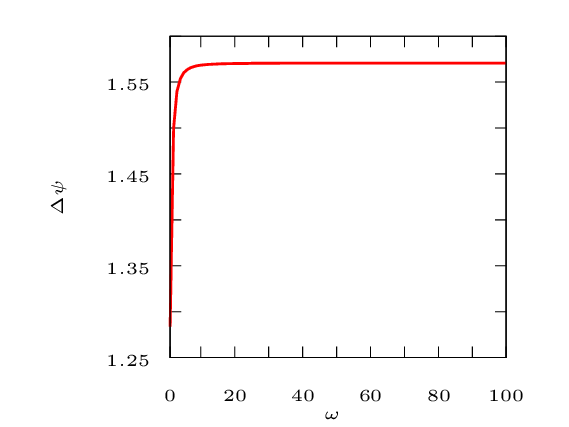}
\end{minipage}
\caption{Plot of $\Delta\chi$ vs $\omega$  in the left panel and $\Delta\psi$ vs $\omega$ in the right panel. The x-axes are in unit of KeV. The parameters for these plots are same as used in fig.[\ref{f312m}].}  
\label{f312m_diff}   
\end{figure}
\indent
Assuming the physical scenario mentioned already, 
we have estimated  the Stokes parameters, polarization and ellipticity angles for each type of
interaction (i.e., $g_{\phi\gamma\gamma}\phi F_{\mu\nu} F^{\mu\nu}$/ $g_{\phi^{\prime}\gamma\gamma} \phi^{\prime} F_{\mu\nu} F^{\mu\nu}$) keeping $\omega_{p}$, path length $ z$, magnetic field $B$,
$g_{\phi_{i}\gamma\gamma}$ and $m_{\phi_i}$ fixed, using the same parameters. We further have estimated the
 difference in the polarization angle, $\Delta \psi =(\psi_{\phi^{\prime}_{i}} - \psi_{\phi_{i}})$ and the ellipticity angle
 $ \Delta\chi =(\chi_{\phi^{\prime}_{i}} - \chi_{\phi_{i}} )$  acquired  by a light beam, due to  pseudoscalar $g_{\phi^{\prime}\gamma\gamma} \phi^{\prime} F_{\mu\nu} F^{\mu\nu}$ and  scalar $g_{\phi\gamma\gamma}\phi F_{\mu\nu} F^{\mu\nu}$ interactions, as the EM beam passes through the magnetosphere of the companion star of a binary system.\\
\indent
The coupling constant $g_{\phi_i\gamma\gamma}$ is considered to be
$g_{\phi_i\gamma\gamma }=1.0\times10^{-11}$ GeV$^{-1}$ for  the scalar ($\phi$) and the
pseudoscalar ($\phi^{\prime}$) interactions and  their mass
 $m_{\phi_{i}} = 1.0 \times 10^{-11}$ GeV. 
We have considered the path length of the photon to be twelve meter, keeping  magnetic field  ($B$) $\sim 1.0 \times 10^{12}$ Gauss
and plasma frequency $  {\bf \omega_p} = 1.0 \times10^{-12} $ GeV. The energy band
of investigation has been kept at soft X-ray region ($ 1-100$ KeV ).
The evolution of Stokes parameters with  energy of the photon $\omega$ for both the interactions has been plotted  in fig.[\ref{Stokes_SC}]  and in fig.[\ref{Stokes_AX}] respectively.\\
\indent
The individual plots of the ellipticity angle ($\chi_{\phi_{i}}$) for scalar photon and pseudoscalar photon interaction can be found in left panel of  fig.[\ref{f312m}] plotted as a function of $\omega$  and the same for polarization angle ($\psi_{\phi_{i}}$) have been plotted in right panel of the  fig.[\ref{f312m}].
The  magnitudes of the two initial amplitudes of  the  orthogonal modes of the photon beam 
are considered to be identical, which is not too drastic an assumption if the system is isotropic and homogeneous. In a more 
realistic situation, same can be extracted from the direct beaming phase 
of the radiation. \\
\indent
The plots are obtained by estimating the respective quantities numerically
maintaining certain identities  (e.g. ${\bf I}^2= {\bf Q}^2+{\bf U}^2+{\bf V}^2$ ) to accuracies of
one part in  $10^{-7}$ to $10^{-5}$, at least. The angles in the plots are in radians.\\
\indent 
The important features in the plots for $\Delta\psi$ vs $\omega$
and $\Delta\chi$ vs $\omega$, are  
(a) their spectral dependence (b) the variations in  magnitudes of the  difference of the ellipticity angle $\Delta\chi$ and polarization angle difference  $\Delta\psi$ -- that is  the variation in angle-difference 
(AD) with $\omega$. It is note worthy that, although  $\Delta\psi$ may get affected 
due to geometrical effects like rotation of the coordinate axis of the 
observer frame etc.; however,  $\Delta\chi$ is free from that problem 
-- because -- $\chi_{\phi_{i}}$ remains invariant under rotation about the propagation direction.
The variation of AD, particularly $\Delta\chi$ with  $\omega$ may 
 be  used for making compact-star models {\bf and}  DM identification
\footnote{The technology to detect circular polarization for
 high energy photons in the KeV-MeV region needs to be developed. The 
space observatories -- currently existing and the planned ones are 
capable of detecting only linear polarization \cite{ASTRO-2020}.} through 
multi-wavelength spectro-polarimetric 
studies, This is  because of the  distinct variational pattern in AD with energy
as is evident from figure [\ref{f312m_diff}].\\
\indent
It may be noted in the passing  that an identity involving  ellipticity angle  $\chi_{\phi_{i}}$, the polarization angle $\psi_{\phi_{i}}$ and polarization fraction $\Pi_{\phi_{i}}^P$, can be derived from their constitutive relations. The same turns out to be,
\begin{eqnarray}
  \cos^{2}2\chi_{\phi_{i}}\cos^{2}2\psi_{\phi_{i}}& =& \Pi_{\phi_{i}}^{P}, \label{identity2}\\
  \cos2\chi_{\phi_i}& =&P_{L\phi_i}.
\label{identity3}
\end{eqnarray}
When the expressions of the $\chi_{\phi_{i}}$, $\psi_{\phi_{i}}$, $\Pi_{\phi_{i}}^P$ and $P_{L\phi_i}$ can be found in eqns. (\ref{psi}) to (\ref{pl}). One can in principle identify the interaction type
by using the observational data in models of compact
stars and verifying eqns. (\ref{identity2}) and (\ref{identity3}). Though this identity in principle should be satisfied for each $\omega$,   but the  observed data seems to  fall short of that expectation. One of the reason
behind this may be use of  energy band spectra 
by the detectors, when this relation is supposed to be satisfied for line spectrum. This brings us to explore the use of numerical data in   statical form and use the same for drawing conclusions.\\
\indent
However for correct identification of the nature  of the interaction, 
the magnitude of the variations in  $\Delta\psi_P$  and  $\Delta\chi$  
should  be greater than the  half power diameter (HPD) or the angular resolution  of the detectors in a given energy band. \\
\indent
If the angular resolution  of the detectors  in the soft  X-ray region (1 to 100 KeV) following \cite{HPD_resolution01, {HPD_resolution02}} falls at $4^{\prime\prime}$, they could in principle be  detected with the current
available resolution of onboard detectors.  Furthermore,  the difference in polarization angles $\Delta \psi =(\psi_{\phi^{\prime}} - \psi_{\phi})$   and in ellipticity angles
$\Delta\chi =(\chi_{\phi^{\prime}} - \chi_{\phi} )$, due to  two different interactions,   also seem to  come in the observable range in some energy intervals
as may be  noticed from the plot in  fig.[\ref{f312m_diff}].\\
\indent
On the face of the discrepancy between observed data and the identities presented in eqns. (\ref{identity2}) and (\ref{identity3}), we have tried to estimate the confidence level (C.L) on the numerical size of the observables obtained from the exact numerical solutions. This analysis is relegated to the end of the supplementary section \cite{Supp}.
  
\section{ Conclusions}
In this paper we have compared the mixing dynamics of  
 $\frac{1}{M}\phi FF$  with  $\frac{1}{M} \phi' F \tilde{F}$ 
with the inclusion of the photon-self-energy-correction (PSEC) evaluated 
in a weakly-magnetized-finite density-medium evaluated at linear order in field strength ($eB$). Inclusion of 
the same modifies their mixing-dynamics of the available degrees of freedom among themselves;
as a result, the mixing matrix for $\frac{1}{M}\phi FF$ becomes $3 \times 3$ 
 and the same for $\frac{1}{M} \phi' F \tilde{F}$  
becomes  $4 \times 4 $ from $2 \times 2$, that is realised in ${\bf MV}$. 
That is, for the $\phi\gamma$ system, there is a mixing between
only three degrees of freedom but for $\phi'\gamma$ system there 
is a mixing among all four available degree of freedom; i.e., the pseudoscalar, 
the longitudinal and the two transverse degrees of freedom of the photon.
 Hence the spectro-polarimetric signals, due to each of them, once PSEC
is considered -- undergoes sizeable amount of modification.\\
\indent
Here we have also estimated difference in the polarimetric observables due to change in the mixing pattern  
for both cases i.e., $\phi\gamma$ and $\phi'\gamma$ system numerically 
and have plotted  in figures [\ref{f312m_diff}]. Our results 
indicate that the differences in the observables are within the detector resolution limit.\\
\indent
Our findings indicate that, magnetized environment of compact astrophysical systems may 
be a good place to look for the DM signatures, coming from scalars as well as pseudoscalars. Such activity  complimentary  to the searches going on 
else-where \cite{subir-axion, servant}.\\

\pagebreak

\section*{Supplementary Document}

\renewcommand{\thefigure}{S\arabic{figure}}
\renewcommand{\theequation}{S\arabic{equation}}
\section{Introduction}

The particles (beyond the standard model) associated with the breaking of chiral symmetry(pseudoscalars)
$\phi'(x)$ \cite{Peccei,Weinberg,Wilczek,Mohapatra}, through quantum effects and the Goldstone boson of a spontaneously broken scale symmetry (dilaton) $\phi(x)$ \cite{Kim, donoghue}, have remained possible candidates of Dark matter for some times now.\\
\indent
 According to our current understanding, about  27 $\%$ of the total matter-energy density is in the form of  Dark matter.  If $\phi $ or $\phi'$ constitutes the  dark-matter that is being  talked about, then their  cosmological  relic density  $\rho(t)$ ( as $\phi$ or $\phi'$) at some epoch $t$, in terms of their lifetime $\tau_{\phi_{i}}$ in $\phi\to \gamma\gamma$ channel, would  be  given by \cite{Kim},
\begin{eqnarray}
\rho(t) &=& \rho_{d} \left(\frac{\eta_d}{\eta}\right)^3 e^{-\frac{t}{\tau_{\phi_i}}}, \nonumber \\
        &=& \frac{\zeta(3)}{\pi^2}\frac{g*(t)}{g_{*d}}T^3(t) e^{-\frac{t}{\tau_{\phi_i}}}.
\label{relic-density}
\end{eqnarray}
In equation (\ref{relic-density})  $\rho_{d}$ corresponds to 
the density and $\eta_d$ the magnitude of the scale factor of Friedman-Robertson-Walker metric. 
And $g_{*d}$ along with $g_{*}$ are the number of degrees of freedom available at the time of 
decoupling and the same at the epoch t, respectively. The last line in 
equation (\ref{relic-density}) has been obtained demanding entropy conservation 
in the co-moving volume of the universe, from the time of decoupling to the epoch $t$. \\
\indent
Their life-time $\tau_{\phi_{i}}$, in turn depends on the parameters of the theory, i.e., scalar-photon  coupling constant   $g_{\phi\gamma\gamma}$ and scalar  mass $m_{\phi}$ or    pseudoscalar-photon coupling constant  $g_{\phi'\gamma\gamma}$ and pseudoscalar's $m_{\phi'}$;  those  regulate the interaction dynamics of these exotic particles 
with photon $(\gamma)$. Hence  estimations of these are intimately related to the identifications of these particles.\\
\indent
 The    Dim-5 operator   $g_{\gamma\gamma\phi}\phi F ^{\mu\nu} F_{\mu\nu}$   that regulates  the 
  massive  scalar - photon  interaction also 
induces  optical activity   to the vacuum 
in a quasi-static or static magnetic field B\cite{Miani,Raffelt}.
 The same is also  true  for  pseudo-scalar- photon system. These interactions  open up the 
possibility of determination of  $g_{\phi\gamma\gamma }    ,~ m_{\phi'} $ or   $g_{\phi'\gamma\gamma }, ~ m_{\phi}$ by studying their  polarimetric signatures.  However, if there are degeneracies in their  masses 
($m_{\phi}$ or $m_{\phi'}$) as well as in  coupling constants  ($g_{\phi\gamma\gamma}$  or $g_{\phi'\gamma\gamma}$), identification would not be so  easy. \\
\indent
In the following  sections,  we have proposed a model to  resolve the issue of indistinguishable nature of these particles (scalars and pseudoscalars ) by studying  their mixing  dynamics in presence of magnetised plasma.  
The mixing matrix for $\phi F_{\mu\nu}F^{\mu\nu}$ interaction ,  turns out to be $3\times 3$ instead of $2\times 2$, usually encountered in ${\bf MV}$, or magnetized plasma.  For  $\phi' F_{\mu\nu}F^{\mu\nu}$  interaction the situation turns out to be   more interesting.  In this situation, mixing of all four degrees of freedom  --three 
degrees of freedom of a photon and one degree of freedom of pseudoscalar --
takes place . This   results in   he mixing matrix   being   $4 \times 4$.\\
\subsection{Electromagnetic form-factors for $A_{\mu}(k)$}
\indent 
In contrast to vacuum --the gauge fields $A_{\mu}(k)$-- for in medium photons, 
can be written in momentum space,  in terms of three electromagnetic form factors: $  A_{\parallel}(k)$, $  A_{\perp}(k)  $, $ A_{L}(k) $ along with the
gauge fixing term $A_{gf}(k)$ and four orthonormal four vectors constructed out of
the available 4-vectors and tensors for the system  in hand:
\begin{eqnarray}
A_{\alpha}(k)= A_{\parallel}(k)\rm{N}_1 b^{(1)}_{\alpha}+ A_{\perp}(k) \rm{N}_2I_{\alpha}+ A_{L}(k)
\rm{N}_L\tilde{u}_{\alpha} + \frac{k_{\alpha}}{k^2}A_{gf}(k).
\label{gauge-pot2}
\end{eqnarray}
Here $N_i$s are the normalization constants and $A_i$s are the form factors already defined above.
We next set  $A_{gf}=0$ , to comply with the Lorentz gauge condition $k_{\mu}. A^{\mu}(k)= 0$.
These vectors, introduced in equation (\ref{gauge-pot2}), are defined as,
\begin{eqnarray}
 \hat{b}^{(1)\nu} \!\!\! &=& \!\!\! N_1  k_\mu\bar{F}^{\mu\nu},  b^{(2)\nu} = k_\mu \tilde{\bar{F}}^{\mu\nu} ,~ \hat{\tilde{u}}^\nu = N_L \left( g^{\mu\nu} - \frac{k^\mu k^\nu}{k^2}\right)u_\mu , 
\hat{I}^\nu = N_2\left(b^{(2)^\nu} - \frac{(\tilde{u}^\mu b^{(2)}_\mu)}{\tilde{u}^2}\tilde{u}^\nu \right) \nonumber \\
\tilde{\bar{F}}^{\mu\nu}\!\!\!&=& \!\!\!\frac{1}{2}\epsilon^{\mu\nu\lambda\rho}\bar{F}_{\lambda\rho}.
\label{ortho-vectors}
\end{eqnarray}
The normalization constants, $N_{1}$ $N_{2}$ and $N_{L}$ in equation 
(\ref{ortho-vectors}) are given by,
\begin{eqnarray}
N_{1} = \frac{1}{\sqrt{-b^{(1)}_\mu b^{(1)^\mu}}}  
=\frac{1}{K_{\perp}B},~~
N_{2} = \frac{1}{\sqrt{-I_\mu I^\mu}}=\frac{K}{\omega K_{\perp}B}    
\mbox{~~~and~~ }
N_{L} = \frac{1}{\sqrt{-\tilde{u}_\mu \tilde{u}^\mu}}=\frac{\sqrt{(k^{\mu}k_{\mu})}}{\vec{|k|}},
\end{eqnarray}
where $ K_{\perp} = (k^2_1 + k^2_2)^{\frac{1}{2}}$. 

In the table below, we provide the transformation properties of different vectors and the associated form factors under the charge conjugation (${\bf C}$), Time reversal (${\bf T}$)  and Parity inversion (${\bf P}$) transformations. Although  ${\bf P}$  and ${\bf T}$  transformation are a part of  Lorentz transformation but they are not orthogonal proper Lorentz transformation (OPLT). Therefore the variables transforming under discrete transformations may remain inert under the action of the generators of symmetries of the OPLT.  
\begin{table}[h!]
\centering
\begin{tabular}{||c | c c c c c c c c c c c c c ||} 
 \hline
         & $F_{\mu\nu}$     & $k_{\mu}$     &   $u_{\mu}$ &    $\tilde{u}_{\mu}$ & $b^{(1)}_{\mu}$ & $b^{(2)}_{\mu}$ & $ I_{\mu}$ & $A_{||}$ & $ A_{\perp} $ &  $A_{L}$  &$f_{\mu\nu}$ & $i$ & $\epsilon_{\mu\nu\rho\sigma}$\\ [0.5ex] 
 \hline\hline
 {\bf C} & $-F_{\mu\nu}$    & $k_{\mu}$     & $-u_{\mu}$ &  $ -\tilde{u}_{\mu}$   & $-b^{(1)}_{\mu}$  & $-b^{(2)}_{\mu}$  &   $-I_{\mu}$  &  $A_{||}$    & $  A_{\perp}  $ &  $A_{L}$ & $-f_{\mu\nu}$ & $i$ & $\epsilon_{\mu\nu\rho\sigma}$    \\ 
 {\bf P} & $ F^{\mu\nu}$    & $k^{\mu}$     &   $u^{\mu}$ &  $ ~\tilde{u}^{\mu}$   & $- b^{(1)}_{\mu}$ & $ b^{(2)\mu}$ &    $I^{\mu}$  & $ A_{||}$    & $  A_{\perp}  $ &  $ A_{L}$   & $f_{\mu\nu}$ & $i$ & $-\epsilon_{\mu\nu\rho\sigma}$  \\
 {\bf T} & $-F^{\mu\nu}$    & $k^{\mu}$     &  $-u^{\mu}$ &    $-\tilde{u}^{\mu}$  & $ b^{(1)}_{\mu}$ & $ -b^{(2)\mu}$   &   $-I^{\mu}$  & $ - A_{||} $ &  $ - A_{\perp}$ &  $ - A_{L} $ &$-f_{\mu\nu}$ & $-i$ & $-\epsilon_{\mu\nu\rho\sigma}$ \\ [1ex]
 
\hline
 \hline
\end{tabular}
\caption{ Tranformation properties for the vectors, tensors and the electromagnetic form factors used to describe $A_{\mu}(k)$ in equation (\ref{gauge-pot.}), under ${\bf C}$, ${\bf P}$ and ${\bf T}$.}
\label{table:1}
\end{table}

\subsection{Gauge field degrees of freedom}
In a material medium the effective Lagrangian in momentum space will be given by, 
\begin{eqnarray}
L_{eff(m)} = \frac{-1}{4} F_{\mu\nu}F^{\mu\nu} + \frac{1}{2}A_{\mu}(-k)\Pi^{\mu\nu}(k,\mu,T)A_{\mu}(k).
\end{eqnarray}
Where, $\Pi_{\mu\nu}(k,\mu, T)$ is the photon self energy tensor including medium corrections. The number of effective dof for the system can be
estimated by estimating the number of first class constraints (FCC) and second class constraints (SCC), that the Lagrangian $L_{eff(m)}$ have. Due to complicated 
structure of $\Pi_{\mu\nu}(k,\mu,T)$ in momentum space, the evaluation of the analytically without any approximation is bit difficult.
However, in the infrared limit the same takes the form,
\begin{eqnarray}
L_{eff(m)} = \frac{-1}{4} F_{\mu\nu}F^{\mu\nu} + \frac{1}{2}\omega_{p}^{2}(u.A)(u.A).
\end{eqnarray}
 This form of the effective Lagrangian, when $u^{\mu}= (1,0,0,0)$ turns out to be
 very close to the Proca Lagrangian. So following the analysis of Proca Lagrangian, this one
  also would have  two SCC and no FCC. Hence the number of degrees  N of freedom would be,
  \begin{eqnarray}
  N= \frac{(N_{PSV}-n_{SCC})}{2} = 3,
  \end{eqnarray}
 where $N_{PSV}$ is number of phase space variables. Therefore, one can now set $A_{gf}$ equal to zero in equation (\ref{gauge-pot2}). As a consequence the $U(1)$ gauge field is seen to obey the condition,
 \begin{eqnarray}
 k_{\mu}A^{\mu}=0.
 \end{eqnarray}
  That resembles the Lorentz gauge condition. Introduction of the contribution from magnetized medium does not alter this conclusion.
\section{Incorporation of matter effects}
\indent
Matter effects are incorporated    through the inclusion of a term    of the following form, $A^{\mu}\Pi_{\mu\nu}(k,\mu, T)A^{\nu}$ in 
the effective  Lagrangian ($L_{eff}$) of the system  \cite{Raffelt-magnetized,Pankaj-Ralston,Ganguly-ann,Ganguly-jain,Tercas,subir-axion,choi-kim}.
 This  is a scalar made up by contracting   in-medium photon self-energy tensor $\Pi_{\mu\nu}(k, \mu, T)$   with gauge fields. Parameters 
 T and $\mu$   stand for 
temperature and chemical potential as arguments  of   $\Pi_{\mu\nu}(k, \mu, T)$.\\
\indent
Such a system can best be described in terms of a set of  photon form-factors 
introduced in \cite{Ganguly-jain}. Unlike \cite{Ganguly-jaiswal} these form 
factors \cite{Ganguly-jain}, turn out to be even or odd under
Time ({\bf T}) reversal or parity ({\bf P}) symmetry transformations (for details consult table [1] in \cite{Supp}. 
The $\phi\gamma\gamma$ Lagrangian  in an external magnetic field, generates couplings between  {\bf PT} violating  $A_{\parallel}$ and {\bf PT} symmetric $\phi$ through  a multiplicative {\bf T} violating factor $i$.
The other two  {\bf PT } violating electromagnetic form factors $A_{\perp}$ and $A_{L}$  (see table [1]) have no coupling with the scalar $\phi$. However, the pseudoscalars (axion) under similar situation, couples to $A_{\perp}$ and $A_{L}$. Therefore, with the inclusion of $\Pi_{\mu\nu}(k, \mu, T)$, when $A_{L}$ becomes non-zero, the mixing matrix for $\phi\gamma$ remains $2\times2$ but the same for  $\phi'\gamma$, turns out to be $3\times3$.

\section{Mixing dynamics of $\phi\gamma$ interaction} In magnetised media
the effective  Lagrangian for $\phi\gamma$  interaction including the faraday effect is given as:
\begin{eqnarray}
L_{eff,\phi}=      \frac{1}{2}\phi[k^2-m_{\phi}^2] -\frac{1}{4}f_{\mu\nu}f^{\mu\nu} +\frac{1}{2} A_{\mu}\Pi^{\mu\nu}(k,\mu,T,eB)A_{\nu}  -\frac{1}{4}g_{\phi\gamma\gamma}\phi{\bar{ F}}^{\mu\nu}f_{\mu\nu}.
\end{eqnarray}
The equations of motion for the electromagnetic form factors for the photon,
in the notation  of \cite{Ganguly-jain}, turn out to be,
\begin{eqnarray}
 (k^2-\Pi_T)A_{\parallel} (k)+ i\Pi^{p}(k)N_{1}N_{2}\left[\epsilon_{\mu\nu\delta\beta}
\frac{k^{\beta}}{\mid{k}\mid}u^{\tilde{\delta}_{\parallel}} b^{(1)\mu} I^{\nu}\right] A_{\perp}(k)
&=& \frac{ig_{\phi\gamma\gamma}\phi(k)}{N_1},
\label{fd1} \\
 (k^2-\Pi_T)A_{\perp} (k) - i\Pi^{p}(k)N_{1}N_{2}\left[\epsilon_{\mu\nu\delta\beta}
\frac{k^{\beta}}{\mid{k}\mid}u^{\tilde{\delta}_{\parallel}} b^{(1)\mu} I^{\nu}\right]A_{\parallel}(k)&= &0,
\label{fd2} \\
(k^2 -\Pi_L)A_L(k) &=& 0. 
\label{fd3}
\end{eqnarray}   
\noindent
The three equations (i.e., (\ref{fd1})-(\ref{fd3})), describe the dynamics of the 
three degrees of freedom of the photon.   And the 
equation of motion for $\phi$ is given by,
\begin{equation}
(k^2 - m^2)\phi(k) = -\frac{ig_{\phi\gamma\gamma} A_{\parallel}(k)}{N_1}. 
\end{equation}
For notational convenience, we, introduce further the new variables $F$ and $G$, defined 
as,
\begin{eqnarray}
F =N_{1}\,N_{2}\, \Pi^{p}(k)\left[\epsilon_{\mu\nu\delta\beta}
\frac{k^{\beta}}{\mid{k}\mid}u^{\tilde{\delta}_{\parallel}} b^{(1)\mu} I^{\nu}\right] 
\mbox{~~and~~}  G = \frac{g_{\phi\gamma\gamma}}{N_1}.
\label{F-G}
\end{eqnarray}
\indent
 As stated already, in the long wavelength limit, we 
consider $\Pi_{T}=\omega^{2}_{p}$, where 
$\omega_p = \sqrt{\frac{4\pi\alpha n_e}{m_e}}$, is the plasma frequency, 
$\alpha$  is electromagnetic coupling constant and $n_e$ is the density of 
electrons.
Other terms, $F$ and $G$ introduced in equation (\ref{F-G}) can be simplified to 
yield  $F=\frac{\omega^{2}_{p}}{\omega}\frac{eBcos{{\theta}}}{m_e}$ and 
$G= - g_{\phi\gamma\gamma}B sin\theta \omega$, where $\theta$ is angle between the 
photon propagation vector $\vec{k}$ and the magnetic field B.
 Here $m_e$ is the mass of 
electron and $e$ is the electronic charge. The same  can be cast
in a compact matrix form, as:
\begin{eqnarray}
\left[\begin{array}{c} k^2 {\bf I} - {\bf M }
 \end{array} \right]
\left(\begin{array}{c} A_{\parallel}(k) \\ A_{\perp}(k)\\ A_{L}(k) \\\phi(k) \end{array} \right)=0.
\label{photon-scalar-mixing-matixx1}
\end{eqnarray}
\indent
As we have already mentioned in section two that for $\phi\gamma$ in presence of magnetized media, the longitudinal degree of freedom of photon does not mix with rest others, therefore we will exclude this term from the mixing matrix. Hence, the above $4 \times 4 $ mixing matrix {\bf M} , can be  reduce in $3\times 3$ matrix. For the sake of brevity, using  shorthand notations the equations of motions can be written as follows:
\begin{equation}
\left[\begin{array}{c} k^2 {\bf I} - {\bf M }
 \end{array} \right]
\left(\begin{array}{c} A_{\parallel}(k) \\ A_{\perp}(k) \\\phi(k) \end{array} \right)=0,
\label{photon-scalar-mixing-matixx2}
\end{equation} 
\noindent
where the $3\times3$ mixing matrix for scalar-photon interaction is given as; 

\begin{equation}
{\bf M} =
    \left( \begin{array}{ccc}              
 \omega^2_p   &  \;\;  iF           & \;\; -iG        \\
-iF           & \;\; \omega^{2}_{p} & \;\;    0       \\
iG            &  \;\;  0            & \;\;    m^2_{\phi}  
   \end{array} \right).
\label{mixing-mat1}   
\end{equation}
%
%
In order to get the stoke parameters $(\bf I(\omega,z), Q(\omega,z), U(\omega,z), V(\omega,z))$ of the radiations, we need to get the solutions of {\bf M}. This can be achieved by diagonalizing the same.

\subsection{Diagonalizing the 3 $\times$ 3 mixing matrix}
Our objective here is to obtain the analytical expression for the diagonalizing  (unitary) matrix {\bf U}, for the  hermitian matrix {\bf M}. We express the elements of the same ({\bf U}), in terms of algebraic expressions for maintaining the desired numerical accuracy (O(10$^{-15}$)). In order  to achieve  that, we need to solve for the characteristic equation, 
obtained from $\rm{~Det~}({\bf M}- \lambda{\bf I}) = 0$ to find the eigen values(roots) of {\bf M}; then  use the same to find the 
corresponding eigen vectors. Finally, using the eigen vectors, construct the unitary matrix  {\bf U} 
the diagonalizes {\bf M}. The characteristic equation for this $3 \times 3$  hermitian matrix {\bf M}, for obvious reasons  
turns out to be a cubic equation, having real roots. \\
\indent
The cubic equation, that follows from the characteristic equation can be  written, in 
terms of parameters b, c and d as,  
\begin{eqnarray}
\lambda^3 + b \lambda^2 +c \lambda + d = 0,
\end{eqnarray}
where  the parameters b, c and d are functions of the elements of mixing matrix {\bf M},  
denoted by:
\begin{eqnarray}  
 b &=& -(2\omega^2_p + m^2_{\phi}) \\
 c &=& \omega^4_p + 2 \omega^2_p m^2_{\phi}- \left( \frac{e{ B}_{\parallel}}{m_e}\frac{\omega^2_p}{\omega}\right)^2 - (g_{\gamma\gamma\phi}{ B}_{\perp}\omega)^2 \\
 d &=& -\left[\omega^4_p m^2_{\phi} - \left( \frac{e{B}_{\parallel}}{m_e}\frac{\omega^2_p}{\omega}\right)^2 m^2_{\phi}  -  (g_{\gamma\gamma\phi}{ B}_{\perp}\omega)^2 \omega^2_p \right]. 
\end{eqnarray}  
Next we introduce the variables p and Q, when $p= \left(\frac{3c-b^2}{9}\right)$ and 
$2Q = \left(\frac{2b^3}{27} -\frac{bc}{3} +d \right)$; in terms of them,
the roots turn out to be, \\
\begin{equation}
\begin{array}{cc}
     \lambda_{1} =&   {\bf R} \cos \alpha + \sqrt{3} {\bf R} \sin \alpha -{ b}/3, \\
     \lambda_{2} =&   {\bf R} \cos \alpha - \sqrt{3} {\bf R }\sin \alpha -{ b}/3, \\
     \lambda_{3} =& \hskip -1.5cm -2{\bf R} \cos \alpha  -{ b}/3.
\end{array} 
\mbox{~~With~~} 
  \left\{  \begin{array}{c}
                 \alpha  =\frac{1}{3} \cos^{-1} \left( \frac{Q}{{\bf R}^3} \right )\\
                \,\,\,\,\, {\bf R} = \sqrt{\left(-p \  \right )} {\bf sgn} \left( Q \right) 
     \end{array}
  \right. 
\label{exact-roots}
\end{equation}
\indent
The  orthonormal eigenvectors  $\bf X_j$ of  {\bf M} are  to be found from the matrix relation,  
$\left[ {\bf M}-\lambda_j\right] \left[ {\bf X_{j}}\right] = 0.$ In terms of its elements,
the normalised column vector  $\left[ {\bf X_{j}}\right]$,  can be denoted as,
\begin{equation}
 \left[ {\bf X_j}\right] =  \left[\begin{array}{c}
                                         \bar{u}_{j} \nonumber \\
                                         \bar{v}_{j} \nonumber \\
                                         \bar{w}_{j} \nonumber \\
                                 \end{array}
                          \right].
\end{equation}
\indent
Following standard methods, one can evaluate these elements, in terms of the roots $\lambda_j$ and elements of the 
mixing matrix {\bf M}. The same, once evaluated turns out to be, 
\begin{equation}
\begin{array}{cc}
     \bar{u}_{j} =&  ({\omega^2_p} -\lambda_j)(m_{\phi}^2 -\lambda_j) \times  \rm{\cal{N}}^{(j)}_{vn}, \\
    \bar{v}_{j} =&  i\frac{e{ B}_{\parallel}}{m_e}\frac{\omega^2_p}{\omega}(m_{\phi}^2 -\lambda_j)  \times  \rm{\cal{N}}^{(j)}_{vn}, \\
     \bar{w}_{j} =&  ig_{\gamma\gamma\phi}{ B}_{\perp}\omega({\omega^2_p} -\lambda_j)  \times  \rm{\cal{N}}^{(j)}_{vn}.
\end{array} 
{\mbox{~~when~~}} 
  \left\{  \begin{array}{c}
                 \rm{\cal{N}}^{(j)}_{vn}  = \frac{1}{\sqrt{[|\bar{u}_j|^2 + |\bar{v}_j|^2 +  |\bar{w}_j|^2]}}  \\
                   \end{array}
  \right.
\label{exact-eigen-vectors-3x3}
\end{equation}
Here $\rm{\cal{N}}^{(j)}_{vn}$  is normalisation constant and $j$ can take values from $1$ to $3$. Using these eigenvectors, the unitary matrix {\bf U} turns out to be:
\begin{equation}
{\bf U} =  
               \left( \begin{array}{ccc} 
(\omega^2_p -\lambda_1)(m^2_{\phi} - \lambda_1) \rm{{\cal{N}}}^{(1)}_{vn}                                   &  \hskip .5cm    (\omega^2_p -\lambda_2)(m^2_{\phi} - \lambda_2)  \rm{{\cal{N}}}^{(2)}_{vn}                                      &  \hskip .5cm   (\omega^2_p -\lambda_3)(m^2_{\phi} - \lambda_3)\rm{{\cal{N}}}^{(3)}_{vn}  \\
i\frac{e{B}_{\parallel}}{m_e}\frac{\omega^2_p}{\omega} (m^2_{\phi} - \lambda_1)\rm{{\cal{N}}}^{(1)}_{vn}    &  \hskip .5cm    i\frac{e{B}_{\parallel}}{m_e}\frac{\omega^2_p}{\omega} (m^2_{\phi} - \lambda_2) \rm{{\cal{N}}}^{(2)}_{vn}       & \hskip .5cm  i\frac{e{ B}_{\parallel}}{m_e}\frac{\omega^2_p}{\omega} (m^2_{\phi} - \lambda_3)\rm{{\cal{N}}}^{(3)}_{vn} \\
 ig_{\gamma\gamma\phi}{B}_{\perp}\omega(\omega^2_p -\lambda_1)\rm{\cal{N}}^{(1)}_{vn}                     & \hskip .5cm  ig_{\gamma\gamma\phi}{B}_{\perp}\omega (\omega^2_p -\lambda_2) \rm{\cal{N}}^{(2)}_{vn}                           & \hskip .5cm  ig_{\gamma\gamma\phi}{B}_{\perp}\omega (\omega^2_p -\lambda_3)\rm{\cal{N}}^{(3)}_{vn}
                       \end{array} 
             \right).
 \label{u-mat}
\end{equation} 
\indent
The obtained Unitary matrix (\ref{u-mat}) diagonalises the 3$\times$3 mixing matrix $\bf{M}$. We used the diagonalized mixing matrix ${\bf M}_{D}$ to produce the solution of the field equation. The necessary steps have been stated in succeeding sections. 
\subsection{Field equation : Solutions}
In order to solve the coupled equation ( \ref{photon-scalar-mixing-matixx}) we multiply the same by {\bf{U}$^{-1}$} to get,

\indent
\begin{equation}
{\bf{U}}^{-1}\left[\begin{array}{c} k^2 {\bf I} - {\bf M }
 \end{array} \right]{\bf{UU}}^{-1}
\left(\begin{array}{c} A_{\parallel}(k) \\ A_{\perp}(k) \\\phi(k) \end{array} \right)=
\left[\begin{array}{c} k^2 {\bf I} - {\bf M}_{D} 
 \end{array} \right],
\left(\begin{array}{c} A'_{\parallel}(k) \\ A'_{\perp}(k) \\\phi'(k) \end{array} \right)=0,
\end{equation}

\noindent
where matrix ${\bf U} $ is given in equation (\ref{u-mat}) and $ {\bf U}^{-1}$ is the inverse 
 of the same and {\bf M}$_{D}$ is the diagonal matrix having eigen values as diagonal elements. Here we have denoted
\begin{eqnarray}
 \left[\begin{array}{c} A'_{\parallel}(k) \\ A'_{\perp}(k) \\ \phi'(k) 
\end{array} \right]={\bf U}^{-1}
\left[\begin{array}{c} A_{\parallel}(k) \\ A_{\perp}(k) \\\phi(k) \end{array} \right].
\label{primed-degrees-of-freedom}
\end{eqnarray}

\noindent
For a photon beam, propagating in the z direction, one can Fourier transform the momentum, $k_{3}$ back 
to $z$ and write $k^2 \approx 2\omega(\omega -i\partial_z)$. Recalling that 
${\bf U}^{-1}{\bf M} {\bf U} = {\bf M}_{D}$, 
 equation (\ref{primed-degrees-of-freedom}) 
can further be cast in the form,
\begin{equation}
\left[\begin{array}{c}  (\omega -i\partial_z)  \mathbf{I}  - \left[ \begin{array}{ccc} \frac{\lambda_1}{2\omega} &   0     &    0 \\
0 &  \frac{\lambda_2}{2\omega}  & 0 \\
0 & 0 & \frac{\lambda_3}{2\omega}  \end{array} \right]        \end{array} \right]
\left[\begin{array}{c} {A'}_{\parallel}(z)\\ {A'}_{\perp}(z) \\ {\phi'}(z) \end{array} \right]=0.
\label{dmat}
\end{equation} 
\noindent
The matrix equation (\ref{dmat}) is easy to solve now. Introducing the variables, 
$\Omega_\parallel = \left(\omega - \frac{\lambda_1}{2\omega}\right) $,
$\Omega_\perp = \left(\omega - \frac{\lambda_2}{2\omega}\right)$ 
and $\Omega_\phi = \left(\omega - \frac{\lambda_3}{2\omega}\right)$,
we can now directly write down the solutions for the column vector 
$\left[{\bf A}(z) \right]$; in  matrix form, they are:
%
\begin{equation}
  \left[
\begin{matrix}
{{A}_{\parallel}(z)} \cr
{{A}_{\perp}(z)} \cr   
{\phi(z)}
\end{matrix}
\right] = {\bf U}\left[ \begin{array}{ccc} e^{-i\Omega_\parallel z} &   0    &    0 \\
0 & e^{-i\Omega_\perp z}  & 0 \\
0 & 0 & e^{-i\Omega_\phi z}  \end{array} \right] {\bf U}^{-1} 
\left[
\begin{matrix}
{{A}_{\parallel}(0)} \cr
{{A}_{\perp}(0)} \cr   
{\phi(0)}
\end{matrix}
\right]
\label{solnmat}.
\end{equation}
\noindent
The magnitudes of the elements of column vector $\left[{\bf A}(0) \right]$ in 
equation (\ref{solnmat}), are subject to 
the physical situations/conditions prevailing at the origin. Using the initial conditions stated in section (2.1), 
 one can write down the solution of equation (\ref{solnmat}) for {\bf A$_\parallel$}($\omega,{\bf z}$), and it is as follows:
\begin{eqnarray}
{\bf A_{\parallel}(\omega,z)}&=& 
\left( e^{-i\Omega_\parallel z }{\bar u}_1 { \bar u}^{*}_1  +  e^{-i\Omega_\perp z } {\bar u}_2{ \bar u}^{*}_2   
+ e^{-i\Omega_\phi z } {\bar u}_3{ \bar u}^{*}_3 \right){\bf A_{\parallel}(\omega, 0)}  \nonumber \\
&+& \left(
e^{-i\Omega_\parallel z }{\bar u}_1 { \bar v}^{*}_1  +  e^{-i\Omega_\perp z } {\bar u}_2{ \bar v}^{*}_2   
+ e^{-i\Omega_\phi z } {\bar u}_3{ \bar v}^{*}_3 \right){\bf A_{\perp}(\omega, 0)}. 
\label{soln-a-parallel-SUP}
\end{eqnarray}  
\noindent 
Similarly the perpendicular component $A_{\perp}(\omega,z) $, turns out to be, 
\begin{eqnarray}
{\bf A_{\perp}(\omega,z)}&=& 
\left( e^{-i\Omega_\parallel z }{\bar v}_1 { \bar u}^{*}_1  +  e^{-i\Omega_\perp z } {\bar v}_2{ \bar u}^{*}_2   
+ e^{-i\Omega_\phi z } {\bar v}_3{ \bar u}^{*}_3 \right){\bf A_{\parallel}(\omega, 0)}  \nonumber \\
&+& \left(
e^{-i\Omega_\parallel z }{\bar v}_1 { \bar v}^{*}_1  +  e^{-i\Omega_\perp z } {\bar v}_2{ \bar v}^{*}_2   
+ e^{-i\Omega_\phi z } {\bar v}_3{ \bar v}^{*}_3 \right){\bf A_{\perp}(\omega, 0)}.  
\label{soln-a-perp-SUP}
\end{eqnarray}  

\subsection{Stokes parameters}
\indent
 To find out the Stokes parameters for Scalar-photon mixing, we need to evaluate  ${\bf{|A_{\parallel}(\omega,z)|^2}}$ and ${\bf{|A_{\perp}(\omega,z)|^2}}$, 
using eqns.[\ref{soln-a-parallel-SUP}] and [\ref{soln-a-perp-SUP}]. 
Introducing the new variables, $\mathbb{P} = |{\bar u}_1||{\bar v}_1|$, 
$\mathbb{Q} = |{\bar u}_2||{\bar v}_2|$ and $\mathbb{R} = |{\bar u}_3||{\bar v}_3|$, the expression for 
${\bf{|A_{\parallel}(\omega,z)|^2}}$ in terms of them, turns out to be,
\begin{align}
{\bf{|A_{\parallel}(\omega,z)|^2}} =  \Big[1 
- &4|{\bar{u}_1}|^2|{\bar u}_2|^2\; sin^2 \left(\frac{(\Omega_{\parallel}-\Omega_{\perp})z}{2} \right)
- 4|{\bar u}_2|^2|{\bar u}_3|^2\; sin^2 \left(\frac{(\Omega_{\perp} -\Omega_{\phi})z}{2} \right)\nonumber\\  
- &4|{\bar u}_3|^2|{\bar u}_1|^2\; sin^2 \left(\frac{(\Omega_{\phi} -\Omega_{\parallel})z}{2} \right)\Big]\nonumber\\
\times &{\bf |A_{\parallel}(\omega,0)|^2}\nonumber\\ 
- 4 \Big[&\mathbb{PQ}\; sin^2 \left(\frac{(\Omega_{\parallel} -\Omega_{\perp})z}{2} \right) 
+ \mathbb{QR}\; sin^2 \left(\frac{(\Omega_{\perp} -\Omega_{\phi})z}{2} \right)\nonumber\\  
+ &\mathbb{RP}\; sin^2 \left(\frac{(\Omega_{\phi} -\Omega_{\parallel})z}{2} \right)\Big]\nonumber\\
\times & {\bf |A_{\perp}(\omega,0)|^2 }\nonumber\\
+ \Big[&|{\bar u}_1 {\bar u}_2|(|{\bar u}_1{\bar v}_2|- |{\bar u}_2{\bar v}_1|)sin \left({(\Omega_{\parallel} -\Omega_{\perp})z} \right) \nonumber\\
+& |{\bar u}_2 {\bar u}_3|(|{\bar u}_2{\bar v}_3|-|{\bar u}_3{\bar v}_2|)sin \left({(\Omega_{\perp} -\Omega_{\phi})z} \right) \nonumber\\
+&|{\bar u}_3 {\bar u}_1|(|{\bar u}_3{\bar v}_1|-|{\bar u}_1{\bar v}_3|)sin \left({(\Omega_{\phi} -\Omega_{\parallel})z} \right)\Big]\nonumber\\
\times &2{\bf |A_{\parallel}(\omega,0)|}{\bf |A_{\perp}(\omega,0)|}.
\label{A_parallel-square}
\end{align}
\indent
Similarly we can find  ${\bf{|A_{\perp}(\omega,z)|^2}}$.
The expression for the same, in terms of the variables introduced earlier, 
turns out to be, 
\begin{align}
{\bf{|A_{\perp}(\omega,z)|^2}} =- 4\Big[&\mathbb{PQ}sin^2 \left(\frac{(\Omega_{\parallel} -\Omega_{\perp})z}{2} \right) + \mathbb{QR}\; sin^2 \left(\frac{(\Omega_{\perp} -\Omega_{\phi})z}{2} \right)\nonumber\\ 
+&\mathbb{RP}\; sin^2 \left(\frac{(\Omega_{\phi} -\Omega_{\parallel})z}{2} \right)\Big]\nonumber\\
\times &{\bf |A_{\parallel}(\omega,0)|^2} \nonumber\\
+ \Big[1 -& 4|{\bar v}_1|^2|{\bar v}_2|^2 \;sin^2 \left(\frac{(\Omega_{\parallel} 
-\Omega_{\perp})z}{2} \right)
- 4|{\bar v}_2|^2 |{\bar v}_3|^2\; sin^2 \left(\frac{(\Omega_{\perp} -\Omega_{\phi})z}{2} \right)\nonumber\\  
- &4|{\bar v}_3|^2|{\bar v}_1|^2\; sin^2 \left(\frac{(\Omega_{\phi} -\Omega_{\parallel})z}{2} \right)\Big]\nonumber\\
\times &{\bf |A_{\perp}(\omega,0)|^2}\nonumber\\
+ \Big[&|{\bar v}_1 {\bar v}_2|(|{\bar u}_1{\bar v}_2|-|{\bar u}_2{\bar v}_1|)sin \left({(\Omega_{\parallel} -\Omega_{\perp})z} \right)\nonumber\\
+& |{\bar v}_2 {\bar v}_3|(|{\bar u}_2{\bar v}_3|-|{\bar u}_3{\bar v}_2|)sin \left({(\Omega_{\perp} -\Omega_{\phi})z} \right)\nonumber\\
+&|{\bar v}_3 {\bar v}_1|(|{\bar u}_3{\bar v}_1|-|{\bar u}_1{\bar v}_3|)sin \left({(\Omega_{\phi} -\Omega_{\parallel})z} \right)\Big]\nonumber\\
\times & 2{\bf |A_{\parallel}(\omega,0)|}{\bf |A_{\perp}(\omega,0)|}.
\label{A_perp-square}
\end{align}

\noindent
To express the Stokes parameters given by equations ({\ref{I}-\ref{V}}), obtained from (\ref{A_parallel-square}) and (\ref{A_perp-square}) we had introduced variables ${\cal {I}_{\parallel }}$, ${\cal {I}_{\perp }}$ and ${\cal {I}_{\parallel \perp }}$ to express {\bf I}. They are given by:
 
\begin{align}
{\cal {I}_{\parallel }}&= 1 - 4(|{\bar u}_1 {\bar u}_2|^2 + \mathbb{PQ}) sin^2 \left(\frac{(\Omega_{\parallel} -\Omega_{\perp})z}{2} \right)  
-4(|{\bar u}_2{\bar u}_3|^2 +\mathbb{QR})sin^2 \left(\frac{(\Omega_{\perp} -\Omega_{\phi})z}{2} \right)\nonumber\\
&\hskip0.62cm-4(|{\bar u}_3 {\bar u}_1|^2 +\mathbb{RP})sin^2 \left(\frac{(\Omega_{\phi} -\Omega_{\parallel})z}{2} \right) .
\end{align}
\begin{align}
{\cal {I}_{\perp }} &= 1 - 4(|{\bar v}_1 {\bar v}_2|^2 + \mathbb{PQ}) sin^2 \left(\frac{(\Omega_{\parallel} -\Omega_{\perp})z}{2} \right) 
- 4(|{\bar v}_2{\bar v}_3|^2 +\mathbb{QR})sin^2 \left(\frac{(\Omega_{\perp} -\Omega_{\phi})z}{2} \right)\nonumber\\ 
&\hskip0.62cm-4(|{\bar v}_3 {\bar v}_1|^2 +\mathbb{RP})sin^2 \left(\frac{(\Omega_{\phi} -\Omega_{\parallel})z}{2} \right). 
\end{align}

\begin{align}
{\cal {I}_{\parallel \perp }}&=(|{\bar u}_1{\bar v}_2| -|{\bar u}_2{\bar v}_1|)|{\bar w}_1{\bar w}_2|sin \left({(\Omega_{\parallel} -\Omega_{\perp})z} \right)
+(|{\bar u}_2{\bar v}_3| -|{\bar u}_3{\bar v}_2|)|{\bar w}_2{\bar w}_3|sin \left({(\Omega_{\perp} -\Omega_{\phi})z} \right)
\nonumber\\ 
&+ (|{\bar u}_3{\bar v}_1|-|{\bar u}_1{\bar v}_3|)|{\bar w}_3{\bar w}_1|sin \left({(\Omega_{\phi} -\Omega_{\parallel})z} \right). 
\end{align}
\noindent
Similarly, to express Stokes parameter {\bf Q}, we had introduced the variables  ${\cal {Q}_{\parallel }}$, ${\cal {Q}_{\perp }} $ and ${\cal {Q}_{\parallel \perp }}$. Their actual forms are,
\begin{align}
{\cal {Q}_{\parallel}}& = 
1 - 4(|{\bar u}_1{\bar u}_2|^2-\mathbb{PQ}) sin^2 \left(\frac{(\Omega_{\parallel} -\Omega_{\perp})z}{2} \right)
- 4(|{\bar u}_2{\bar u}_3|^2 -\mathbb{QR})sin^2 \left(\frac{(\Omega_{\perp} -\Omega_{\phi})z}{2} \right)\nonumber\\
&\hskip0.62cm-  4 (|{\bar u}_3{\bar u}_1|^2-\mathbb{RP})sin^2 \left(\frac{(\Omega_{\phi} -\Omega_{\parallel})z}{2} \right). 
\end{align}
\begin{align}
{\cal {Q}_{\perp}}&= 1-4(|{\bar v}_1 {\bar v}_2|^2-\mathbb{PQ}) sin^2 \left(\frac{(\Omega_{\parallel} -\Omega_{\perp})z}{2} \right)
- 4(|{\bar v}_2 {\bar v}_3|^2-\mathbb{QR})sin^2 \left(\frac{(\Omega_{\perp} -\Omega_{\phi})z}{2} \right)\nonumber\\ 
& \hskip0.62cm -4 (|{\bar v}_3 {\bar v}_1|^2-\mathbb{RP})sin^2 \left(\frac{(\Omega_{\phi} -\Omega_{\parallel})z}{2} \right). 
\end{align}
\begin{align}
{\cal {Q}_{\parallel \perp}}&= (|{\bar u}_1{\bar v}_2| -|{\bar u}_2{\bar v}_1|)(|{\bar u}_1{\bar u}_2| -|{\bar v}_1{\bar v}_2|)sin \left({(\Omega_{\parallel} -\Omega_{\perp})z} \right) \nonumber\\
&\hskip-0.04cm+ (|{\bar u}_2{\bar v}_3| -|{\bar u}_3{\bar v}_2|)(|{\bar u}_2{\bar u}_3| -|{\bar v}_2{\bar v}_3|)sin \left({(\Omega_{\perp} -\Omega_{\phi})z} \right)\nonumber\\
&\hskip-.04cm+ (|{\bar u}_3{\bar v}_1| -|{\bar u}_1{\bar v}_3|)(|{\bar u}_3{\bar u}_1| -|{\bar v}_3{\bar v}_1|)sin \left({(\Omega_{\phi} -\Omega_{\parallel})z} \right).
\end{align}
 In the same sprit, the expanded expressions for the variables ${\cal {U}_{\parallel }}$, ${\cal {U}_{\perp }} $ and ${\cal {U}_{\parallel \perp }}$, already  introduced in the text to express Stokes parameter {\bf U}, are given by:
\begin{align}
{\cal {U}_{\parallel}} &= |{\bar u}_1{\bar u}_2|(|{\bar v}_1{\bar u}_2| -|{\bar u}_1{\bar v}_2|)
sin \left({(\Omega_{\parallel} -\Omega_{\perp})z} \right) 
+|{\bar u}_2{\bar u}_3|(|{\bar v}_2{\bar u}_3| -|{\bar u}_2{\bar v}_3|)
sin \left({(\Omega_{\perp} -\Omega_{\phi})z}\right)  \nonumber\\
&+|{\bar u}_3{\bar u}_1|(|{\bar v}_3{\bar u}_1| -|{\bar u}_3{\bar v}_1|)
sin \left({(\Omega_{\phi} -\Omega_{\parallel})z} \right). 
\end{align}
\begin{align}
{\cal {U}_{\perp}}&= |{\bar v}_1{\bar v}_2|(|{\bar v}_1{\bar u}_2| - |{\bar u}_1{\bar v}_2|)
sin \left({(\Omega_{\parallel} -\Omega_{\perp})z} \right)
+|{\bar v}_2{\bar v}_3|(|{\bar v}_2{\bar u}_3| -|{\bar u}_2{\bar v}_3|)
sin \left({(\Omega_{\perp} -\Omega_{\phi})z} \right) \nonumber\\
&+ |{\bar v}_3{\bar v}_1|(|{\bar v}_3{\bar u}_1| -|{\bar u}_3{\bar v}_1|)
sin \left({(\Omega_{\phi} -\Omega_{\parallel})z} \right).
\end{align}
\begin{align}
\hskip -4cm{\cal {U}_{\parallel \perp}}& =  (|{\bar v}_1{\bar u}_2| - |{\bar u}_1{\bar v}_2|)^2 \;cos \left({(\Omega_{\parallel} -\Omega_{\perp})z} \right)
+(|{\bar v}_2{\bar u}_3| - |{\bar u}_2{\bar v}_3|)^2\;cos \left({(\Omega_{\perp} -\Omega_{\phi})z} \right)\nonumber\\ 
&+ (|{\bar v}_3{\bar u}_1| -|{\bar u}_3{\bar v}_1|)^2\;cos \left({(\Omega_{\phi} -\Omega_{\parallel})z} \right). 
\end{align}
  And lastly, the expressions for ${\cal {V}_{\parallel }}$, ${\cal {V}_{\perp }} $ and ${\cal {V}_{\parallel \perp }}$, introduced to express the measure of circular polarisation {\bf V}  are:
\begin{align}
{\cal {V}_{\parallel}}&=  [(|{\bar u}_1 {\bar v}_1||{\bar u}_1|^2 + |{\bar u}_2 {\bar v}_2||{\bar u}_2|^2 + |{\bar u}_3{\bar v}_3||{\bar u}_3|^2) 
+ |{\bar u}_1{\bar u}_2|({|\bar v}_1{\bar u}_2| + |{\bar u}_1{\bar v}_2|)\;cos \left({(\Omega_{\parallel} -\Omega_{\perp})z} \right)\nonumber\\ 
&+ |{\bar u}_2{\bar u}_3|(|{\bar v}_2{\bar u}_3| + |{\bar u}_2{\bar v}_3|)\;cos \left({(\Omega_{\perp} -\Omega_{\phi})z} \right)
+|{\bar u}_3 {\bar u}_1|(|{\bar v}_3{\bar u}_1| +|{\bar u}_3{\bar v}_1|)\;cos \left({(\Omega_{\phi} -\Omega_{\parallel})z} \right)].
\end{align}
\begin{align}
{\cal {V}_{\perp}}&= [(|{\bar u}_1 {\bar v}_1||{\bar v}_1|^2 + |{\bar u}_2 {\bar v}_2||{\bar v}_2|^2 + |{\bar u}_3 {\bar v}_3||{\bar v}_3|^2) 
+{|\bar v}_1{\bar v}_2|({|\bar u}_1{\bar v}_2| + |{\bar v}_1{\bar u}_2|)\;cos \left({(\Omega_{\parallel} -\Omega_{\perp})z} \right)\nonumber\\ 
&+ {\bar v}_2{\bar v}_3|(|{\bar u}_2{\bar v}_3| + |{\bar u}_3{\bar v}_2|)\;cos \left({(\Omega_{\perp} -\Omega_{\phi})z} \right)
+|{\bar v}_3 {\bar v}_1|(|{\bar u}_3{\bar v}_1| + |{\bar u}_1{\bar v}_3|)\;cos \left({(\Omega_{\phi} -\Omega_{\parallel})z} \right)]. 
\end{align}
\begin{align}
\hskip-.7cm{\cal {V}_{\parallel \perp}}& = [(|{\bar u}_1|^2|{\bar v}_2|^2 - |{\bar u}_2|^2|{\bar v}_1|^2)sin \left({(\Omega_{\parallel} -\Omega_{\perp})z} \right)
+(|{\bar u}_2|^2|{\bar v}_3|^2 - |{\bar u}_3|^2|{\bar v}_2|^2)
sin \left({(\Omega_{\perp} -\Omega_{\phi})z} \right)\nonumber\\
&+(|{\bar u}_3|^2|{\bar v}_1|^2 - |{\bar u}_1|^2|{\bar v}_3|^2)
sin \left({(\Omega_{\phi} -\Omega_{\parallel})z} \right)].
\end{align}


\section{Mixing dynamics of $\phi'\gamma$ interaction in magnetized medium }

The effective Lagrangian for photon-pseudoscalar interaction including Faraday term is given as: 
\begin{eqnarray}
L_{eff,\phi'}=      \frac{1}{2}\phi'[k^2-m_{\phi'}^2] -\frac{1}{4}f_{\mu\nu}f^{\mu\nu} +\frac{1}{2} A_{\mu}\Pi^{\mu\nu}(k,\mu,T,eB)A_{\nu}  -\frac{1}{4}g_{\phi'\gamma\gamma}\phi'\tilde{{\bar{ F}}}^{\mu\nu}f_{\mu\nu}.
\end{eqnarray}

%
The dynamics of the gauge potential in terms, for coupled pseudoscalar-photon system, 
following \cite{Ganguly-jain} turns out to be, 
\begin{eqnarray}
\left[ k^2g_{\mu\nu}-\Pi_{\mu\nu}(k)- \Pi_{p}(k)P_{\mu\nu} \right]
A^{\nu}(k)= ig_{\phi'\gamma\gamma}\tilde{\bar{F}}_{\mu\nu} k^{\nu}\phi'(k),
\label{eom-1}
\end{eqnarray}
\noindent
where the terms have their usual meaning.
The resulting equations of motion for the form factors are given by: 
\begin{eqnarray}
(k^2 - \Pi_T(k))A_{\perp}(k) - \Pi_{p}N_1N_2\left[
P_{\mu\nu}b^{(1)\mu} I^{\nu}
\right]A_{\parallel}(k) + {
\left(i g_{\phi'\gamma\gamma} N_2b^{(2)}_{\mu}I^{\mu}\right)\phi'(k)}\, &=& 0,
\nonumber \\
(k^2 -\Pi_T(k))A_{\parallel}(k)+
\Pi_{p}N_1N_2\left[P_{\mu\nu}b^{(1)\mu}
I^{\nu} \right]A_{\perp}(k) &=& 0, \nonumber \\ 
\left(k^2 - \Pi_L \right)A_L(k)+
{i {g_{\phi'\gamma\gamma}} N_L\left(b^{(2)}_\mu\tilde{u}^{\mu}
\right)\phi'(k)} &= & 0. 
\label{photon-eom} 
\end{eqnarray}
And the  equation of motion (EOM) for the pseudoscalar field  is,
\begin{equation}
\left(k^2 -m^2 \right)\phi'(k) =\left[{\left(i{g_{\phi'\gamma\gamma}}b^{(2)}_{\mu}I^{\mu}\right)}{N}_2 A_{\perp}(k)
+ {\left( i  {g_{\phi'\gamma\gamma}} b^{(2)}_{\mu} \tilde{u}^{\mu} \right)} {N_L} {A_L}(k) \right].
\label{scalar-eom}
\end{equation}
%
%


\noindent
Introducing $\Pi_T = \omega^2_p$, and the following notations for
the elements of ${\bf M^{\prime}}$,  i.e., ${\bf M}^{\prime}_{12}= -{\bf M}^{\prime}_{21}= iF = i\frac{\omega_B\omega^2_p}{\omega}cos\theta$ 
followed by  ${\bf M}^{\prime}_{24} =- {\bf M}^{\prime}_{42} = iG =ig_{\phi'\gamma\gamma}\omega Bsin\theta$ and ${\bf M}^{\prime}_{34} = -{\bf M}^{\prime}_{43}
= -iL = -i g_{\phi'\gamma\gamma}\omega_p Bsin\theta $,  the EOMs i.e. eqns. (\ref{photon-eom}) and equation. (\ref{scalar-eom}), can be combined to be expressed in a compact matrix notation as,

\begin{equation}
\left[k^{2}{\bf I} -{\bf M^{\prime} }\right]
\left( \begin{array}{c}
A_{\parallel}(k)   \\
A_{\perp}(k)   \\
A_{L}(k)  \\
\phi' (k)  \\
\end{array} \right)=0,
\label{axion_field}
\end{equation}
where ${\bf I}$ is an identity matrix and   matrix ${\bf M^{\prime}}$ is 
the  $4\times 4$ mixing matrix, given as follows : 

\begin{equation}
{\bf M^{\prime}} = \left( \begin{array}{cccc}
\omega^2_p  & \;\; iF  & \;\; 0  & \;\;0 \\
-iF  & \;\; \omega^2_p &\;\;0  &\;\; iG \\
0  & \;\; 0  & \;\;{\Pi_L}   & \;\;  -iL\\
0  &  \; -iG  & \;\;iL   & \;\; m^2_{\phi'}
\end{array} \right).
\label{mat-axion2}
\end{equation}
The angle $\theta$, in (\ref{mat-axion2})  is the angle between  vector $\vec{k}$ and $\vec{B} $. 

\subsection{Diagonalizing the $4\times4$ mixing matrix} 
Diagonalising the matrix ${\bf M^{\prime}}$ given in (\ref{mat-axion-photon}), with its full generality
is a complex task, involving finding the roots of a fourth order polynomial.
Therefore the same was performed perturbatively in \cite{Ganguly-jain}.
In the following we discuss that procedure briefly.\\
\indent
In order to find the approximate eigen values and eigen vectors we 
can write ${\bf M^{\prime}}$ as a sum of two parts,
\begin{equation}
{\bf M}^{\prime}  = \left[ \begin{array}{cccc}
\omega^2_p & \;\; iF  & \;\; 0  & \;\;0 \\
-iF  & \;\; \omega^2_p & \;\;0  &\;\; iG \\
0 & \;\; 0  & \;\; {\Pi_L}  & \;\;  0\\
0  &  \;\; -iG  &\;\; 0 & \;\; m^2_{\phi'}
\end{array} \right]
+
\left[ \begin{array}{cccc}
0 \;\; & \;\; 0  & \;\; 0  & \;\;0 \\
0\;\;  & \;\; 0  & \;\; 0  &\;\; 0 \\
0 \;\; & \;\; 0  & \;\; 0  & \;\;-iL\\
0 \;\; & \;\; 0  & \;\; iL & \;\; 0
\end{array} \right].
\label{pert-matrix}
\end{equation}
\noindent
The  first matrix on the rhs of equation (\ref{pert-matrix}) is
 ${\bf M}^{ \prime 0}$ and the second one is  ${\bf M}^{ \prime \delta} $,
constructed with the $M^{\prime}_{34}$ and $M^{\prime}_{43}$. 
%
%
%
In the limit $m_{\phi'} \to 0$, the eigenvalues of the  ${\bf M}^{ \prime 0} $
 matrix can be expressed--in terms of 
the variables X and Y ( defined as: $X = \frac{F}{(m^2_{\phi'} - \omega^2_p)}$ and $Y = \frac{G}{(m^2_{\phi'} - \omega^2_p)}$ ), in the following way,
\begin{equation}
\begin{array}{cc}
     \lambda^{0}_{1} = & \frac{G^2}{2\omega^2_p}+ \omega^2_p - \sqrt{F^2 + \left(\frac{G}{2\omega_p}\right)^2} , \\
     \lambda^{0}_{2} = & \frac{G^2}{2\omega^2_p}+ \omega^2_p + \sqrt{F^2 + \left(\frac{G}{2\omega_p}\right)^2} , \\
     \lambda^{0}_{3} = & \hskip -3.5cm \Pi_L ,\\
     \lambda^{0}_{4} = & \hskip -3.5cm -\frac{G^2}{\omega^2_p}.   
\end{array} 
\label{zeroth-eigenvectors}
\end{equation}

\noindent
The eigenvectors  
of  ${\bf M}^{ \prime 0}$, defined in terms of ${\tilde{\lambda}_1} = \sqrt{X^2 +\frac{Y^2}{4}}- \frac{Y^2}{2}$, ${\tilde{\lambda}_2} 
= -\sqrt{X^2 +\frac{Y^2}{4}}- \frac{Y^2}{2}$ and ${\tilde{\lambda}_3} = 1 + {Y^2}$ turn out to  be,
 
\noindent\begin{minipage}{.5\linewidth}

\begin{equation*}
                                  |\lambda^0_1> ={\frac{1}{d_1}}    \left( \begin{array}{c}

         						  -1 \\

          						 i\frac{\tilde{\lambda}_1}{X}   \\

          						 0 \\

         						 Y\frac{\tilde{\lambda}_1}{X} 
\end{array} \right),
\label{d1}
\end{equation*}
\end{minipage}%
\begin{minipage}{.5\linewidth}
\begin{equation*}
                           |\lambda^0_2> ={\frac{1}{d_2}}         \left( \begin{array}{c}

         						      -1   \\
                                   			       i\frac{\tilde{\lambda}_2}{X}   \\
                                                          	      0 \\
                                                                Y\frac{\tilde{\lambda}_2}{X} 
\end{array} \right),
\label{d2}
\end{equation*}

\end{minipage}
\vskip.5cm
\noindent\begin{minipage}{.5\linewidth}

\begin{equation*}
                            |\lambda^0_3> ={\frac{1}{d_3}}         \left( \begin{array}{c}
                            
                            0  \\
                            0 \\
                            1 \\
                            0 
         						 
\end{array} \right),
\label{d3}
\end{equation*}
\end{minipage}%
\begin{minipage}{.5\linewidth}
\begin{equation}
                         |\lambda^0_4> ={\frac{1}{d_4}}\left( \begin{array}{c}
					0   \\
					-iY   \\
					0 \\
					1 

\end{array} \right).
\label{d4}
\end{equation}

\end{minipage}

\noindent
In eqns. of  (\ref{d4}) the variables $d_1 = \sqrt{1 + \frac{\tilde{\lambda}^2_1}{X^2}}$,  $d_2 = \sqrt{1 + \frac{\tilde{\lambda}^2_2}{X^2}}$
and $d_3 = d_4 =1.$ \\
\indent
As mentioned before, we estimated the approximate eigen-values and 
eigen-vectors of matrix ${\bf M}^{\prime}$, by treating ${\bf M}^{\prime \delta}$ as 
perturbation over  ${\bf M}^{\prime 0}$ . The eigenvectors, under this 
approximation turn out to be,  
 \begin{equation*}
 |\lambda_1> =\left( \begin{array}{c}
-{\frac{1}{d_1}}\left(1 - \frac{L^2 w^2_1}{d^2_1(\lambda^{0}_3 - \lambda^{0}_1)^2}\right)^\frac{1}{2}  \\
\frac{v_1}{d_1}\left(1 - \frac{L^2 w^2_1}{d^2_1(\lambda^{0}_3 - \lambda^{0}_1)^2}\right)^\frac{1}{2}  \\
-i\frac{ L w_1}{d_1(\lambda^{0}_3 - \lambda^{0}_1)}\\
 \frac{w_1}{d_1}\left(1 - \frac{L^2 w^2_1}{d^2_1(\lambda^{0}_3 - \lambda^{0}_1)^2}\right)^\frac{1}{2} 
\end{array} \right),
\hskip 3.20cm
 |\lambda_2> =\left( \begin{array}{c}
-{\frac{1}{d_2}}\left(1 - \frac{L^2 w^2_2}{d^2_2(\lambda^{0}_3 - \lambda^{0}_2)^2}\right)^\frac{1}{2}  \\
\frac{v_2}{d_2}\left(1 - \frac{L^2 w^2_2}{d^2_2(\lambda^{0}_3 - \lambda^{0}_2)^2}\right)^\frac{1}{2}  \\
-i\frac{ L w_2}{d_2(\lambda^{0}_3 - \lambda^{0}_2)}\\
 \frac{w_2}{d_2}\left(1 - \frac{L^2 w^2_2}{d^2_2(\lambda^{0}_3 - \lambda^{0}_2)^2}\right)^\frac{1}{2} 
\end{array} \right)
\end{equation*} 
\\
 \begin{equation}
 |\lambda_3> =\left( \begin{array}{c}
i\left[\frac{Lw_1}{d^2_1(\lambda_3^{0} - \lambda^{0}_1)} + \frac{Lw_2}{d^2_2(\lambda^{0}_3 - \lambda^{0}_2) }    \right]  \\
- i\left[\frac{Lw_1 v_1}{d^2_1(\lambda^{0}_3 - \lambda^{0}_1)} + \frac{Lw_2 v_2}{d^2_2(\lambda^{0}_3 - \lambda^{0}_2) } +  \frac{Lw_4 v_4}{d^2_4(\lambda^{0}_3 - \lambda^{0}_4)}  \right]    \\
\left[1 - \frac{L^2 w^2_1}{d^2_1(\lambda^{0}_3 - \lambda^{0}_1)^2} - \frac{L^2 w^2_2}{d^2_2(\lambda^{0}_3 - \lambda^{0}_2)^2}- \frac{L^2 w^2_4}{d^2_4(\lambda^{0}_3 - \lambda^{0}_4)^2}\right]^\frac{1}{2}  \\
 -i  \left[\frac{L w^2_1}{d^2_1(\lambda^{0}_3 - \lambda^{0}_1)} + \frac{Lw^2_2}{d^2_2(\lambda^{0}_3 - \lambda^{0}_2)}+ \frac{Lw^2_4}{d^2_4(\lambda^{0}_3 - \lambda^{0}_4)}\right] 
\end{array} \right),
 \hskip .3cm
 |\lambda_4> =\left( \begin{array}{c}
0  \\
\frac{v_4}{d_4}\left(1 - \frac{L^2 w^2_4}{d^2_4(\lambda^{0}_3 - \lambda^{0}_4)^2}\right)^\frac{1}{2}  \\
-i\frac{ L w_4}{d_4(\lambda^{0}_3 - \lambda^{0}_4)}\\
 \frac{w_4}{d_4}\left(1 - \frac{L^2 w^2_4}{d^2_4(\lambda^{0}_3 - \lambda^{0}_4)^2}\right)^\frac{1}{2} 
\end{array} \right)
\end{equation} 
\\
In the above equations, $v_1 = i \frac{\tilde{\lambda}_1}{X} $, $v_2 = i \frac{\tilde{\lambda}_2}{X} $, $v_4 = -iY $,
and $w_1 = Y\frac{\tilde{\lambda}_1}{X} $, $w_2 = Y\frac{\tilde{\lambda}_2}{X} $, $w_4 = 1$. We use these eigenvectors to construct the Unitary matrix {\bf U} to diagonalize the mixing matrix {\bf M'} by unitary transformation. The constructed  Unitary matrix is given as follows:

\begin{equation}
\begin{scriptsize}
{\bf U} =  
                \left( \begin{array}{cccc} 
-\frac{1}{d_{1}}\sqrt{1-\xi_1^{2}} \rm{\cal{N}}^{(1)}_{vn}   &  \hskip .1cm    -\frac{1}{d_{2}}\sqrt{1-\xi_2^{2}} \rm{\cal{N}}^{(2)}_{vn} &  \hskip .1cm   i(\frac{\xi_1}{d_{1}}+\frac{\xi_2}{d_{2}})\rm{\cal{N}}^{(3)}_{vn}  &  0 \\

\frac{v_{1}}{d_{1}}\sqrt{1-\xi_1^{2}} \rm{\cal{N}}^{(1)}_{vn} &  \hskip .1cm    \frac{v_{2}}{d_{2}}\sqrt{1-\xi_2^{2}}\rm{\cal{N}}^{(2)}_{vn}& \hskip .1cm   -i(\xi_1\frac{v_{1}}{d_{1}}+\xi_2\frac{v_{2}}{d_{2}}+\xi_4\frac{v_{4}}{d_{4}}) \rm{\cal{N}}^{(3)}_{vn}&    \frac{v_{4}}{d_{4}}\sqrt{1-\xi_4^{2}}\rm{\cal{N}}^{(4)}_{vn} \\

 -i\xi_1\rm{\cal{N}}^{(1)}_{vn}  & \hskip .1cm -i\xi_2\rm{\cal{N}}^{(2)}_{vn} & \hskip .1cm    \sqrt{1-\xi_1^{2}-\xi_2^{2}-\xi_4^{2}} \rm{\cal{N}}^{(3)}_{vn} & -i\xi_4\rm{\cal{N}}^{(4)}_{vn} \\

 \frac{w_{1}}{d_{1}}\sqrt{1-\xi_1^{2}}\rm{\cal{N}}^{(1)}_{vn}  &\frac{w_{2}}{d_{2}}\sqrt{1-\xi_2^{2}}\rm{\cal{N}}^{(2)}_{vn}  &\hskip .1cm -i(\xi_1\frac{w_{1}}{d_{1}}+\xi_2\frac{w_{2}}{d_{2}}+\xi_4\frac{w_{4}}{d_{4}})\rm{\cal{N}}^{(3)}_{vn} & \frac{w_{4}}{d_{4}}\sqrt{1-\xi_4^{2}}\rm{\cal{N}}^{(4)}_{vn}
                       \end{array} 
             \right)
 \label{u-mat_axion}
 \end{scriptsize}
\end{equation} \\
 Variables introduced in (\ref{u-mat_axion}) are defined as; $\xi_i = \frac{Lw_i}{d_i(\lambda_{3}^0-\lambda_{i}^0)}$ ($i$ can take values 1,2,4). 
 Here, u$_j$, v$_j$ ,w$_j$ and x$_j$ are the components of the $j^{th}$ column vectors  of {\bf U}, and 
 \begin{equation*}
                 \rm{\cal{N}}^{(j)}_{vn}  = \frac{1}{\sqrt{[|u_j|^2 + |v_j|^2 +  |w_j|^2+|x_j|^2]}}  
                     \end{equation*}
                      is the corresponding normalisation constant ($j=1, 2, 3, 4$).

\subsection{Field equation : Solutions}
To get the solution of the coupled equation (\ref{axion_field}), we multiply the same by {\bf{U}$^{-1}$} , hence:

\indent
\begin{equation}
{\bf{U}}^{-1}\left[\begin{array}{c} k^2 {\bf I} - {\bf M' } 
 \end{array} \right]{\bf{UU}}^{-1}
\left(\begin{array}{c} A_{\parallel}(k) \\ A_{\perp}(k)\\A_L(k) \\\phi'(k) \end{array} \right)=
\left[\begin{array}{c} k^2 {\bf I} - {\bf M}_{D} 
 \end{array} \right]
\left(\begin{array}{c} A'_{\parallel}(k) \\ A'_{\perp}(k)\\A'_L(k) \\\phi''(k) \end{array} \right)=0,
\label{M_D_axion}
\end{equation} 

\noindent

\begin{eqnarray}
{\bf[ C_{A,\phi'}]}=
\left[\begin{array}{c} A_{\parallel}(k) \\ A_{\perp}(k)\\A_L(k) \\\phi'(k) \end{array} \right]= {\bf U} \left[\begin{array}{c} A'_{\parallel}(k) \\ A'_{\perp}(k) \\A'_L(k)\\ \phi''(k) 
\end{array} \right].
\label{primed-degrees-of-freedom_axion}
\end{eqnarray}

\noindent
\noindent
Following {\cite {Raffelt}}, one can write partial of  $k_{3} $(Fourier transform) for a beam propagating in z direction, equation (\ref{primed-degrees-of-freedom_axion})
can further be written explicitly in the form,

\begin{equation}
\left[\begin{array}{c}  (\omega -i\partial_z)  \mathbf{I}  - \left[ \begin{array}{cccc} \frac{\lambda_1}{2\omega} &   0     &    0 &0\\
0 &  \frac{\lambda_2}{2\omega}  & 0&0 \\
0&0&\frac{\lambda_3}{2\omega}&0\\  
0&0 & 0 & \frac{\lambda_4}{2\omega}  \end{array} \right]        \end{array} \right]
\left[\begin{array}{c} {A'}_{\parallel}(z)\\ {A'}_{\perp}(z)\\A'_L(k) \\ {\phi''}(z) \end{array} \right]=0.
\label{dmat_axion}
\end{equation} 
\noindent
The solutions for the column vector ${\bf[ C_{A,\phi'}]}$, can be written  in  matrix form as was performed in the scalar- photon case with the  introduction of  the variables, 
$\Omega_\parallel = \left(\omega - \frac{\lambda_1}{2\omega}\right) $,
$\Omega_\perp = \left(\omega - \frac{\lambda_2}{2\omega}\right)$ $\Omega_L = \left(\omega - \frac{\lambda_3}{2\omega}\right)$
and $\Omega_{\phi'} = \left(\omega - \frac{\lambda_4}{2\omega}\right)$,

%
\begin{equation}
  \left[
\begin{matrix}
{{A}_{\parallel}(z)} \cr
{{A}_{\perp}(z)} \cr   
{{A}_{L}(z)} \cr
{\phi'(z)}
\end{matrix}
\right] = {\bf U}\left[ \begin{array}{cccc} e^{-i\Omega_\parallel z} &   0    &    0 &0\\
0 & e^{-i\Omega_\perp z}  & 0&0 \\
0 & 0 & e^{-i\Omega_L z}  &0\\
0 & 0 &0& e^{-i\Omega_{\phi'} z} 
\end{array} \right] {\bf U}^{-1} 
\left[
\begin{matrix}
{{A}_{\parallel}(0)} \cr
{{A}_{\perp}(0)} \cr  
 {{A}_{L}(0)} \cr  
{\phi'(0)}
\end{matrix}
\right]
\label{solnmat_axion}.
\end{equation}
\noindent
The elements of the column vector, $\left[{\bf A}(0) \right]$ appearing  in 
equation (\ref{solnmat_axion}), are the initial values whose magnitudes are subject to 
the physical conditions considered  at  z = 0. For the boundary conditions for pseudoscalar field,  
 the solution of equation (\ref{solnmat_axion}) for {\bf A$_\parallel$}($\omega,{\bf z}$), turns out to be:
\begin{eqnarray}
A_{\parallel}(\omega,z)&=& 
\left( e^{i\Omega_\parallel z}{\hat u}_1 { \hat u}^{*}_1  +  e^{i\Omega_\perp z } {\hat u}_2{ \hat u}^{*}_2 +  e^{i\Omega_L z } {\hat u}_3{ \hat u}^{*}_3 
+ e^{i\Omega_{\phi'} z } {\hat u}_4{ \hat u}^{*}_4 \right)A_{\parallel}(\omega, 0)  \nonumber \\
&+& \left(
e^{i\Omega_\parallel z }{\hat u}_1 { \hat v}^{*}_1  +  e^{i\Omega_\perp z } {\hat u}_2{ \hat v}^{*}_2   
+ e^{i\Omega_L z } {\hat u}_3{ \hat v}^{*}_3 + e^{i\Omega_{\phi'} z } {\hat u}_4{ \hat v}^{*}_4  \right)A_{\perp}(\omega, 0)\nonumber\\
&+& \left(
e^{i\Omega_\parallel z }{\hat u}_1 { \hat w}^{*}_1  +  e^{i\Omega_\perp z } {\hat u}_2{ \hat w}^{*}_2   
+ e^{i\Omega_L z } {\hat u}_3{ \hat w}^{*}_3 + e^{i\Omega_{\phi'} z } {\hat u}_4{ \hat w}^{*}_4 \right)A_{L}(\omega, 0),
\label{soln-parallel}
\end{eqnarray}  
\noindent 
and the  component $A_{\perp}(\omega,z)$ turns out to be, 
\begin{eqnarray}
A_{\perp}(\omega,z)&=& 
\left( e^{i\Omega_\parallel z}{\hat v}_1 { \hat u}^{*}_1  +  e^{i\Omega_\perp z } {\hat v}_2{ \hat u}^{*}_2 +  e^{i\Omega_L z } {\hat v}_3{ \hat u}^{*}_3 
+ e^{i\Omega_{\phi'} z } {\hat v}_4{ \hat u}^{*}_4 \right)A_{\parallel}(\omega, 0)  \nonumber \\
&+& \left(
e^{i\Omega_\parallel z }{\hat v}_1 { \hat v}^{*}_1  +  e^{i\Omega_\perp z } {\hat v}_2{ \hat v}^{*}_2   
+ e^{i\Omega_L z } {\hat v}_3{ \hat v}^{*}_3 + e^{i\Omega_{\phi'} z } {\hat v}_4{ \hat v}^{*}_4     \right)A_{\perp}(\omega, 0)\nonumber\\
&+& \left(
e^{i\Omega_\parallel z }{\hat v}_1 { \hat w}^{*}_1  +  e^{i\Omega_\perp z } {\hat v}_2{ \hat w}^{*}_2   
+ e^{i\Omega_L z } {\hat v}_3{ \hat w}^{*}_3 + e^{i\Omega_{\phi'} z } {\hat v}_4{ \hat w}^{*}_4     \right)A_{L}(\omega, 0).
\label{soln-perp}
\end{eqnarray}

\subsection{ Stokes parameters }

For $\phi'\gamma$ mixing, the Stokes parameter ({\bf{I}}, {\bf{Q}}, {\bf{U}} and {\bf{V}}) is determined by the set of equations appearing in (\ref{I-axion} - \ref{V-axion}) by using (\ref{soln-parallel}) and (\ref{soln-perp}). 
The variables introduced in (\ref{I-axion}) for the expression for Stokes parameter {\bf{I}} can be expressed as:
\begin{align}
I_{\parallel} &= \Big[1 - 4|{\hat u}_1 {\hat u}_2|\mathbb{V}_{12} sin^2 \left(\frac{(\Omega_{\parallel} -\Omega_{\perp})z}{2} \right)  
- 4|{\hat u}_2{\hat u}_3\mathbb{V}_{23} sin^2 \left(\frac{(\Omega_{\perp} -\Omega_{L})z}{2} \right)\nonumber\\
&\hskip 0.8cm-4|{\hat u}_3 {\hat u}_1|\mathbb{V}_{31}sin^2 \left(\frac{(\Omega_{L} -\Omega_{\parallel})z}{2} \right)\Big]. 
\end{align}
\begin{align}
I_{\perp}& = \Big[1 -4|{\hat v}_1 {\hat v}_2|\mathbb{V}_{12} sin^2 \left(\frac{(\Omega_{\parallel} -\Omega_{\perp})z}{2} \right) 
- 4|{\hat v}_2{\hat v}_3|\mathbb{V}_{23}sin^2 \left(\frac{(\Omega_{\perp} -\Omega_{L})z}{2} \right)\nonumber\\ 
&\hskip 0.8cm- |{\hat v}_3 {\hat v}_1|\mathbb{V}_{31}sin^2 \left(\frac{(\Omega_{L} -\Omega_{\parallel})z}{2} \right)
-4|{\hat v}_3|^2|{\hat v}_4|^2 sin^2 \left(\frac{(\Omega_{L}-\Omega_{\phi'})z}{2} \right) \nonumber\\
&\hskip 0.8cm-4|{\hat v}_4|^2|{\hat v}_1|^2 sin^2 \left(\frac{(\Omega_{\phi'}-\Omega_{\parallel})z}{2} \right)
-4|{\hat v}_2|^2|{\hat v}_4|^2 sin^2 \left(\frac{(\Omega_{\perp}-\Omega_{\phi'})z}{2} \right)\Big]. 
\end{align}
\begin{align}
\hskip 1cmI_{L}&= \Big[(4|{\hat w}_1{\hat w}_2|\mathbb{V}_{12} sin^2 \left(\frac{{(\Omega_{\parallel} -\Omega_{\perp})z}}{2} \right) 
+ 4|{\hat w}_2{\hat w}_3|\mathbb{V}_{23} cos^2 \left(\frac{{(\Omega_{\perp} -\Omega_{L})z}}{2} \right) \nonumber\\
&\hskip 0.0cm+ 4|{\hat w}_3{\hat w}_1|\mathbb{V}_{31} cos^2 \left(\frac{{(\Omega_{L} -\Omega_{\parallel})z}}{2} \right) 
+4 |{\hat v}_3{\hat v}_4||{\hat w}_3{\hat w}_4|cos^2 \left(\frac{{(\Omega_{L} -\Omega_{\phi'})z}}{2} \right)\nonumber\\
 &\hskip 0.0cm+ 4 |{\hat v}_4{\hat v}_1||{\hat w}_4{\hat w}_1|sin^2 \left(\frac{{(\Omega_{\phi'} -\Omega_{\parallel})z}}{2} \right)
+4|{\hat v}_2{\hat v}_4||{\hat w}_2{\hat w}_4|sin^2 \left(\frac{{(\Omega_{\phi'} -\Omega_{\perp})z}}{2} \right)\Big]. 
\end{align}
\begin{align}
I_{\parallel \perp}& = \Big[2(|{\hat u}_2|^2|{\hat u}_1{\hat v}_1|
-|{\hat v}_2|^2|{\hat u}_1{\hat v}_1| - |{\hat u}_1|^2|{\hat u}_2{\hat v}_2| 
+ |{\hat v}_1|^2|{\hat u}_2{\hat v}_2|)\;sin ({(\Omega_{\parallel} -\Omega_{\perp})z})\nonumber\\
&+ 2(|{\hat u}_3|^2|{\hat u}_2{\hat v}_2|
-|{\hat v}_3|^2|{\hat u}_2{\hat v}_2| - |{\hat u}_2|^2|{\hat u}_3{\hat v}_3| 
+ |{\hat v}_2|^2|{\hat u}_3{\hat v}_3|)\;sin ({(\Omega_{\perp} -\Omega_{L})z})\nonumber\\
&+2(|{\hat u}_3|^2|{\hat u}_1{\hat v}_1|
-|{\hat v}_3|^2|{\hat u}_1{\hat v}_1| - |{\hat u}_1|^2|{\hat u}_3{\hat v}_3| 
+ |{\hat v}_1|^2|{\hat u}_3{\hat v}_3|)\;sin ({(\Omega_{\parallel} -\Omega_{L})z})\nonumber\\
&+2|{\hat v}_4|^2|{\hat u}_1{\hat v}_1|)\;sin ({(\Omega_{\phi'} -\Omega_{\parallel})z})
-2|{\hat v}_4|^2|{\hat u}_2{\hat v}_2|)\;sin ({(\Omega_{\perp} -\Omega_{\phi'})z})\nonumber\\
&-2|{\hat v}_4|^2|{\hat u}_3{\hat v}_3|)\;sin ({(\Omega_{L} -\Omega_{\phi'})z})\Big].   
\end{align}
\begin{align}
I_{\parallel L} &=  \Big[2(|{\hat u}_1|^2|{\hat u}_2{\hat w}_2|
-|{\hat u}_2|^2|{\hat u}_1{\hat w}_1| + |{\hat u}_1{\hat v}_1| |{\hat v}_2{\hat w}_2| 
-  |{\hat u}_2{\hat v}_2| |{\hat v}_1{\hat w}_1|) sin ({(\Omega_{\parallel} -\Omega_{\perp})z})\nonumber\\
&-2(|{\hat u}_2|^2|{\hat u}_3{\hat w}_3|
+|{\hat u}_3|^2|{\hat u}_2{\hat w}_2| + |{\hat u}_2{\hat v}_2||{\hat v}_3{\hat w}_3| 
+|{\hat u}_3 {\hat v}_3||{\hat v}_2{\hat w}_2|) sin ({(\Omega_{\perp} -\Omega_{L})z})\nonumber\\
&-2(|{\hat u}_3|^2|{\hat u}_1{\hat w}_1|
+|{\hat u}_1|^2|{\hat u}_3{\hat w}_3| + |{\hat u}_3{\hat v}_3| |{\hat v}_1{\hat w}_1| 
+  |{\hat u}_1{\hat v}_1| |{\hat v}_3{\hat w}_3|) sin ({(\Omega_{\parallel} -\Omega_{L})z})\nonumber\\
&+2|{\hat u}_3{\hat v}_3||{\hat v}_4{\hat w}_4|)\;sin ({(\Omega_{L} -\Omega_{\phi'})z})
-2|{\hat u}_1 {\hat v}_1||{\hat v}_4{\hat w}_4|)\;sin ({(\Omega_{\phi'} -\Omega_{\parallel})z})\nonumber\\
&+2|{\hat u}_2{\hat v}_2||{\hat v}_4{\hat w}_4|)\;sin ({(\Omega_{\perp} -\Omega_{\phi'})z})\Big]. 
\end{align}
\begin{align}
I_{\perp L} &= \Big[2(|{\hat u}_3{\hat v}_3||{\hat u}_3{\hat w}_3|
-|{\hat u}_1{\hat v}_1||{\hat u}_1{\hat w}_1|
-|{\hat u}_2{\hat v}_2||{\hat u}_2{\hat w}_2|
-|{\hat v}_1|^2|{\hat v}_1{\hat w}_1| - |{\hat v}_2|^2 |{\hat v}_2{\hat w}_2| \nonumber\\
&+ |{\hat v}_3|^2 |{\hat v}_3{\hat w}_3|-|{\hat v}_4|^2 |{\hat v}_4{\hat w}_4|)
+2(|{\hat v}_4|^2|{\hat v}_3{\hat w}_3| -|{\hat v}_3|^2|{\hat v}_4{\hat w}_4|)\;cos ({(\Omega_{L} -\Omega_{\phi'})z})\nonumber\\
&-2(|{\hat v}_4|^2|{\hat v}_1{\hat w}_1| +|{\hat v}_1|^2|{\hat v}_4{\hat w}_4|)\;cos ({(\Omega_{\phi'} -\Omega_{\parallel})z})\nonumber\\
&-2(|{\hat v}_2|^2|{\hat v}_4{\hat w}_4| +|{\hat v}_4|^2|{\hat v}_2{\hat w}_2|)\;cos ({(\Omega_{\perp} -\Omega_{\phi'})z})\nonumber\\
&-2(|{\hat u}_1{\hat v}_1||{\hat u}_2{\hat w}_2|
+|{\hat u}_2 {\hat v}_2||{\hat u}_1{\hat w}_1| + |{\hat v}_1|^2|{\hat v}_2{\hat w}_2| 
+|{\hat v}_2|^2|{\hat v}_1{\hat w}_1|) cos ({(\Omega_{\parallel} -\Omega_{\perp})z})\nonumber\\
&+2(|{\hat u}_2{\hat v}_2||{\hat u}_3{\hat w}_3|
-|{\hat u}_3 {\hat v}_3||{\hat u}_2{\hat w}_2| + |{\hat v}_2|^2|{\hat v}_3{\hat w}_3| 
-|{\hat v}_3|^2|{\hat v}_2{\hat w}_2|) cos ({(\Omega_{\perp} -\Omega_{L})z})\nonumber\\
&+2(|{\hat u}_1{\hat v}_1||{\hat u}_3{\hat w}_3|
-|{\hat u}_3 {\hat v}_3||{\hat u}_1{\hat w}_1| + |{\hat v}_1|^2|{\hat v}_3{\hat w}_3| 
-|{\hat v}_3|^2|{\hat v}_1{\hat w}_1|) cos ({(\Omega_{\parallel} -\Omega_{L})z})
\Big]. 
\end{align}
Also,  the variables introduced in expressing the Stokes parameter {\bf{Q}}, are expressed as follows,
\begin{align}
 Q_{\parallel} &= \Big[1 - 4|{\hat u}_1 {\hat u}_2|\mathbb{\tilde{V}}_{12} sin^2 \left(\frac{(\Omega_{\parallel} -\Omega_{\perp})z}{2} \right)  
- 4|{\hat u}_2{\hat u}_3\mathbb{\tilde{V}}_{23} sin^2 \left(\frac{(\Omega_{\perp} -\Omega_{L})z}{2} \right)\nonumber\\
&-4|{\hat u}_3 {\hat u}_1|(\mathbb{\tilde{V}}_{31})sin^2 \left(\frac{(\Omega_{L} -\Omega_{\parallel})z}{2} \right)\Big].
\end{align}
\begin{align}
Q_{\perp} &= \Big[1 + 4|{\hat v}_1 {\hat v}_2|\mathbb{\tilde{V}}_{12} sin^2 \left(\frac{(\Omega_{\parallel} -\Omega_{\perp})z}{2} \right) 
+ 4|{\hat v}_2{\hat v}_3|\mathbb{\tilde{V}}_{23}sin^2 \left(\frac{(\Omega_{\perp} -\Omega_{L})z}{2} \right)\nonumber\\ 
&+ 4|{\hat v}_3 {\hat v}_1|\mathbb{\tilde{V}}_{31}sin^2 \left(\frac{(\Omega_{L} -\Omega_{\parallel})z}{2} \right)
-4|{\hat v}_3|^2|{\hat v}_4|^2 sin^2 \left(\frac{(\Omega_{L}-\Omega_{\phi'})z}{2} \right) \nonumber\\
&-4|{\hat v}_4|^2|{\hat v}_1|^2 sin^2 \left(\frac{(\Omega_{\phi'}-\Omega_{\parallel})z}{2} \right)
-4|{\hat v}_2|^2|{\hat v}_4|^2 sin^2 \left(\frac{(\Omega_{\perp}-\Omega_{\phi'})z}{2} \right)\Big]. 
\end{align}
\begin{align}
Q_{L} &=  \Big[(4|{\hat w}_1{\hat w}_2|\mathbb{\tilde{V}}_{12} sin^2 \left(\frac{{(\Omega_{\parallel} -\Omega_{\perp})z}}{2} \right) 
+ 4|{\hat w}_2{\hat w}_3|\mathbb{\tilde{V}}_{23} cos^2 \left(\frac{{(\Omega_{\perp} -\Omega_{L})z}}{2} \right) \nonumber\\
 &+ 4|{\hat w}_3{\hat w}_1|\mathbb{\tilde{V}}_{31} cos^2 \left(\frac{{(\Omega_{L} -\Omega_{\parallel})z}}{2} \right) 
-4 |{\hat v}_3{\hat v}_4||{\hat w}_3{\hat w}_4|cos^2 \left(\frac{{(\Omega_{L} -\Omega_{\phi'})z}}{2} \right)\nonumber\\
 &-4 |{\hat v}_4{\hat v}_1||{\hat w}_4{\hat w}_1|sin^2 \left(\frac{{(\Omega_{\phi'} -\Omega_{\parallel})z}}{2} \right)
-4|{\hat v}_2{\hat v}_4||{\hat w}_2{\hat w}_4|sin^2 \left(\frac{{(\Omega_{\phi'} -\Omega_{\perp})z}}{2} \right)\Big]. 
\end{align}
\begin{align}
 Q_{\parallel \perp} &= \Big[2(|{\hat u}_2|^2|{\hat u}_1{\hat v}_1|
+|{\hat v}_2|^2|{\hat u}_1{\hat v}_1| - |{\hat u}_1|^2|{\hat u}_2{\hat v}_2| 
- |{\hat v}_1|^2|{\hat u}_2{\hat v}_2|)\;sin ({(\Omega_{\parallel} -\Omega_{\perp})z})\nonumber\\
&+2(|{\hat u}_3|^2|{\hat u}_2{\hat v}_2|
+|{\hat v}_3|^2|{\hat u}_2{\hat v}_2| - |{\hat u}_2|^2|{\hat u}_3{\hat v}_3| 
- |{\hat v}_2|^2|{\hat u}_3{\hat v}_3|)\;sin ({(\Omega_{\perp} -\Omega_{L})z})\nonumber\\
&+2(|{\hat u}_3|^2|{\hat u}_1{\hat v}_1|
+|{\hat v}_3|^2|{\hat u}_1{\hat v}_1| - |{\hat u}_1|^2|{\hat u}_3{\hat v}_3| 
- |{\hat v}_1|^2|{\hat u}_3{\hat v}_3|)\;sin ({(\Omega_{\parallel} -\Omega_{L})z})\nonumber\\
&+2|{\hat v}_4|^2|{\hat u}_1{\hat v}_1|)\;sin ({(\Omega_{\phi'} -\Omega_{\parallel})z})
-2|{\hat v}_4|^2|{\hat u}_2{\hat v}_2|)\;sin ({(\Omega_{\perp} -\Omega_{\phi'})z})\nonumber\\
&-2|{\hat v}_4|^2|{\hat u}_3{\hat v}_3|)\;sin ({(\Omega_{L} -\Omega_{\phi'})z})\Big]. 
\end{align}
\begin{align}
Q_{\parallel L}& =  \Big[2(|{\hat u}_1|^2|{\hat u}_2{\hat w}_2|
-|{\hat u}_2|^2|{\hat u}_1{\hat w}_1| - |{\hat u}_1{\hat v}_1| |{\hat v}_2{\hat w}_2| 
+ |{\hat u}_2{\hat v}_2| |{\hat v}_1{\hat w}_1|) sin ({(\Omega_{\parallel} -\Omega_{\perp})z})\nonumber\\
&-2(|{\hat u}_2|^2|{\hat u}_3{\hat w}_3|
+|{\hat u}_3|^2|{\hat u}_2{\hat w}_2| - |{\hat u}_2{\hat v}_2||{\hat v}_3{\hat w}_3| 
-|{\hat u}_3 {\hat v}_3||{\hat v}_2{\hat w}_2|) sin ({(\Omega_{\perp} -\Omega_{L})z})\nonumber\\
&-2(|{\hat u}_3|^2|{\hat u}_1{\hat w}_1|
+|{\hat u}_1|^2|{\hat u}_3{\hat w}_3| - |{\hat u}_3{\hat v}_3| |{\hat v}_1{\hat w}_1| 
-|{\hat u}_1{\hat v}_1| |{\hat v}_3{\hat w}_3|) sin ({(\Omega_{\parallel} -\Omega_{L})z})\nonumber\\
&-2|{\hat u}_3{\hat v}_3||{\hat v}_4{\hat w}_4|)\;sin ({(\Omega_{L} -\Omega_{\phi'})z})
+2|{\hat u}_1 {\hat v}_1||{\hat v}_4{\hat w}_4|)\;sin ({(\Omega_{\phi'} -\Omega_{\parallel})z})\nonumber\\
&-2|{\hat u}_2{\hat v}_2||{\hat v}_4{\hat w}_4|)\;sin ({(\Omega_{\perp} -\Omega_{\phi'})z})\Big]. 
\end{align}
\begin{align}
Q_{\perp L}&=  \Big[2(|{\hat u}_3{\hat v}_3||{\hat u}_3{\hat w}_3|
-|{\hat u}_1{\hat v}_1||{\hat u}_1{\hat w}_1|
-|{\hat u}_2{\hat v}_2||{\hat u}_2{\hat w}_2|
+|{\hat v}_1|^2|{\hat v}_1{\hat w}_1| + |{\hat v}_2|^2 |{\hat v}_2{\hat w}_2| \nonumber\\
&- |{\hat v}_3|^2 |{\hat v}_3{\hat w}_3|+|{\hat v}_4|^2 |{\hat v}_4{\hat w}_4|)
-2(|{\hat v}_4|^2|{\hat v}_3{\hat w}_3| -|{\hat v}_3|^2|{\hat v}_4{\hat w}_4|)\;cos ({(\Omega_{L} -\Omega_{\phi'})z})\nonumber\\
&+2(|{\hat v}_4|^2|{\hat v}_1{\hat w}_1| +|{\hat v}_1|^2|{\hat v}_4{\hat w}_4|)\;cos ({(\Omega_{\phi'} -\Omega_{\parallel})z})\nonumber\\
&+2(|{\hat v}_2|^2|{\hat v}_4{\hat w}_4| +|{\hat v}_4|^2|{\hat v}_2{\hat w}_2|)\;cos ({(\Omega_{\perp} -\Omega_{\phi'})z})\nonumber\\
&-2(|{\hat u}_1{\hat v}_1||{\hat u}_2{\hat w}_2|
+|{\hat u}_2 {\hat v}_2||{\hat u}_1{\hat w}_1| - |{\hat v}_1|^2|{\hat v}_2{\hat w}_2| 
-|{\hat v}_2|^2|{\hat v}_1{\hat w}_1|) cos ({(\Omega_{\parallel} -\Omega_{\perp})z})\nonumber\\
&+2(|{\hat u}_2{\hat v}_2||{\hat u}_3{\hat w}_3|
-|{\hat u}_3 {\hat v}_3||{\hat u}_2{\hat w}_2| - |{\hat v}_2|^2|{\hat v}_3{\hat w}_3| 
+|{\hat v}_3|^2|{\hat v}_2{\hat w}_2|) cos ({(\Omega_{\perp} -\Omega_{L})z})\nonumber\\
&+2(|{\hat u}_1{\hat v}_1||{\hat u}_3{\hat w}_3|
-|{\hat u}_3 {\hat v}_3||{\hat u}_1{\hat w}_1| - |{\hat v}_1|^2|{\hat v}_3{\hat w}_3| 
+|{\hat v}_3|^2|{\hat v}_1{\hat w}_1|) cos ({(\Omega_{\parallel} -\Omega_{L})z})
\Big],
\end{align}
where $\mathbb{V}_{ij} =|{\hat u}_i {\hat u}_j| + |{\hat v}_i {\hat v}_j|$ and $\mathbb{\tilde{V}}_{ij} =|{\hat u}_i {\hat u}_j| - |{\hat v}_i {\hat v}_j|$( $i$ and $j$ may take value from 1 to 3).
In similar way, variables used in the expression  of  Stokes parameters {\bf {U}} and {\bf{V}}, written as: 
\begin{align}
U_{\parallel} &= 2\Big[|{\hat u}_1 {\hat u}_2|( |{\hat u}_1 {\hat v}_2| -|{\hat u}_2 {\hat v}_1|) sin \left((\Omega_{\parallel} -\Omega_{\perp})z \right)  
+|{\hat u}_2 {\hat u}_3|( |{\hat u}_2 {\hat v}_3| -|{\hat u}_3 {\hat v}_2|) sin \left((\Omega_{\perp} -\Omega_{L})z \right) \nonumber\\ 
&+|{\hat u}_3 {\hat u}_1|( |{\hat u}_3 {\hat v}_1| -|{\hat u}_1 {\hat v}_3|) sin \left((\Omega_{L} -\Omega_{\parallel})z \right) \Big]. \tag {B.13}
\end{align}
\begin{align}
U_{\perp}& =2\Big[|{\hat v}_1 {\hat v}_2|( |{\hat u}_1 {\hat v}_2| -|{\hat u}_2 {\hat v}_1|) sin \left((\Omega_{\parallel} -\Omega_{\perp})z \right)  
+|{\hat v}_2 {\hat v}_3|( |{\hat u}_2 {\hat v}_3| -|{\hat u}_3 {\hat v}_2|) sin \left((\Omega_{\perp} -\Omega_{L})z \right) \nonumber\\ 
&+|{\hat v}_3 {\hat v}_1|( |{\hat u}_3 {\hat v}_1| -|{\hat u}_1 {\hat v}_3|) sin \left((\Omega_{L} -\Omega_{\parallel})z \right)  
-|{\hat u}_1 {\hat v}_1||{\hat v}_4|^2 sin \left((\Omega_{\phi'} -\Omega_{\parallel})z \right)\nonumber\\
 &+ |{\hat u}_3 {\hat v}_3||{\hat v}_4|^2 sin ((\Omega_{L} -\Omega_{\phi'})z)\Big]. 
\end{align}
\begin{align}
U_{L}&=  2\Big[|{\hat w}_1 {\hat w}_2|( |{\hat u}_1 {\hat v}_2| -|{\hat u}_2 {\hat v}_1|) sin \left((\Omega_{\parallel} -\Omega_{\perp})z \right)  
-|{\hat w}_2 {\hat w}_3|( |{\hat u}_2 {\hat v}_3| -|{\hat u}_3 {\hat v}_2|) sin \left((\Omega_{\perp} -\Omega_{L})z \right) \nonumber\\ 
&+|{\hat w}_3 {\hat w}_1|( |{\hat u}_3 {\hat v}_1| -|{\hat u}_1 {\hat v}_3|) sin \left((\Omega_{\parallel} -\Omega_{L})z \right)  
 + |{\hat u}_1 {\hat w}_1||{\hat v}_4{\hat w}_4| sin \left((\Omega_{\parallel} -\Omega_{\phi'})z \right) \nonumber\\ 
&+|{\hat u}_2 {\hat w}_2||{\hat v}_4{\hat w}_4| sin \left((\Omega_{\perp} -\Omega_{\phi'})z \right)
-|{\hat u}_3 {\hat w}_3||{\hat v}_4 {\hat w}_4| sin \left((\Omega_{L} -\Omega_{\phi'})z \right)\Big]. 
\end{align}
\begin{align}
U_{\parallel \perp}&=  2\Big[(|{\hat u}_1|^2|{\hat v}_2|^2 + |{\hat u}_2|^2|{\hat v}_1|^2)\;cos ({(\Omega_{\parallel} -\Omega_{\perp})z}) 
+ (|{\hat u}_2|^2|{\hat v}_3|^2 + |{\hat u}_3|^2|{\hat v}_2|^2)\;cos ({(\Omega_{\perp} -\Omega_{L})z})\nonumber\\
&+ (|{\hat u}_3|^2|{\hat v}_1|^2 + |{\hat u}_1|^2|{\hat v}_3|^2)\;cos ({(\Omega_{L} -\Omega_{\parallel})z})
+ |{\hat u}_1|^2|{\hat v}_4|^2\;cos ({(\Omega_{\phi'} -\Omega_{\parallel})z})\nonumber\\
&+ |{\hat u}_2|^2|{\hat v}_4|^2\;cos ({(\Omega_{\perp} -\Omega_{\phi'})z})
+ |{\hat u}_2|^2|{\hat v}_4|^2\;cos ({(\Omega_{L} -\Omega_{\phi'})z})
-2|{\hat u}_1 {\hat u}_2||{\hat v}_1{\hat v}_2|) cos ({(\Omega_{\parallel} -\Omega_{\perp})z})\nonumber\\
&-2|{\hat u}_2 {\hat u}_3||{\hat v}_2{\hat v}_3|) cos ({(\Omega_{\perp} -\Omega_{L})z})
-2|{\hat u}_3 {\hat u}_1||{\hat v}_3{\hat v}_1|) cos ({(\Omega_{\parallel} -\Omega_{L})z})\Big]. 
\end{align}
\begin{align}
U_{\parallel L} &=  2 \Big[|{\hat u}_1 {\hat u}_2|(|{\hat w}_1{\hat v}_2|+|{\hat w}_2{\hat v}_1|) 
 cos ({(\Omega_{\parallel} -\Omega_{\perp})z})
+ |{\hat u}_2 {\hat u}_3|(|{\hat w}_2{\hat v}_3|-|{\hat w}_3{\hat v}_2|) 
 cos ({(\Omega_{\perp} -\Omega_{L})z})\nonumber\\
&-  |{\hat u}_3 {\hat u}_1|(|{\hat w}_3{\hat v}_1|-|{\hat w}_1{\hat v}_3|) 
 cos ({(\Omega_{\parallel} -\Omega_{L})z})- (|{\hat u}_1|^2|{\hat v}_2 {\hat w}_2| + |{\hat u}_2|^2|{\hat v}_1 {\hat w}_1|)\;cos ({(\Omega_{\parallel} -\Omega_{\perp})z})\nonumber\\ 
 &+ (|{\hat u}_2|^2|{\hat v}_3 {\hat w}_3| - |{\hat u}_3|^2|{\hat v}_2 {\hat w}_2|)\;cos ({(\Omega_{\perp} -\Omega_{L})z})
-  (|{\hat u}_3|^2|{\hat v}_1 {\hat w}_1| - |{\hat u}_1|^2|{\hat v}_3 {\hat w}_3|)\;cos ({(\Omega_{\perp} -\Omega_{L})z})\nonumber\\ 
 &- |{\hat u}_1|^2|{\hat v}_4{\hat w}_4 \;cos ({(\Omega_{L} -\Omega_{\phi'})z})
- |{\hat u}_2|^2|{\hat v}_4{\hat w}_4 \;cos ({(\Omega_{\perp} -\Omega_{\phi'})z}) 
-|{\hat u}_3|^2|{\hat v}_4{\hat w}_4 \;cos ({(\Omega_{\phi'} -\Omega_{L})z})\Big].
 \nonumber 
\end{align}
\begin{align}
U_{\perp L}&= \Big[ (|{\hat v}_1|^2|{\hat u}_2 {\hat w}_2| - |{\hat v}_2|^2|{\hat u}_1 {\hat w}_1|)\;sin ({(\Omega_{\parallel} -\Omega_{\perp})z}) - (|{\hat v}_2|^2|{\hat u}_3 {\hat w}_3| + |{\hat v}_3|^2|{\hat u}_2 {\hat w}_2|)\;sin ({(\Omega_{\perp} -\Omega_{L})z}) \nonumber\\
&-  (|{\hat v}_3|^2|{\hat u}_1 {\hat w}_1| + |{\hat v}_1|^2|{\hat u}_3 {\hat w}_3|)\;sin ({(\Omega_{\parallel} -\Omega_{L})z}) + |{\hat u}_1 {\hat w}_1 ||{\hat v}_4|^2 \;sin ({(\Omega_{\phi'} -\Omega_{\parallel})z})\nonumber\\ 
&-|{\hat u}_2 {\hat w}_2 ||{\hat v}_4|^2 \;sin ({(\Omega_{\perp} -\Omega_{\phi'})z}) + |{\hat u}_3 {\hat w}_3 ||{\hat v}_4|^2 \;sin ({(\Omega_{L} -\Omega_{\phi'})z})\nonumber\\   
&- (|{\hat v}_1 {\hat v}_2|(|{\hat u}_1{\hat w}_2|-|{\hat u}_2{\hat w}_1|) 
 sin ({(\Omega_{\parallel} -\Omega_{\perp})z})
+ (|{\hat v}_2 {\hat v}_3|(|{\hat u}_2{\hat w}_3|+|{\hat u}_3{\hat w}_2|) 
 sin ({(\Omega_{\perp} -\Omega_{L})z}) \nonumber\\
 &+ (|{\hat v}_3 {\hat v}_1|(|{\hat u}_3{\hat w}_1|+|{\hat u}_1{\hat w}_3|) 
 sin ({(\Omega_{\parallel} -\Omega_{L})z}) 
 +|{\hat u}_1 {\hat v}_1 ||{\hat v}_4 {\hat w}_4| sin ({(\Omega_{\phi'} -\Omega_{\parallel})z}) \nonumber\\
 &-|{\hat u}_2 {\hat v}_2 ||{\hat v}_4 {\hat w}_4| sin ({(\Omega_{\perp} -\Omega_{\phi'})z}) 
-|{\hat u}_3 {\hat v}_3 ||{\hat v}_4 {\hat w}_4| sin ({(\Omega_{L} -\Omega_{\phi'})z})\Big]. 
\end{align}

\begin{align}
V_{\parallel} &= 2\Big[|{\hat u}_1|^2|{\hat u}_1 {\hat v}_1| + |{\hat u}_2|^2|{\hat u}_2 {\hat v}_2| + |{\hat u}_3|^2|{\hat u}_3 {\hat v}_3| 
+ |{\hat u}_1||{\hat u}_2|(|{\hat u}_1 {\hat v}_2| + |{\hat u}_2 {\hat v}_1|)cos((\Omega_{\parallel} -\Omega_{\perp})z)\nonumber\\
&+ |{\hat u}_2||{\hat u}_3|(|{\hat u}_2 {\hat v}_3| + |{\hat u}_3 {\hat v}_2|)cos((\Omega_{\perp} -\Omega_{L})z)
+ |{\hat u}_3||{\hat u}_1|(|{\hat u}_3 {\hat v}_1| + |{\hat u}_1 {\hat v}_3|)cos((\Omega_{\parallel} -\Omega_{L})z) \Big].\nonumber
\end{align} 
\begin{align}
V_{\perp}&=2\Big[|{\hat v}_1|^2|{\hat u}_1 {\hat v}_1| + |{\hat v}_2|^2|{\hat u}_2 {\hat v}_2| + |{\hat v}_3|^2|{\hat u}_3 {\hat v}_3|+ |{\hat v}_1 {\hat v}_2|(|{\hat u}_1 {\hat v}_2|+|{\hat u}_2 {\hat v}_1|) cos ((\Omega_{\parallel} -\Omega_{\perp})z)\nonumber\\  
&+|{\hat v}_2 {\hat v}_3|(|{\hat u}_2 {\hat v}_3|+|{\hat u}_3 {\hat v}_2|)cos ((\Omega_{\perp} -\Omega_{L})z) 
+|{\hat v}_3 {\hat v}_1|(|{\hat u}_3 {\hat v}_1|+|{\hat u}_1 {\hat v}_3|) cos((\Omega_{L} -\Omega_{\parallel})z)  \nonumber\\ 
&+|{\hat u}_1 {\hat v}_1||{\hat v}_4|^2 cos \left((\Omega_{\phi'} -\Omega_{\parallel})z \right)
 +|{\hat u}_3 {\hat v}_3||{\hat v}_4|^2 cos ((\Omega_{L} -\Omega_{\phi'})z)\Big].
\end{align}
\begin{align}
V_{L} &= 2\Big[|{\hat w}_1|^2|{\hat u}_1 {\hat v}_1| + |{\hat w}_2|^2|{\hat u}_2 {\hat v}_2| + |{\hat w}_3|^2|{\hat u}_3 {\hat v}_3|+ |{\hat w}_1 {\hat w}_2|(|{\hat u}_1 {\hat v}_2|+|{\hat u}_2 {\hat v}_1|) cos ((\Omega_{\parallel} -\Omega_{\perp})z)\nonumber\\  
&-|{\hat w}_2 {\hat w}_3|(|{\hat u}_2 {\hat v}_3|+|{\hat u}_3 {\hat v}_2|)cos ((\Omega_{\perp} -\Omega_{L})z) 
-|{\hat w}_3 {\hat w}_1|(|{\hat u}_3 {\hat v}_1|+|{\hat u}_1 {\hat v}_3|) cos((\Omega_{L} -\Omega_{\parallel})z)\nonumber\\ 
&+|{\hat v}_4{\hat w}_4|\big[|{\hat u}_1 {\hat w}_1|cos \left((\Omega_{\phi'} -\Omega_{\parallel})z \right)
 +|{\hat u}_2 {\hat w}_2|cos ((\Omega_{\perp} -\Omega_{\phi'})z)
 -|{\hat u}_3 {\hat w}_3|cos ((\Omega_{L}-\Omega_{\phi'})z)
 \big]\Big].\nonumber 
\end{align}
\begin{align}
V_{\parallel \perp} &= 2\Big[(|{\hat u}_2|^2|{\hat v}_1|^2 - |{\hat u}_1|^2|{\hat v}_2|^2)\;sin ({(\Omega_{\parallel} -\Omega_{\perp})z}) 
+ (|{\hat u}_3|^2|{\hat v}_2|^2 - |{\hat u}_2|^2|{\hat v}_3|^2)\;sin({(\Omega_{\perp} -\Omega_{L})z})\nonumber\\
&+ (|{\hat u}_1|^2|{\hat v}_3|^2 - |{\hat u}_3|^2|{\hat v}_1|^2)\;sin ({(\Omega_{L} -\Omega_{\parallel})z})
+ |{\hat u}_1|^2|{\hat v}_4|^2\;sin ({(\Omega_{\phi'} -\Omega_{\parallel})z})\nonumber\\
&-|{\hat u}_2|^2|{\hat v}_4|^2\;sin ({(\Omega_{\perp} -\Omega_{\phi'})z})
- |{\hat u}_3|^2|{\hat v}_4|^2\;sin ({(\Omega_{L} -\Omega_{\phi'})z})
\Big].
\end{align}
\begin{align}
V_{\parallel L} &=2 \Big[|{\hat u}_1 {\hat u}_2|(|{\hat w}_2{\hat v}_1|-|{\hat w}_1{\hat v}_2|) 
 sin ({(\Omega_{\parallel} -\Omega_{\perp})z})
- |{\hat u}_2 {\hat u}_3|(|{\hat w}_3{\hat v}_2|+|{\hat w}_2{\hat v}_3|) 
 sin ({(\Omega_{\perp} -\Omega_{L})z})\nonumber\\
&- |{\hat u}_3 {\hat u}_1|(|{\hat w}_1{\hat v}_3|+|{\hat w}_3{\hat v}_1|) 
 sin ({(\Omega_{\parallel} -\Omega_{L})z})- |{\hat u}_1|^2|{\hat v}_4 {\hat w}_4| sin ({(\Omega_{\parallel} -\Omega_{\phi'})z})\nonumber\\ 
 &+ |{\hat u}_2|^2|{\hat v}_4 {\hat w}_4|\;sin ({(\Omega_{\perp} -\Omega_{\phi'})z})
- (|{\hat u}_3|^2|{\hat v}_4 {\hat w}_4|\;sin ({(\Omega_{L} -\Omega_{\phi'})z})\nonumber\\ 
&+(|{\hat u}_1|^2|{\hat v}_2 {\hat w}_2| - |{\hat u}_2|^2|{\hat v}_1 {\hat w}_1|)\;sin ({(\Omega_{\parallel} -\Omega_{\perp})z})-(|{\hat u}_2|^2|{\hat v}_3 {\hat w}_3| + |{\hat u}_3|^2|{\hat v}_2 {\hat w}_2|)\;sin ({(\Omega_{\perp} -\Omega_{L})z})  \nonumber\\ 
 &- (|{\hat u}_3|^2|{\hat v}_1 {\hat w}_1| + |{\hat u}_1|^2|{\hat v}_3 {\hat w}_3|)\;sin ({(\Omega_{\parallel} -\Omega_{L})z}) \Big]. 
\end{align}
\begin{align}
V_{\perp L}&=2 \Big[2|{\hat v}_3|^2|{\hat u}_3 {\hat w}_3|-2|{\hat v}_2|^2|{\hat u}_2 {\hat w}_2|-2|{\hat v}_1|^2|{\hat u}_1 {\hat w}_1| - (|{\hat v}_2|^2|{\hat u}_1 {\hat w}_1| + |{\hat v}_1|^2|{\hat u}_2 {\hat w}_2|)\;cos ({(\Omega_{\parallel} -\Omega_{\perp})z}) \nonumber\\
&- (|{\hat v}_3|^2|{\hat u}_2 {\hat w}_2| - |{\hat v}_2|^2|{\hat u}_3 {\hat w}_3|)\;cos ({(\Omega_{\perp} -\Omega_{L})z}) 
+ (|{\hat v}_1|^2|{\hat u}_3 {\hat w}_3| - |{\hat v}_3|^2|{\hat u}_1 {\hat w}_1|)\;cos ({(\Omega_{\parallel} -\Omega_{L})z})\nonumber\\
 &- |{\hat u}_1 {\hat w}_1 ||{\hat v}_4|^2 \;cos ({(\Omega_{\phi'} -\Omega_{\parallel})z}) 
-|{\hat u}_2 {\hat w}_2 ||{\hat v}_4|^2 \;cos ({(\Omega_{\perp} -\Omega_{\phi'})z}) + |{\hat u}_3 {\hat w}_3 ||{\hat v}_4|^2 \;cos ({(\Omega_{L} -\Omega_{\phi'})z})\nonumber\\   
&-(|{\hat v}_1 {\hat v}_2|(|{\hat u}_1{\hat w}_2|+|{\hat u}_2{\hat w}_1|) 
 cos ({(\Omega_{\parallel} -\Omega_{\perp})z})
+ (|{\hat v}_2 {\hat v}_3|(|{\hat u}_2{\hat w}_3|-|{\hat u}_3{\hat w}_2|) 
 cos ({(\Omega_{\perp} -\Omega_{L})z}) \nonumber\\
&-  (|{\hat v}_3 {\hat v}_1|(|{\hat u}_3{\hat w}_1|-|{\hat u}_1{\hat w}_3|) 
 cos ({(\Omega_{\parallel} -\Omega_{L})z}) 
 -|{\hat u}_1 {\hat v}_1 ||{\hat v}_4 {\hat w}_4|cos ({(\Omega_{\phi'} -\Omega_{\parallel})z}) \nonumber\\
 &- |{\hat u}_2 {\hat v}_2 ||{\hat v}_4 {\hat w}_4|cos ({(\Omega_{\perp} -\Omega_{\phi'})z}) 
-|{\hat u}_3 {\hat v}_3 ||{\hat v}_4 {\hat w}_4| cos ({(\Omega_{L} -\Omega_{\phi'})z})\Big]. 
\end{align}

\indent
It is worth noting that, due to the appearance of the longitudinal degree of photon in mixing, the Stokes parameters for this case, augmented with extra longitudinal components, denoted with suffix L for each of the variables {\bf I, Q, U,V}. 
{\em Primarily}, it is due to this reason, that the Lorentz and ${\bf C}{\bf P} {\bf T}$ symmetry of each degree of 
freedom (i.e., $A_{\parallel}, A_{\perp}, A_{L}, \phi^{\prime}$) -- that constitutes every term in any equation  of motion--follows  
from the same (symmetry) enjoyed by the interaction vertex (IL)
\footnote{ The word symmetries here mean the discrete and continuous symmetries, internal or space time ( those following  from the actions of the generators of the symmetry group: 
constituting the set of orthogonal proper Lorentz transformations (OPLT )) that the  associated  dynamical and background-fields respect}. That is, the transformation properties of a vertex, under a set of 
discrete symmetry transformations, is decided by 
the actions of the same, i.e., --- on the fields that constitute
the vertex. In other words, under the actions of  {\bf C},  {\bf P} and {\bf T} on the  background  field ($\bar{F}^{\mu\nu}$), dynamical photons field 
($f_{\mu\nu}$) the scalar field $\phi$  or the pseudoscalar field $\phi'$ -- 
 that collectively constitutes the vertex.

\section{Statistical  analysis}
\subsection{Normal distributions from numerical solutions }
We have plotted the Stokes parameters in figures [\ref{I1}-\ref{V1}]  and  the ellipticity angle $\chi_{\phi_{i}}$ and polarization angle $ \psi_{\phi_{i}}$ (in radians) in figures [\ref{E1}-\ref{P1}] using the {\it exact numerical solutions} obtained by  diagonalizing  (\ref{mat}) and  (\ref{mat-axion-photon}) exactly for the following system parameters  $m_{\phi} = m_{\phi^{\prime}} = 1.0\times10^{-11}$ GeV$^{-1}$, $g_{\phi,\phi^{\prime}\gamma\gamma} = 1.0\times10^{-11}$  GeV$^{-1}$, B $\sim$ 10$^{12}$ Gauss, $\omega_{p} = 1.0 \times 10^{-10}$ GeV, in energy range (1-100) KeV, path length $z = 22.3$ meters,  for both eqns.[\ref{mat}] and  [\ref{mat-axion-photon}]. 
\begin{figure}[h!]
\begin{minipage}[b]{.48\textwidth}  
\hskip .5 cm \includegraphics[width=1\linewidth]{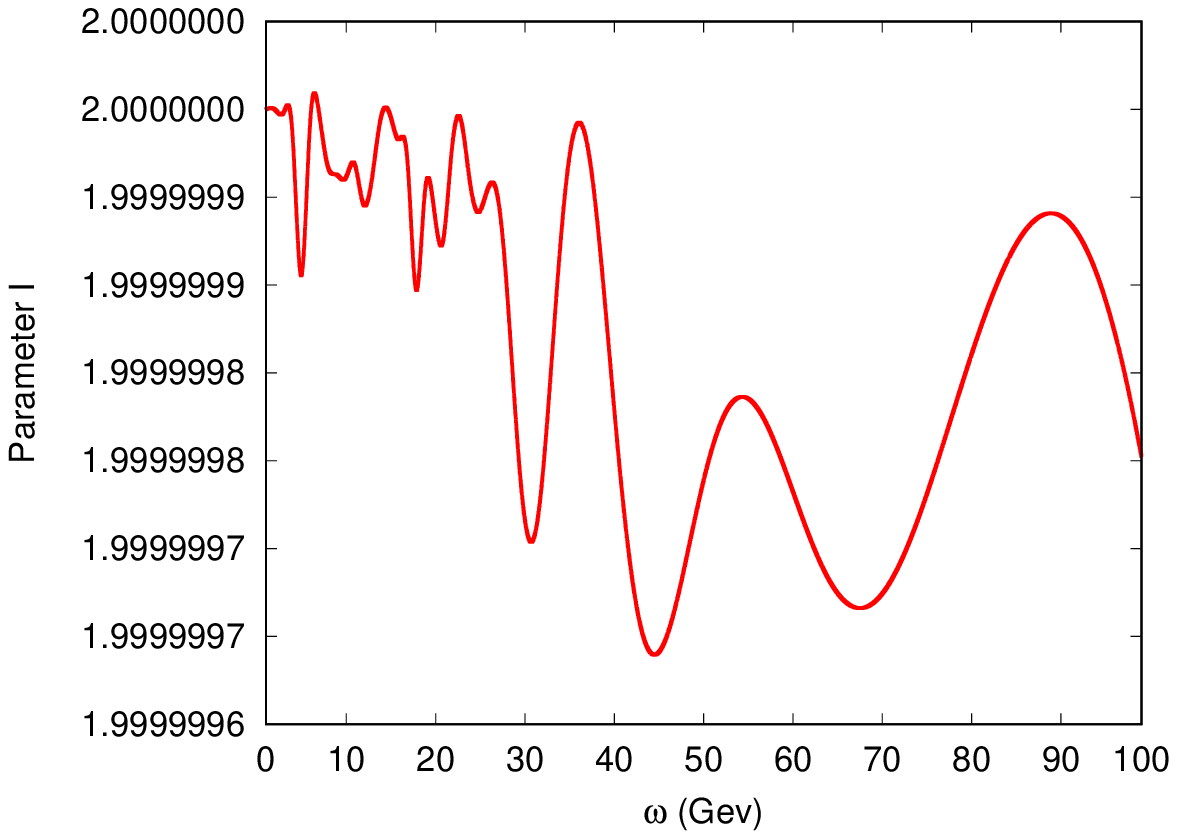}
\end{minipage}
\begin{minipage}[b]{.48\textwidth}  
\hskip .5 cm\includegraphics[width=1\linewidth]{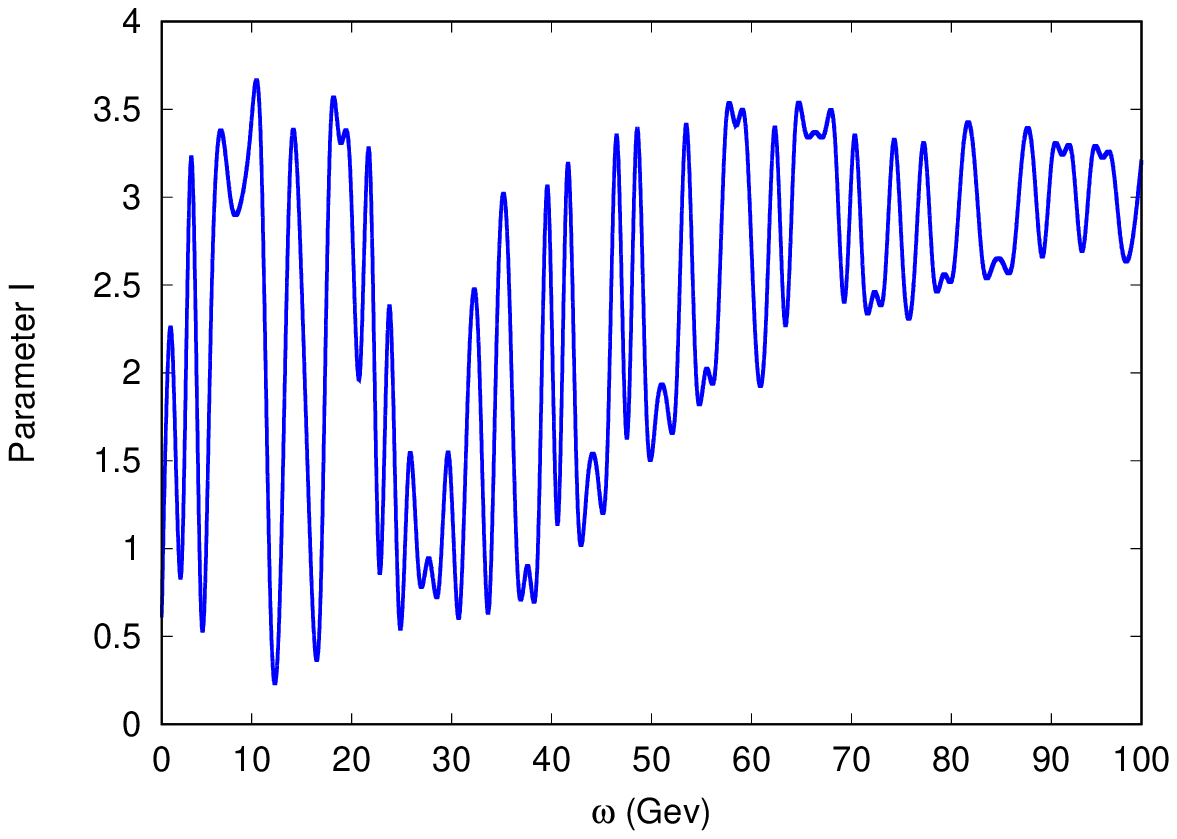}
\end{minipage}
\vskip 1.0cm
\caption{Stokes parameter I for  $\phi\gamma$ and $\phi^{\prime}\gamma$ .}   
\label{I1}
\end{figure}

\begin{figure}[h!]
\begin{minipage}[b]{.48\textwidth}  
\hskip .5 cm\includegraphics[width=1\linewidth]{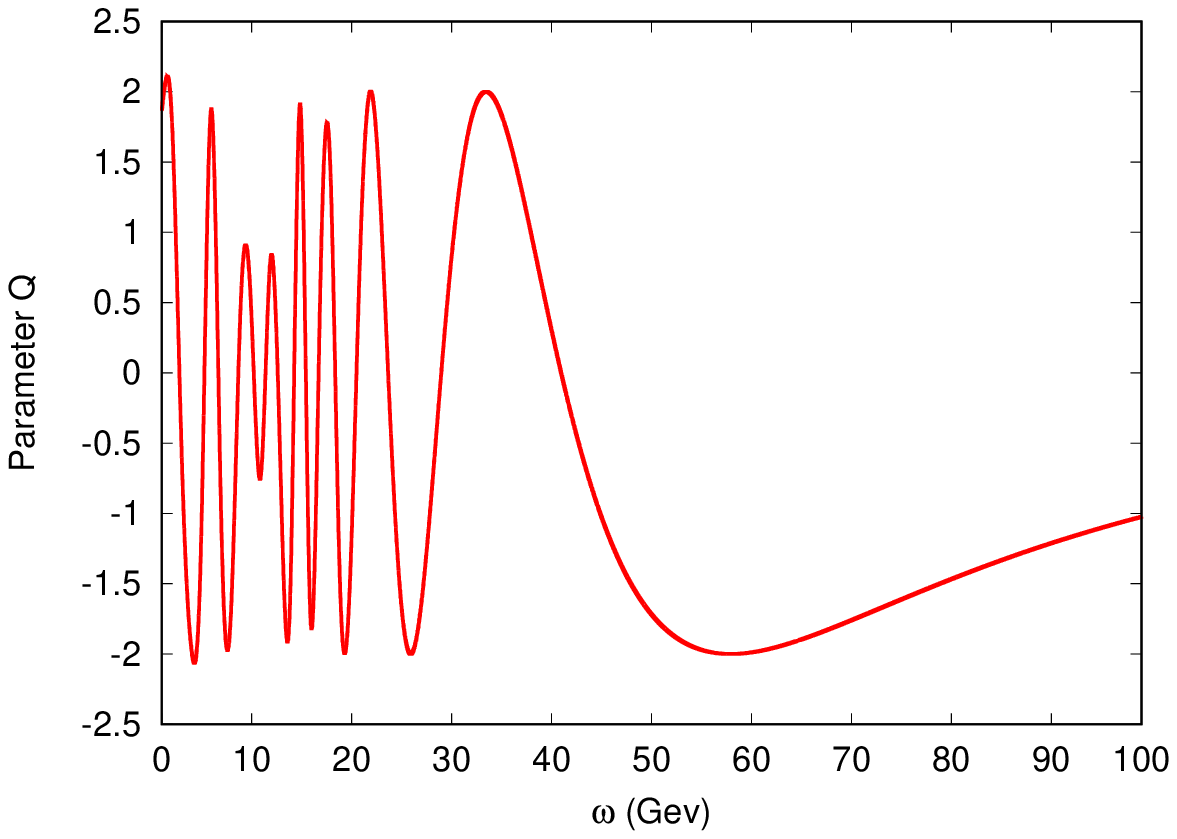}
\end{minipage}
\begin{minipage}[b]{.48\textwidth}  
\hskip .5 cm\includegraphics[width=1\linewidth]{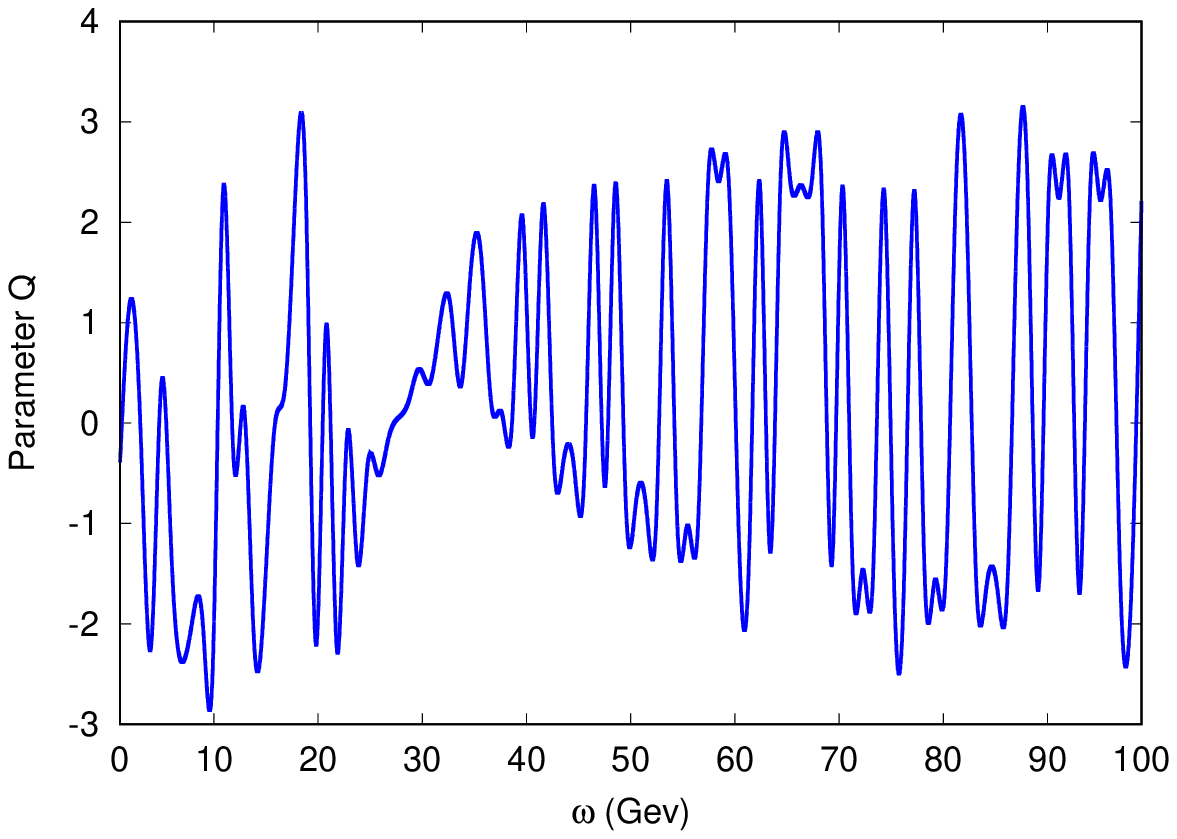}
\end{minipage}
\vskip 1.0 cm\caption{Stokes parameter Q for  $\phi\gamma$ and $\phi^{\prime}\gamma$ .}   
\label{Q1}
\end{figure}

\begin{figure}[h!]
\begin{minipage}[b]{.48\textwidth}  
\hskip .5 cm\includegraphics[width=1\linewidth]{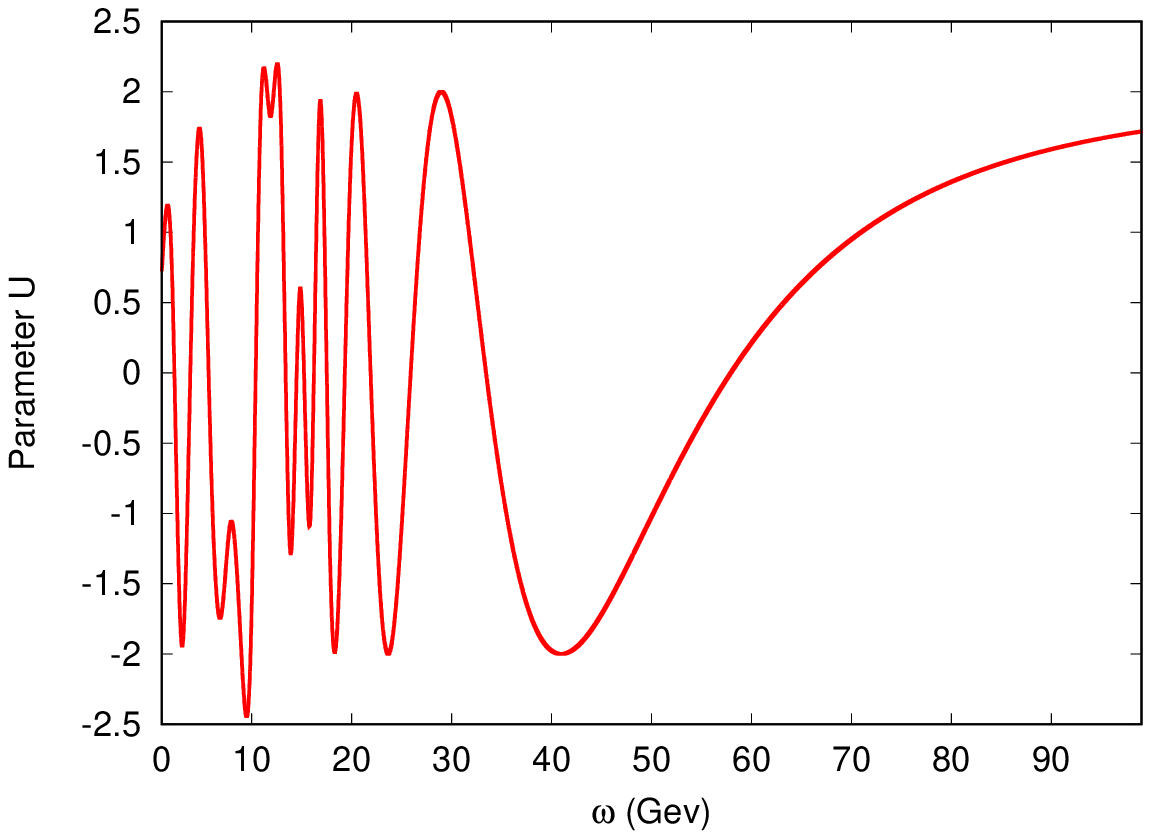}
\end{minipage}
\begin{minipage}[b]{.48\textwidth}  
\hskip .5 cm\includegraphics[width=1\linewidth]{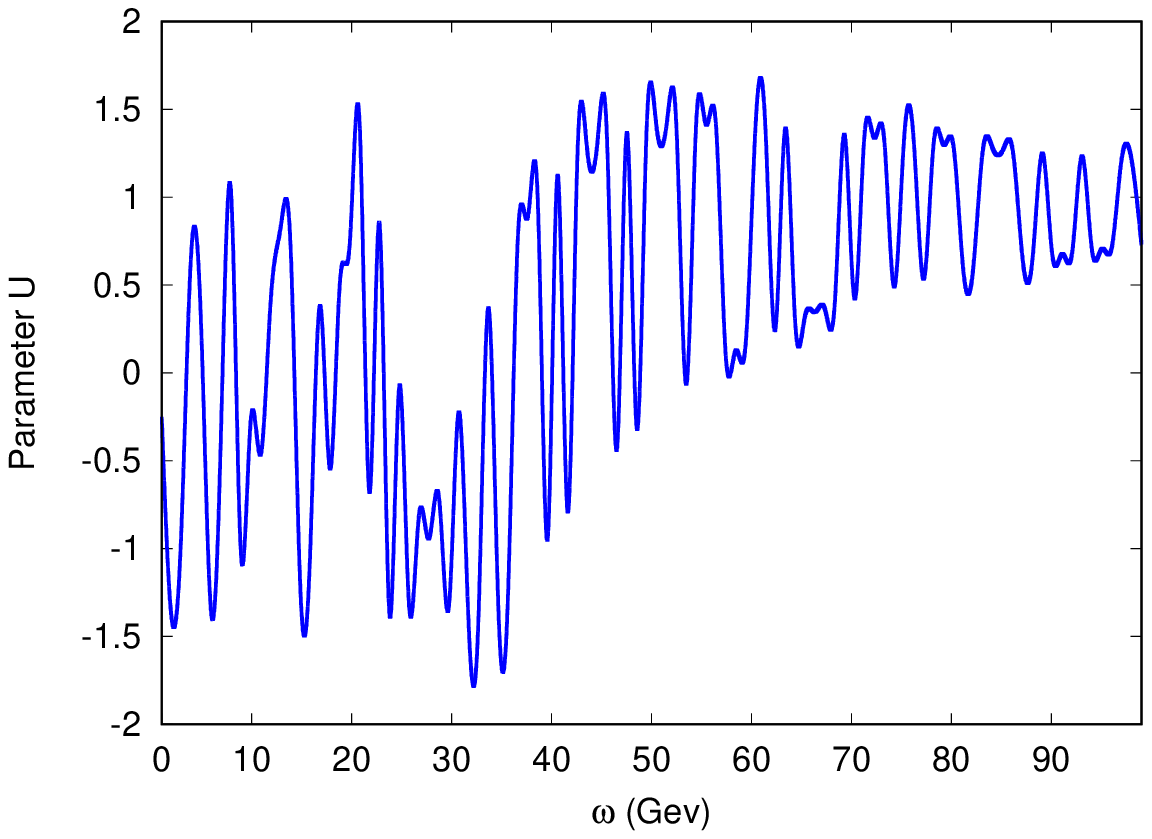}
\end{minipage}
\vskip 1.0 cm\caption{Stokes parameter U for  $\phi\gamma$ and $\phi^{\prime}\gamma$ .}
\label{U1}   
\end{figure}

\begin{figure}[h!]
\begin{minipage}[b]{.48\textwidth}  
\hskip .5 cm\includegraphics[width=1\linewidth]{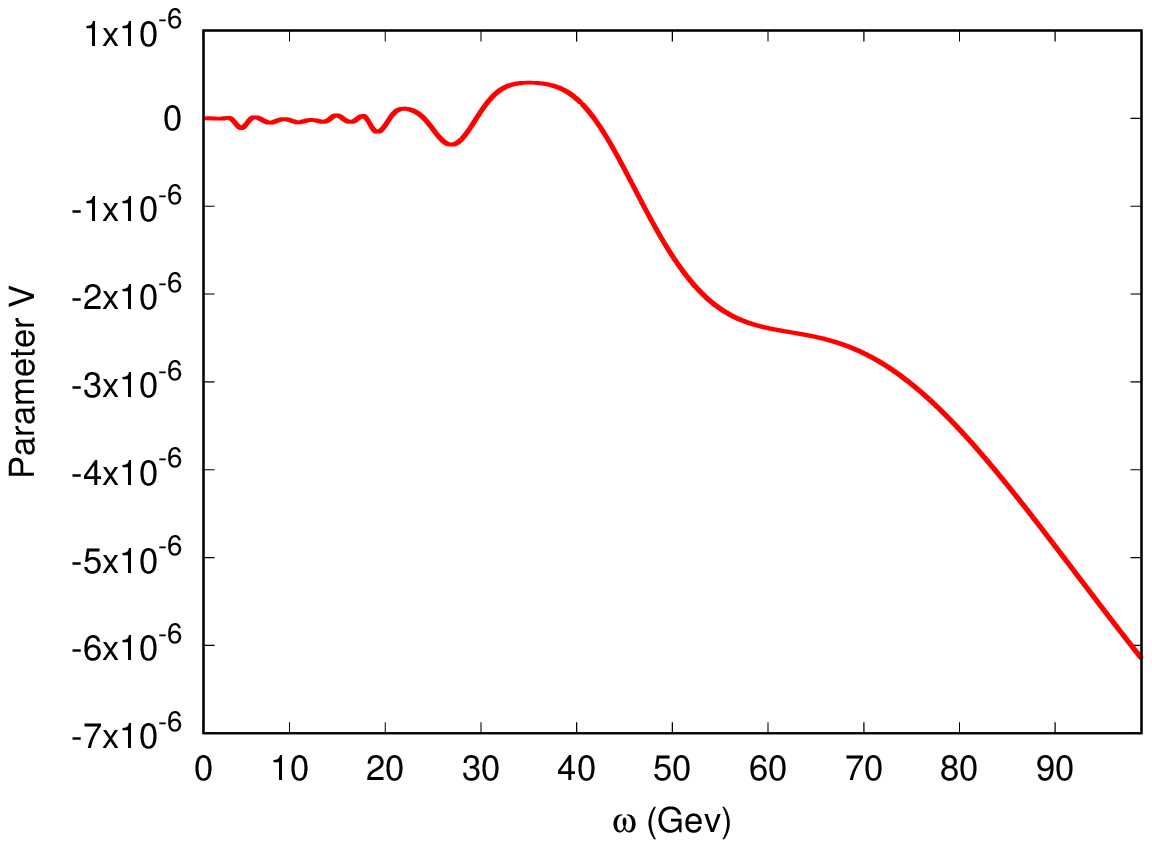}
\end{minipage}
\begin{minipage}[b]{.48\textwidth}  
\hskip .5 cm\includegraphics[width=1\linewidth]{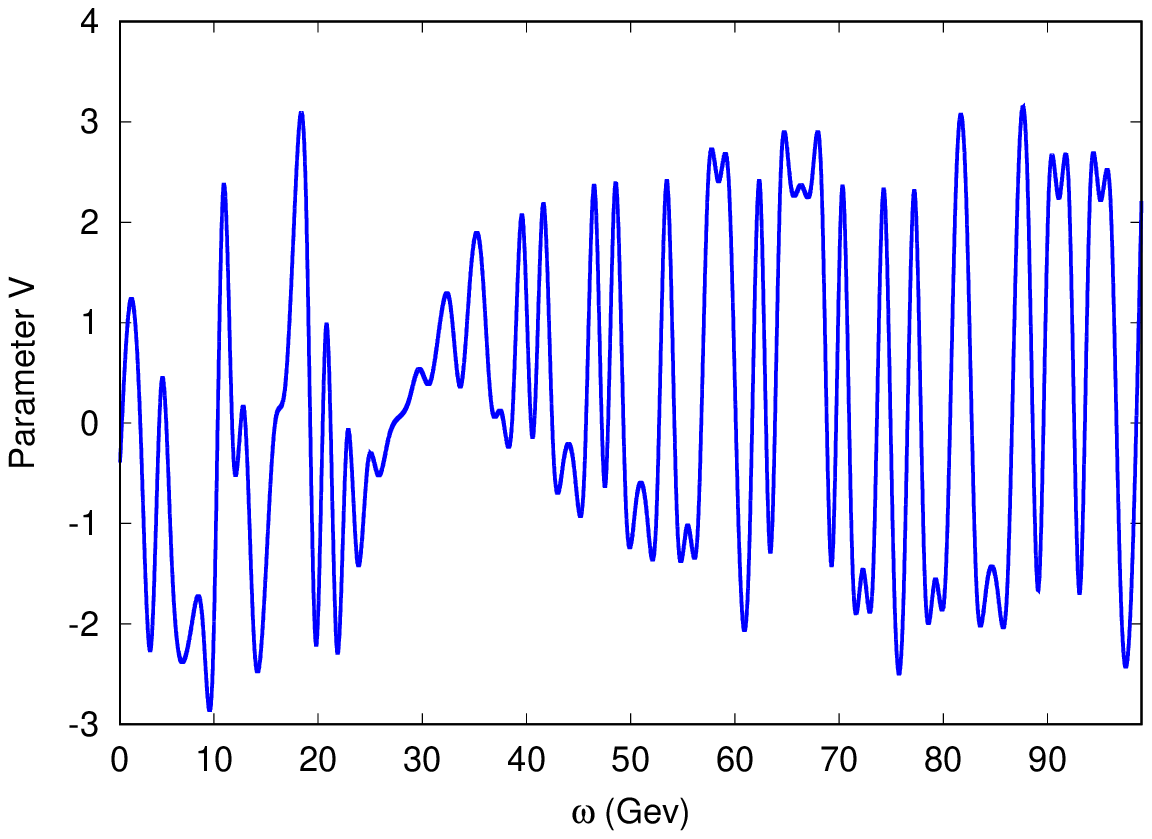}
\end{minipage}
\vskip 1.0 cm\caption{Stokes parameter V for  $\phi\gamma$ and $\phi^{\prime}\gamma$ .}   
\label{V1}
\end{figure}

\begin{figure}[h!]
\begin{minipage}[b]{.48\textwidth}  
\hskip .5 cm\includegraphics[width=1\linewidth]{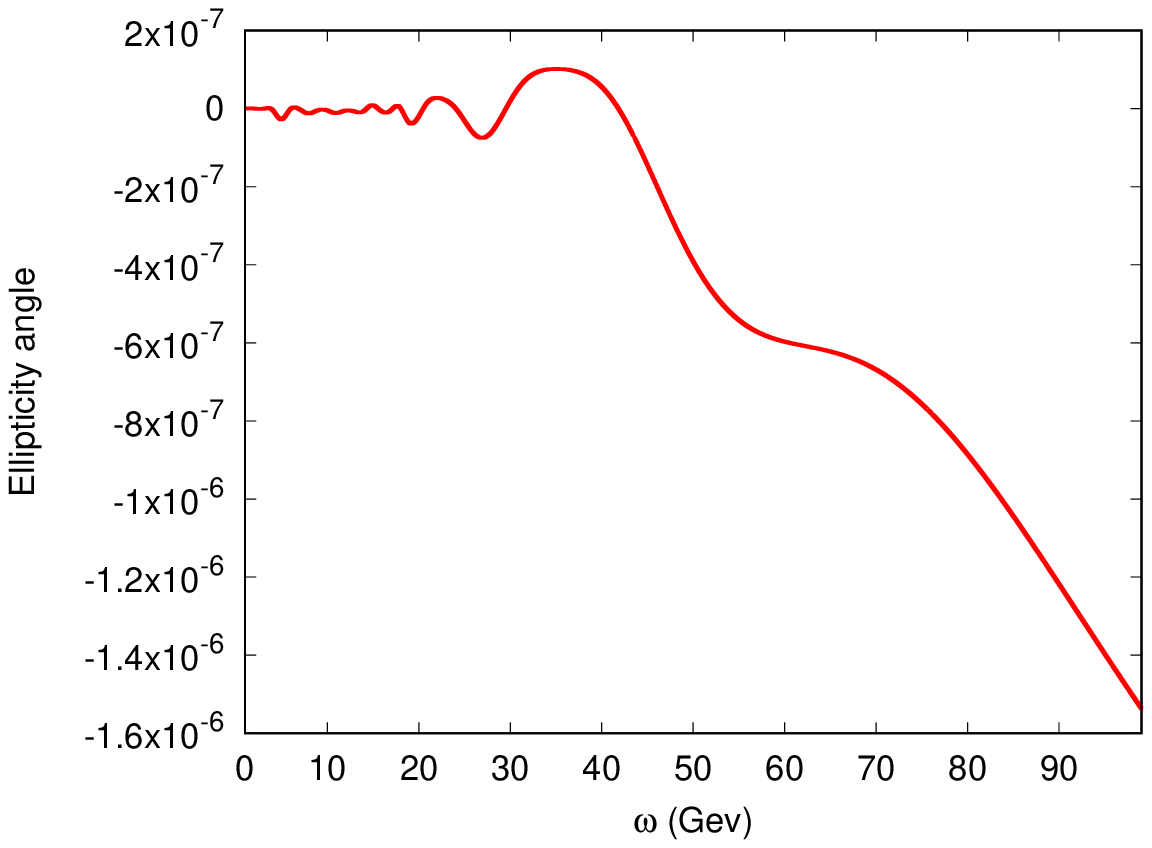}
\end{minipage}
\begin{minipage}[b]{.48\textwidth}  
\hskip .5 cm\includegraphics[width=1\linewidth]{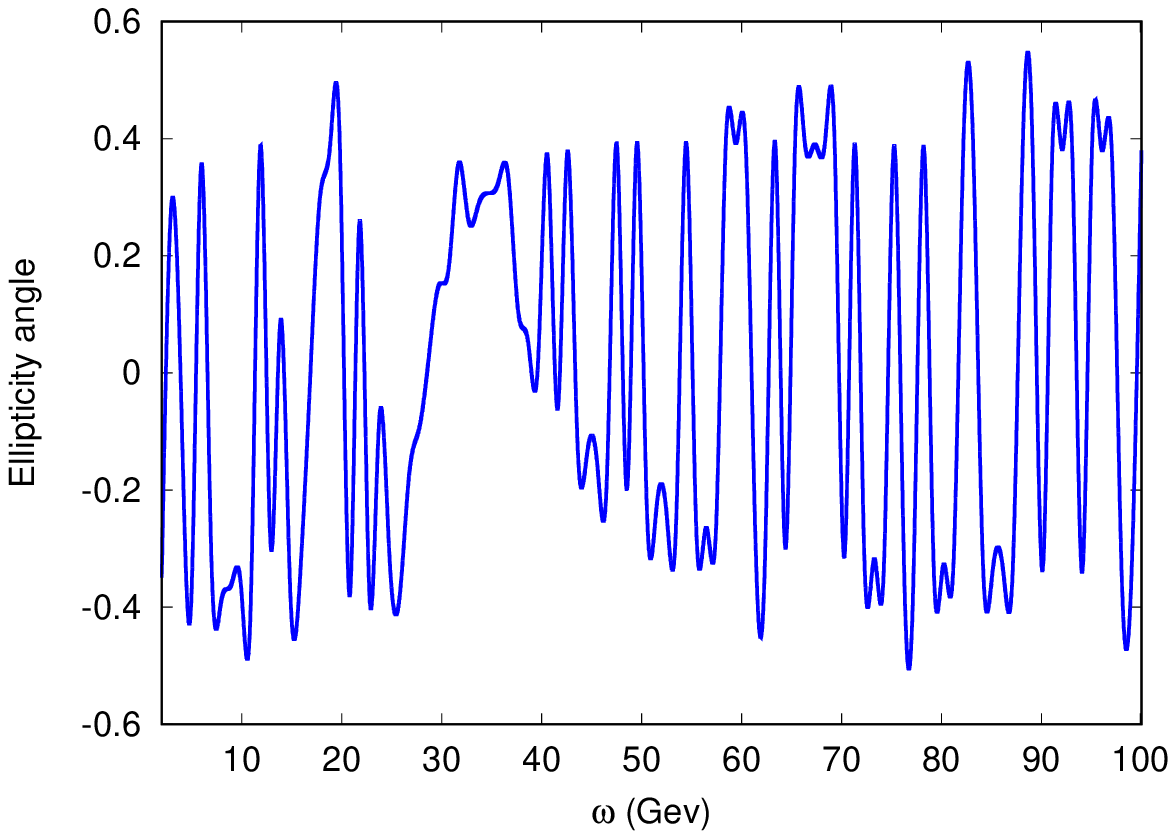}
\end{minipage}
\vskip 1.0 cm\caption{Ellipticity angle  for  $\phi\gamma$ and $\phi^{\prime}\gamma$ .}   
\label{E1}
\end{figure}

\begin{figure}[h!]
\begin{minipage}[b]{.48\textwidth}  
\hskip .5 cm\includegraphics[width=1\linewidth]{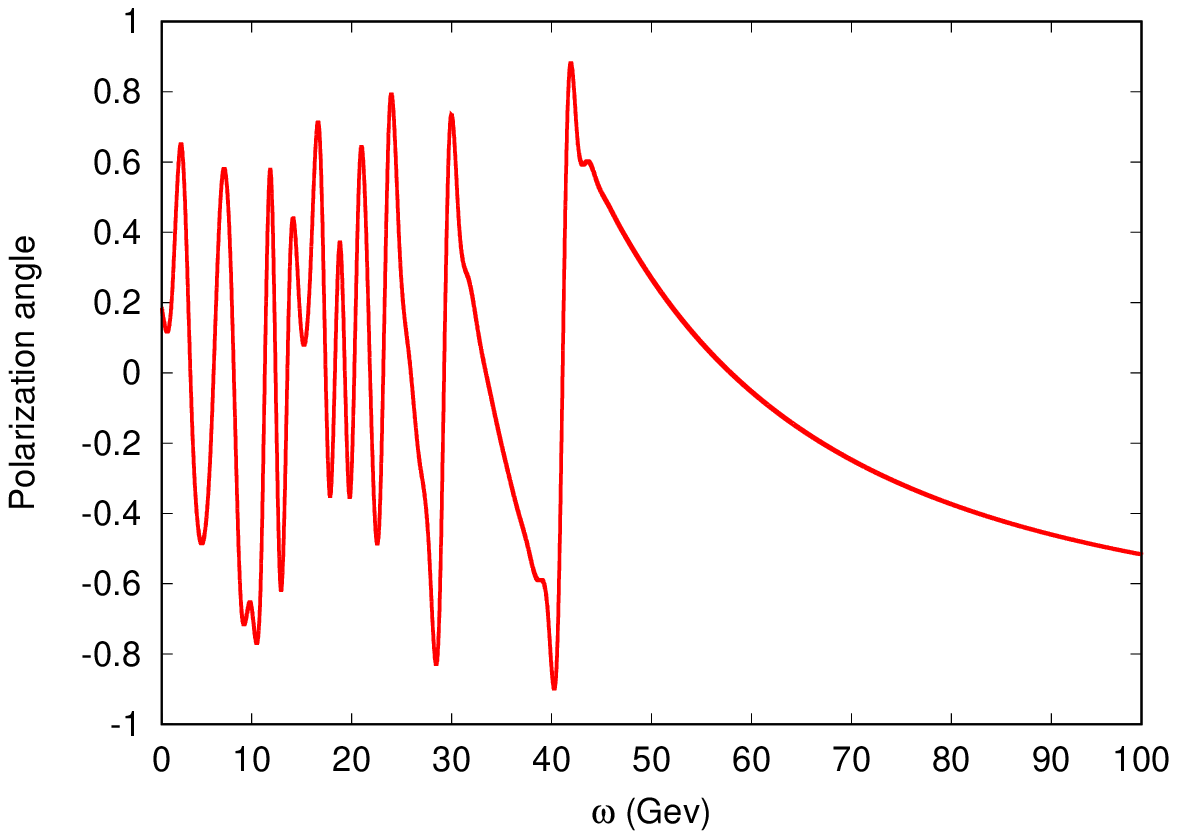}
\end{minipage}
\begin{minipage}[b]{.48\textwidth}  
\hskip .5 cm\includegraphics[width=1\linewidth]{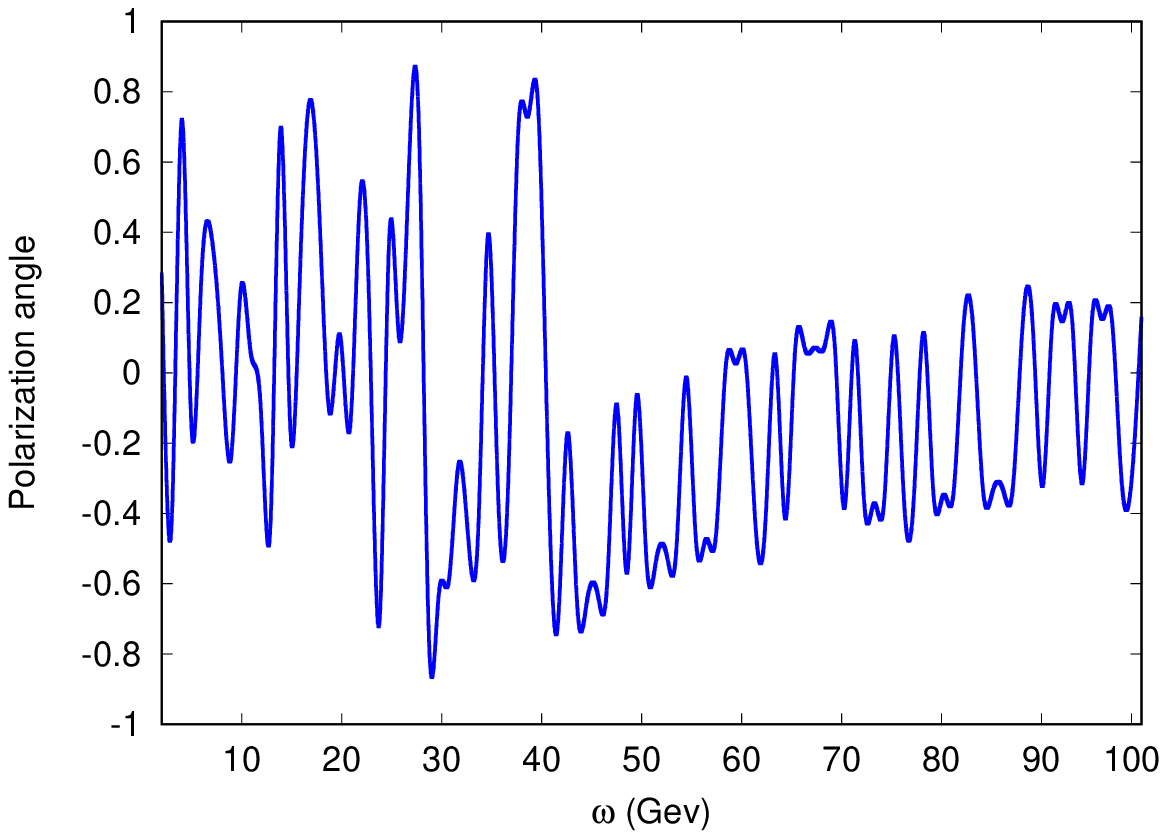}
\end{minipage}
\vskip 1.0 cm\caption{Polarization angle  for  $\phi\gamma$ and $\phi^{\prime}\gamma$ .}  
\label{P1} 
\end{figure}

\begin{figure}[h!]
\begin{minipage}[b]{.48\textwidth}  
\hskip -.5 cm\includegraphics[width=1\linewidth]{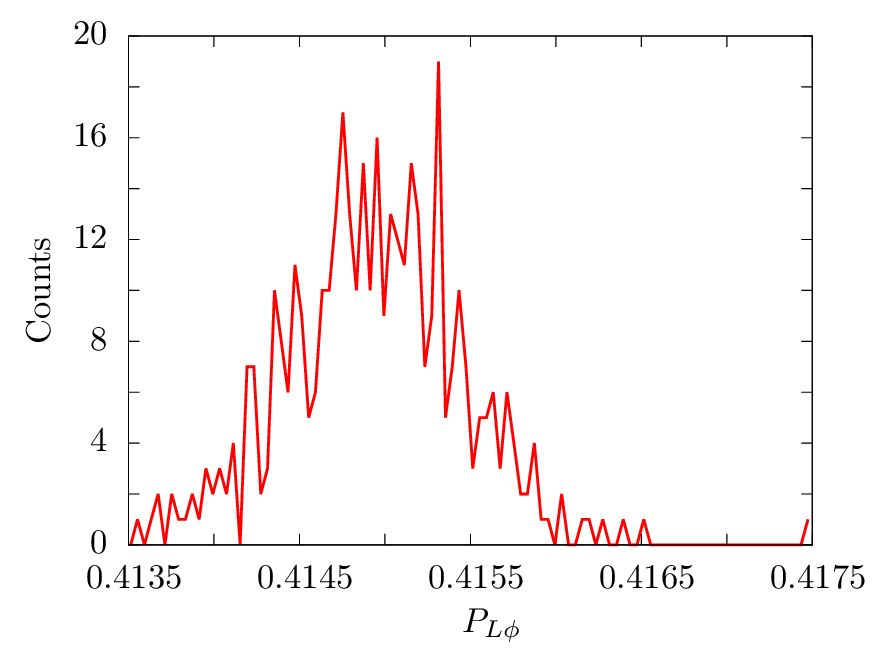}
\end{minipage}
\begin{minipage}[b]{.48\textwidth}  
\hskip -.5 cm\includegraphics[width=1\linewidth]{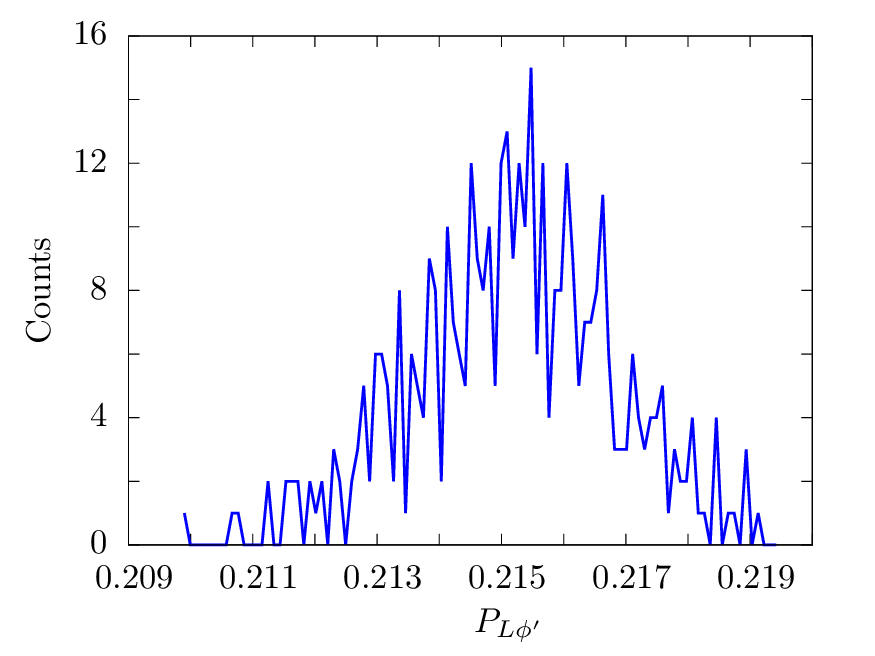}
\end{minipage}
\vskip .10 cm\caption{Plots of degree of linear polarization vs counts. The parameters for this plots are; $m_{\phi} = m_{\phi^{\prime}} = 1.0\times10^{-11}$ GeV$^{-1}$, $g_{\phi,\phi^{\prime}\gamma\gamma} = 1.0\times10^{-11}$  GeV
$^{-1}$, B $\sim$ 10$^{12}$ Gauss, $\omega_{p} = 1.0 \times 10^{-10}$ GeV, in energy range (1-100) KeV, path length $z = 22.3$ meters.}  
\label{gauss-sc-ax} 
\end{figure}

%
%
%
%
The oscillating behaviour of the  Stokes parameters and the observables ($\chi_{\phi_{i}}$ and $\psi_{\phi_{i}}$) as a function of energy is evident from the plots displayed in figures [\ref{I1}-\ref{P1}].
None of the current space based observations confirm the same and we believe the proposed observations would not be able to detect the same. The chief reason behind this being the space based detectors are mostly sensitive over an energy band not a line.\\
\indent
So, in the spirit of the analysis of the observational   X-ray  polarization data, we  divide initially the window of the energy band under consideration i.e., $1-100$ KeV   into $10000$   intervals and evaluated the average of the  parameters ${\bf{ I}}, {\bf{ Q}}$ and ${\bf{U}}$ to get  ${\bf\bar{ I}}, {\bf\bar{ Q}}$ and ${\bf \bar{U}}$ and hence the average of degree of linear polarization ${P_{L\phi_{i}}}$  to get the first set of samples. This procedure was followed by increasing $N$ to $N^\prime$ such that $N^{\prime} = N + \Delta N$, when $\Delta N = 100$. We continued this process until $N^{\prime} =N+ n\Delta N$ when $n = 400$. Thus we generated  a samples of $401$ data points for ${P}_{L\phi_{i}}$. \\
\indent
These samples were further distributed on frequency distribution plots for both the systems ($\gamma\phi$, $\gamma\phi^\prime$) by choosing the appropriate bin size along the x-axis. These frequency distributions were further  plotted in fig.[\ref{gauss-sc-ax}]  for scalar photon (in the left panel) as well as pseudoscalar photon (in the right panel) systems.\\
\indent
Assuming the obtained frequency distributions of the samples as normal distributions, we have estimated the (mean) average value  and the variance of the of the $P_{L\phi_{i}}$ obtained in the  energy interval of $1-100$ KeV. The  mean of  $P_{L\phi_{i}}$  for $\gamma\phi$ and $\gamma\phi^\prime$ system   turns out to be $\mu_{\phi} = 0.414$ and $\mu_{\phi^{\prime}} = 0.215 $ and the corresponding variance turns out to be $\sigma^{2}_{\phi} = 0.275 \times 10^{-6} $  and $\sigma^{2}_{\phi^{\prime}} = 0.264 \times 10^{-5}$ respectively. \\
\indent
Statistically, it means in the one sigma level for $\gamma\phi$ system the range  of  $P_{L\phi_{i}}$ lies between $0.414 - 0.275 \times 10^{-6}$ to $0.414 + 0.275 \times 10^{-6}$  and for  $\gamma\phi^\prime$ system it is  $0.215 + 0.264 \times 10^{-5}$  to $0.215 + 0.264 \times 10^{-5} $. As can be seen from the range that with increase in the numbers of dof the range for the confidence level increases by the factor of 10. The C.L analysis for other interaction parameters should  be performed for designing the detectors for space borne experiments.\\

\subsection{MDP }
Many of the astrophysical polarimeters use  minimum detectable polarization  (MDP) to define detector efficiency.
The probability amplitude of polarization that has $1\%$ chance of being detected is called MDP. Usually the expression 
of the same is given by,
\begin{eqnarray}
  MDP = \frac{4.29}{\sqrt{N}\mu},
\end{eqnarray}
when $\mu$ is the modulation factor that can vary between 0 and 1 and $ N$ is the sample size.
Instruments like  {\bf APEX}  has $MDP$ is $\sim 1\%$  at $5.2$ KeV \cite{APEX}. The $MDP$ in terms of   
mean $\mu_{\phi_{i}}$ and $\sigma_{\phi_{i}}$ of the observed data of $P_{L\phi_{i}}$  when background 
is zero and modulation factor is one, can be cast as:
\begin{eqnarray}
  MDP_{\phi_{i}} = \mu_{\phi_{i}} +  \frac{4.29}{\sqrt{2}} \sigma_{\phi_{i}}.
\end{eqnarray}
In our case of investigation, we have found that $MDP_{\phi} > MDP_{\phi^{\prime}}$ for the parameters range under consideration in this work. Thus in the same parameter space the increase in the number of degrees of freedom of the system, the detectability of $P_{L\phi_{i}}$ seems to be reduced. This information should be considered while searching for ALPs in compact star models using observed data.

%
%
%
%

\end{document}